\documentclass[11pt]{article}
\usepackage{epsfig,amssymb,amsmath,psfrag}
\usepackage{slashed}
\usepackage{graphicx} 
\usepackage{hyperref}

\catcode`\@=11

\def \be  {\begin{equation}}
\def \ee  {\end{equation}}
\def \ba  {\begin{eqnarray}}
\def \ea  {\end{eqnarray}}
\def \baa {\begin{eqnarray*}}
\def \eaa {\end{eqnarray*}}
\def \bb  {\begin {thebibliography} }
\def \eb  {\end{thebibliography}}
\def \lab #1 {\label{#1}}

\newcommand\re[1]{(\ref{#1})}

\def \qqqquad {\qquad\qquad}
\def \matrix #1 {\left(\begin{array}{cc} #1 \end{array}\right)}

\def \tr {\mathop{\rm tr}\nolimits}

\def \e  {\mathop{\rm e}\nolimits}
\newcommand\lr[1]{{\left({#1}\right)}}

\newcommand \ket [1] {|{#1}\rangle}

\newcommand{\insertfig}[2]{\mbox{\epsfysize=#1cm \epsfbox{#2.eps}}}%
\newcommand{\ft}[2]{{\textstyle\frac{#1}{#2}}}

\def\XXint#1#2#3{{\setbox0=\hbox{$#1{#2#3}{\int}$}
     \vcenter{\hbox{$#2#3$}}\kern-.5\wd0}}

\def\numberbysection{\@addtoreset{equation}{section}
                     \def\theequation{\thesection.\arabic{equation}}}
\numberbysection

\begin{document}

\renewcommand{\thefootnote}{\fnsymbol{footnote}}

\begin{titlepage}
\begin{flushright}
\begin{tabular}{l}
\end{tabular}
\end{flushright}

\vskip3cm

\begin{center}
 {\Large \bf L\"uscher formula for GKP string}
\end{center}

\vspace{1cm}

\centerline{\sc B.~Basso,$^{1, }$\footnote{bbasso@princeton.edu} ~A.V.~Belitsky$^{2, }$\footnote{andrei.belitsky@asu.edu}}

\vspace{10mm}

\centerline{\it ${}^1$Princeton Center for Theoretical Science,}
\centerline{\it Jadwin Hall, Princeton University,}
\centerline{\it Princeton, NJ 08544, USA}

\vspace{5mm}

\centerline{\it ${}^2$Department of Physics,}
\centerline{\it Arizona State University,}
\centerline{\it Tempe, AZ 85287-1504, USA}

\vspace{1cm}

\centerline{\bf Abstract}

\vspace{5mm}  
  
We investigate finite-size corrections to anomalous dimensions of  large-spin twist-two operators in the planar maximally supersymmetric 
Yang-Mills theory. We develop a framework for analysis of these corrections, that is complementary to the conventional spin-chain
approach, by making use of the hole rather than the magnon picture. From the dual string theory perspective where the large-spin operator is 
identified with the Gubser-Klebanov-Polyakov (GKP) string, our approach is equivalent to constructing the first L\"uscher correction to the 
energy of the GKP string by incorporating the contribution of virtual excitations propagating on it. It allows us to propose a formula that controls 
a particular class of large-spin corrections to the twist-two anomalous dimension and holds at any value of the coupling constant. Compared to 
wrapping corrections computed with magnons propagating on the spin chain, the finite-size corrections that are encoded in our formalism start 
at a lower loop level. Our formalism thus calls for modification of the asymptotic contributions which are conventionally incorporated within 
the Asymptotic Bethe Ansatz. An educated guess allows us to remedy this pitfall and successfully confront our predictions with known results up 
to five loop accuracy at weak coupling. Finally, our formula sheds light on the weak-to-strong coupling transition for the subleading large-spin 
corrections under study and confirms stringy expectations at strong coupling where they are found to be identical to the first L\"uscher correction 
to the vacuum energy of the O(6) sigma model.

\end{titlepage}

\setcounter{footnote} 0

{\small \tableofcontents}

\newpage

\renewcommand{\thefootnote}{\arabic{footnote}}

\section{Introduction and discussion}\label{ID}

AdS/CFT correspondence~\cite{Mal97} provides a powerful framework for geometrical reformulation of gauge theory dynamics in 
the planar limit in terms of free strings propagating on a nontrivial curved background. For maximally supersymmetric 
Yang-Mills (SYM) theory it allowed one to establish a precise dictionary between field and string theory observables, in particular,
with the spectrum of scaling dimensions of single trace gauge invariant operators in $\mathcal{N}=4$ Yang-Mills theory 
identified with energies of string configurations on AdS$_5 \times$S${}^5$ target space. The weak/strong nature of the 
duality was hampering direct tests of the conjecture since the weak coupling regime on the gauge side corresponded 
to a highly quantum regime for the string sigma model, and inversely the domain when quantum fluctuation on the strings 
are suppressed required techniques for strong coupling expansion of anomalous dimensions of composite operators. The 
difficulty in pursuing the endeavor of testing the duality as a function of parameters defining both sides of the story was 
overcome by identifying hidden integrable structures of the system. On the string theory side, the classical sigma model 
was found to possess an infinite number of conserved charges \cite{KMMZ}. On the gauge theory side, the dilatation operator 
whose eigenvalues define anomalous dimensions was found to correspond to a certain long-range magnet \cite{MZ, BelDerKorMan04, BS05}, 
--- a generalization of an earlier discovery of the XXX Heisenberg spin chain for maximal helicity operators at one loop in 
QCD~\cite{L98, BDM98}.

In spite of this revelation the extraction of exact anomalous dimensions at any value of 't Hooft coupling $g^2 = g_{\rm YM}^2
N_c/(4 \pi)^2$ from the underlying integrable models is still highly nontrivial. However since the energies depend on quantum 
numbers of corresponding states, like spin, one possesses an extra lever arm which makes the calculations manageable in certain 
limits. Namely, significant simplifications of underlying calculations occur when one considers operators with large values 
of spin on either compact or non-compact parts of the AdS$_5 \times$S${}^5$ background, or both~\cite{BMN, GKP02}. Thus one is on the quest 
of finding systematic techniques which provide interpolating formulas between the regimes of weak and strong coupling. 
Being interested in generic properties intrinsic to supersymmetric gauge theories and more importantly not, since 
isometries of the noncompact manifold corresponds to space-time symmetries shared by both types of field theories, 
we will focus on the latter in the present work. This configuration corresponds to the so-called Gubser-Klebanov-Polyakov 
(GKP) string~\cite{GKP02}, which is a classical folded closed string rotating with a large angular momentum $S$ on an AdS$_3$ part of the 
AdS$_5 \times$S${}^5$ space. In this limit, the string of proper length $R = 2 \, \ln S + O(1)$ becomes homogeneous and 
stretches all the way to the boundary when $S \to \infty$. Then its proper energy, as measured in the rotating frame of the 
string, scales with the length $R$,
\be
\label{EgkpScaling}
E_{\rm GKP} = E_{\rm GKP}(g, S) =  w (g) R + \dots
\, .
\ee
where the energy density $w (g)$ is determined classically by the string tension $g$, $w (g)  = 2 g + \dots$. This scaling behavior
is akin the large-spin asymptotics of scaling dimensions of Wilson operators in all gauge theories. While the GKP 
string determines the leading Regge trajectory of the string with a given spin $S$, its gauge theory dual corresponds 
to the twist-two operators, schematically,
\be
\label{Twist2}
\mathcal{O}_{2,S} = {\rm tr} \, Z D_+^S Z + \dots
\, ,
\ee
where $Z$ is one of the complex scalars of $\mathcal{N} = 4$ SYM and $D_+$ is the light-cone projection of the
covariant derivative. The large-spin asymptotics of their scaling dimension $\Delta$ is driven by the cusp 
anomalous dimension~\cite{KR85}%
\footnote{The cusp anomalous dimension $A(g)$ is a function of the coupling constant $g$ only that controls the UV divergences of a cusped Wilson loop~\cite{P80} with a large hyperbolic angle~\cite{KR85}, i.e., in the light-like limit.}
\be
E \equiv \Delta - S = 2 A(g) \ln S + \dots
\, .
\ee
This behavior is known to persist to all orders in 't Hooft coupling constant and reflects the copious emission of soft gluons at the edge of phase space 
\cite{gKM, BGK03}. Thus, the identification $\ket{ \mbox{GKP string} } = \mathcal{O}_{2,S}$, and its corollary $E_{\rm GKP} = E$, provided a stringent 
test for the conjectured duality~\cite{GKP02}. The scaling law \re{EgkpScaling} was shown to hold at higher loop orders within the $1/g$ expansion of 
the string sigma model, with explicit computations performed at one \cite{FT02} and two loops \cite{RTCusp}. 

The interpolating formula arising from the weak coupling expansion on the gauge theory side was deduced by 
studying the large-magnon limit of the long-range spin chain. In the language of the spin chain encoding the integrable 
structure of the dilatation operator in $\mathcal{N} = 4$ SYM, this scaling dimension corresponds to the energy 
of the length $L=2$ chain with $S$ magnon excitations of momenta $p_j$ propagating on it. The integrability implies 
that the total spin-chain energy is a sum of individual excitations with dispersion relation~\cite{BDS} $E(p)$,
\be
\label{ABAdispersion}
 \Delta = 2+ \sum_{j=1}^{S} E(p_j)
\, , \qquad
E(p) = \sqrt{1 + 16 g^2 \sin^2 (p/2)} \, ,
\ee
while the momenta obey Asymptotic Bethe Ansatz (ABA) equations~\cite{BS05}
\be
\label{ABAintro}
{\rm e}^{i L p_k} = \prod_{j \neq k} S (p_j, p_k)
\, ,
\ee
where $S (p_j, p_k)$ is the two-particle scattering matrix. These considerations yielded an integral equation, the so-called Beisert-Eden-Staudacher 
(BES) equation~\cite{BES}, for the density of Bethe roots 
\be
\label{BetheRoots}
u_j = \ft12 \cot (p_j/2)  \sqrt{1 + 16 g^2 \sin^2 (p_j/2)}
\ee
which generates the cusp anomalous dimension. The solution at weak coupling~\cite{BES} matched the result of tedious calculation to fourth order in $g^2$~\cite{BGK03, KLV, KLOV, BernDS, BCDKS}, while much trickier strong-coupling analyses~\cite{AABEK07, BKK07, KSV08, BK09} were found in agreement with the energy density $w(g)$ to two loop order in $1/g$ expansion. Together with the numerical interpolation of 
Ref.~\cite{Benna06}, this confirmed the conjectured duality for non-protected observables, yielding $w (g) =A (g)$.

This success in interpolating between the weak and strong coupling for the cusp anomalous dimension raises the question of how far we can go in 
unravelling the transition between the gauge and string theory at the level of subleading corrections in large spin. It has been known already for 
some time that the approach developed for the cusp anomalous dimension is amenable to generalizations~\cite{FZ09} for the first subleading 
contribution $\sim \log^0{S}$ controlled by the so-called virtual scaling function, $B = 2 + O(g^4)$. The twist-two energy is then found to assume the form
\be\label{ABE}
E = 2A\log{\bar{S}} + B + \ldots\, ,
\ee
where $\log{\bar{S}} \equiv \log{S} + \gamma_{\textrm{E}}$, with $\gamma_{\textrm{E}}$ the Euler-Mascheroni constant, and where both $A$ and $B$ 
are functions of the coupling only, interpolating between the gauge and string theory predictions. The ellipsis in~(\ref{ABE}) stands for the corrections we 
are after in this paper that are explicitly suppressed at large spin. Both at weak and strong coupling, our understanding of their structure is mostly based 
on explicit expressions for the first few orders of corresponding expansions. At weak coupling, for instance, the energy receives an infinite series of 
corrections suppressed by inverse powers of the spin, each possibly receiving logarithmic enhancements from powers of $\log{\bar{S}}$. The details of 
this expansion are not entirely arbitrary and display an interesting pattern~\cite{DMS}. In particular, leaving aside the corrections $\sim 1/S$, which can 
be eliminated by an appropriate redefinition of the expansion parameter $S \rightarrow \eta(S) = S + \ldots\, $, to which we shall come back later, the first 
new, `dynamically independent' contributions appear at order $O(1/S^2)$. One can say to some extent that the typical scale of corrections at weak 
coupling is $\sim 1/S^2$. A similar pattern was observed at strong coupling~\cite{BFTT}, at least classically, suggesting that a direct interpolation along the 
lines as explained before is possible. However, the fact that this program cannot be realized becomes transparent at the one-loop level on the string side. 
Here, the situation changes drastically with corrections of the type $\sim 1/\log{S}$ prevailing over the aforementioned power suppression. This was uncovered 
both from direct string theory computations \cite{SNZ, BDFPT, GRRT} and integrability methods~\cite{GSSV}. To one-loop order, the first stringy corrections 
to the scaling~(\ref{ABE}) is found to be driven by the Casimir energy~\cite{SNZ, BDFPT, GRRT, GSSV}
\be\label{CasimirEnergy}
\delta E \sim -{5\pi \over 12\log{S}}\, ,
\ee
that merely reflects the contributions of the five massless modes of the GKP string. The reconciliation of the string and gauge theory predictions thus 
entails considering an interpolating function that dependents in an entangled manner on both the 't Hooft coupling and the spin.

The weak-to-strong transition \textit{\`a la} BES seems furthermore doomed from the outset as soon as one takes a first glance on the above spin suppressed 
contributions. The reason is that this strategy relies solely on the ABA equations, which, though defined for any value of the coupling, are known not to account 
entirely for all gauge and/or string contributions~\cite{AJK05, KLRSV}. This implies, in particular, that the identities quoted in~(\ref{ABAdispersion}) and 
(\ref{ABAintro}) are not correct beyond certain accuracy which is determined by the size of the so-called wrapping corrections.%
\footnote{The terminology comes from the fact that in the spin-chain language these contributions are associated to wrapping interactions whose range 
exceeds the length of the chain.} The exact energy should then be written as
\be
\label{ABA+wrap}
E =E^{\rm ABA} + E^{\rm wrap}\, ,
\ee
with $E^{\textrm{ABA}}$, only, being determined by~(\ref{ABAdispersion}, \ref{ABAintro}). To account for the new contribution, i.e., $E^{\rm wrap}$, one has to 
start from the integrable model on the string side and invoke the Thermodynamic Bethe Ansatz~\cite{Ztba}. On this side, indeed, the whole problem is
formulated within the framework of a two-dimensional quantum field theory, compactified on a cylinder of circumference $\sim L$, and whose ground state, the  
Berenstein-Maldacena-Nastase (BMN) reference state~\cite{BMN}, is identified with the spin-chain vacuum $\textrm{tr}\, Z^L$. In this set up, the decomposition
(\ref{ABA+wrap}) can be understood qualitatively as the splitting of the energy of an excited state into two pieces: one which, under the assumption of integrability, 
simply sums up the mechanical energies of each particle constituents and the other one incorporating the energy stored in the virtual fluctuations that they source. 
The latter contributions are conveniently understood as originating from the exchange of virtual excitations circulating around the cylinder~\cite{AJK05} and 
scattering with the real particles that build up the state under study. An ingenious trick~\cite{Ztba}, that goes under the name of double Wick rotation and which 
interchanges space and time, allows one to analyze the dynamics of these virtual excitations which mirrors in many aspects the original one. At this end, these 
considerations can be used to advocate the leading wrapping contributions, also known as the (first) L\"uscher correction, to the energy of the twist-two 
operators~\cite{BJ, BJL}. The energy $E^{\rm wrap}$ is then found to admit the form
\be
\label{GammaWrapLeading}
E^{\rm wrap} \sim \int dp \, {\rm e}^{- L E_\ast (p)} (\cdots) \, ,
\ee
exhibiting a typical exponential decay with the system size $L$ and its rate controlled by the energy~\cite{AJK05} $E_\ast = 2 \, {\rm arcsinh} \sqrt{1 + p_\ast^2}/(4 g)$ 
of the particle exchanged in the mirror kinematics. For our discussion, it matters that the wrapping corrections appear delayed at large spin: they are expected 
indeed not to affect the validity of the leading asymptotics~(\ref{ABE}) such that the coefficients $A$ and $B$ are free from wrapping and are governed entirely by 
the ABA equations. This is confirmed by the large spin asymptotics of the leading wrapping correction for the twist-two operators~\cite{BJL} that is found to scale 
like~\cite{BF09}
\be
E^{\rm wrap} \sim - g^8  \lr{128 \zeta_3+\frac{64\pi^2}{3}} \frac{\ln^2 S}{S^2} \, .
\ee
It immediately confirms however that the contributions $\sim 1/S^2$ that we want to elucidate in our current analysis are contaminated by the wrapping effects. 
The same wrapping corrections appear also crucial at strong coupling for the recovery of the stringy expectation~(\ref{CasimirEnergy}) as was shown in~\cite{GSSV}.

The computation of this wrapping energy is amenable to a systematic treatment proposed in \cite{GKV, AF}. But this framework comes with a hefty pricetag:  one 
has to deal with an excited state with a very large number of quanta, namely, $S \gg 1$ magnons accounting for the light-cone derivatives in~(\ref{Twist2}). This 
makes the analysis of the wrapping energy rather difficult, and, so far, only the first two leading contributions, i.e., the four- \cite{BJL} and five-loop~\cite{LRV} 
expressions, have been extracted.%
\footnote{This approach has indeed the drawback of involving the summation over an infinite number of virtual excitations, bound states of elementary ones, 
which scatter with an `equally large' number of real particles. The five-loops expression~\cite{LRV} further requires to into account the incorporation of the 
back-reaction of these virtual contributions on the quantization conditions for the set of momenta~(\ref{ABAintro}).} Though they possess a very remarkable feature of being applicable at any spin, not necessarily large.

The AdS/CFT correspondence challenges us however to unravel the details of the transition between the gauge and string theory and already hints at a possible 
new path. Namely, getting back to the string theory result~(\ref{CasimirEnergy}) and its interpretation as a Casimir energy, one immediately realizes that the
contributions we are after are naturally associated with the finite-size corrections induced by virtual particles now circulating around the GKP string itself, and thus 
over a length $R \sim 2\log{S}$. This shift in paradigm, which places the GKP string at the center of the analysis and treats it as a vacuum on its own, yields at a 
qualitative level an immediate realization of the change of the scale $1/S^2$ to $1/\log{S}$ as one passes from the gauge to string expectations. The main 
observation is the remarkable feature of the dynamics on top of the GKP string, which was elucidated by Alday and Maldacena~\cite{AM07}: the theory is gapped 
at weak coupling with a mass gap $m \sim 1$ but becomes asymptotically gapless at strong coupling with $m \sim \exp{(-\pi g)}$. At strong enough coupling, this 
phenomenon is moreover governed by the O(6) sigma model which stands for the low-energy effective theory on the GKP vacuum. A direct analogy with a 
relativistic QFT would then lead us to conclude that the two previous scales merely reflect these two extreme regimes, with the finite-size corrections estimated 
as $\sim \exp{(-mR)} \sim 1/S^2$ in the massive `phase' and as $\sim 1/R \sim 1/\log{S}$ in the massless one. The validity of this analogy is far from being 
obvious and actually relies on the double Wick rotation invariance of the GKP background, which was first articulated in~\cite{AGMSV10}.

\begin{figure}[t]
\begin{center}
\mbox{
\begin{picture}(0,200)(190,0)
\put(0,0){\insertfig{7}{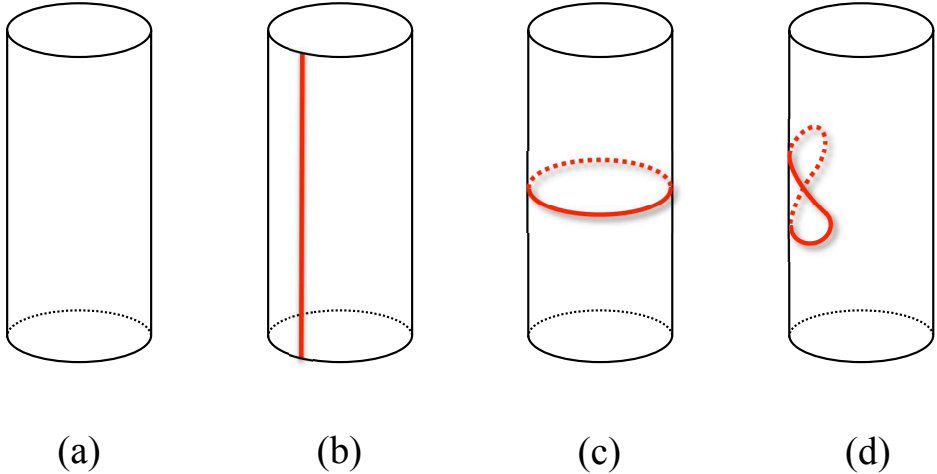}}
\end{picture}
}
\end{center}
\caption{\label{Cylinder} (a) Vacuum corresponds to twist-two operators. (b) Real one-particle excitation represents 
twist-three operators. (c) Corrections from finite size of the system. (d) Virtual bulk correction renormalizing the background.}
\end{figure}

The purpose of this paper is to establish precisely, at a quantitative level, the above picture by advocating and constructing a representation for the leading 
finite-size correction which transparently incorporates the physics of excitations on top of the GKP string. It will imply that the loop-by-loop order large-spin 
predictions on the gauge and string theory side are the two ends of the stick emerging under the appropriate conditions, i.e., $g^2 \ll 1/\log{S}$ and 
$g \gg \ft{1}{\pi}\log{\log{S}}$, respectively. Our first step, however, is to set up the following ansatz for the complete energy of the string. Viewed as a vacuum 
energy, we aim at decomposing it as
\be\label{Ansatz}
E = E^{\rm bulk} + E^{\rm FS}\, ,
\ee
where the finite-size energy $E^{\textrm{FS}}$ is always suppressed with the system's length, in the large volume limit $\log{S} \gg 1$. The details of this 
suppression depend, however, on certain features of the theory, in particular, on its spectrum. The bulk energy, on the other hand, scales with the length 
and incorporates therefore the leading asymptotic behavior~(\ref{ABE}). From the world-sheet point of view, the bulk energy should resum corrections 
associated with diagram of trivial topology contrary to the finite-size corrections which accommodate those with nontrivial windings.%
\footnote{The former contribution can also be understood as renormalizing the GKP background.} They are shown in Fig.\ \ref{Cylinder} (d) and (c), respectively. 
The decomposition~(\ref{Ansatz}) immediately raises several questions: How do we compute the bulk/finite-size energy and what are their relation to the 
ABA/wrapping expression? Probably the most naive and conservative attempt is to set that $E^{\textrm{bulk}} \stackrel{?}{=} E^{\textrm{ABA}}$ and 
$E^{\textrm{FS}} \stackrel{?}{=} E^{\textrm{wrap}}$. We shall see that this is not possible however. To realize why this is not the case, let us focus on the 
finite-size corrections and derive a L\"uscher-like formula that encodes them. 

As we already said, our approach relies on a complimentary formalism which operates with a different set of degrees of freedom than the magnons. 
Information about these dual excitations can be deduced directly from the ABA equations, up to corrections that are suppressed by at least the system 
size $\sim \log{S}$. Integrable models, indeed, are often endowed with a particle-hole transformation which allows one to exchange particle and vacuum 
modes. While the hole degrees of freedom are invisible within the ABA approach they become manifest within the formalism of the Baxter equation 
\cite{Bel06, Bel09}. The long-range Baxter equation is a second-order finite-difference equation for the $Q$-function
\be\label{LRBaxterEq}
\Delta_{+}(u+\ft{i}{2})Q(u+i) + \Delta_{-}(u-\ft{i}{2})Q(u-i) = t(u)Q(u)\, .
\ee
where $\Delta_{\pm}(u)$ are known dressing factors and $t(u)$ as the eigenvalue of an auxiliary transfer 
matrix. The solution to \re{LRBaxterEq} identical to ABA is sought in the polynomial form
\be\label{PolSol}
Q(u) = \prod_{j=1}^{S}(u-u_{j})\, ,
\ee
where the set of rapidities $u_{j}$ coincide with the Bethe roots \re{BetheRoots}. The equivalence between the two formalisms is achieved by setting $u =u_j$ 
in both sides of Eq.\ \re{LRBaxterEq} and observing that the latter admits the form 
\be
1+Y_{\textrm{h}}(u_j) = 0\, , \qquad Y_{\textrm{h}}(u) = {\Delta_{-}(u-\ft{i}{2})Q(u-i) \over \Delta_{+}(u+\ft{i}{2})Q(u+i)}\, ,
\ee
where $Y_{\textrm{h}}(u)$, introduced here for convenience, is a quantity that we shall encounter on many different occasions in the following discussion. An easy 
inspection confirms its agreement with Eq.\ \re{ABA}, as anticipated.

However, now in addition to the Bethe roots, the Baxter equation encodes another set of data that enters through the transfer matrix. Namely, at one-loop 
order $t(u) = t^{(0)} (u) + \mathcal{O} (g^2)$ is parametrized by its roots
\be
t^{(0)} (u) = 2 \prod_{k = 1}^L (u - \delta_k)
\, .
\ee
The idea behind the particle-hole transformation is to trade the dynamics of the Bethe roots $u_{j}$ for the one of the holes $\delta_k$. In the large spin limit, two 
of the holes are large and somehow decouple, leaving $L-2$ (small) holes identified with the dynamical degrees of freedom \cite{BGK06}. These holes encode 
the dynamics of excitations, here, the scalar $Z$-fields propagating on the background of derivatives.

The state with no small holes corresponds to the twist-two operator and is identified with the GKP vacuum, see Fig.\ \ref{Cylinder} (a).%
\footnote{The $SO(6)$ symmetry of the GKP vacuum, which is not obvious for the operator \re{Twist2}, is achieved  by means of isotopic degrees of freedom in 
the form of composite Bethe roots which do carry zero energy and momentum~\cite{B10}.} Adding a hole amounts to inserting a $Z$ field in this background. 
This is not the only flavor of excitations however. To find all of them one should extend consideration to the full system of ABA equations~\cite{BS05}. The whole 
set of elementary excitations was elucidated in Refs.\ \cite{B10, GMSV}: they are associated to the twist-one partons of the gauge theory, i.e., $6$ scalars, $4+4$ 
fermions and $1+1$ gauge fields.%
\footnote{They carry $\mathfrak{so}(6) \simeq \mathfrak{su}(4)$ isotopic degrees of freedom which are easily inferred from their corresponding degeneracy.} For illustration, a one-particle excitation $X$ on top of the ground state is found as
\be
\ket{\mbox{one-particle excitation}} 
= \sum_{s = 0}^S c_{s} {\rm tr} \, Z D_+^{s} X D_+^{S - s} Z + \dots\, ,
\ee
where $X$ is one of the previously mentioned parton inserted in the twist-two background~(\ref{Twist2}). This is a twist-three operator, which on the string theory 
side corresponds to a particle propagating on the closed-string world sheet, see Fig.\ \ref{Cylinder} (b). The generalization to higher twists follows the same line of 
reasoning with possible additional states in the form of twist $>1$ partons which are bound states of gauge fields. This pattern of excitations and their associated 
dispersion relations are in good agreement with string theory predictions~\cite{FT02, DL10, GRRTdisp}, though a few mismatches still await their reconciliation.

Equipped with the spectrum of excitations, one can make an educated guess for the form of the leading finite-size corrections. This one can be argued to take 
a universal form, known as the L\"uscher formula. The logic here is identical to the one for magnons~\cite{AJK05, BJ}. Our proposal for the L\"uscher 
formula reads
\be\label{LuscherIntro}
E^{\textrm{FS}}  = -\sum_{\star}{n_{\star}\over 2\pi}\int dp\,  Y^{\textrm{mirror}}_{\star}(p) + \ldots \, ,
\ee
where the function $Y^{\textrm{mirror}}_{\star}(p)$ controls the amplitude for propagation of a  $\star$-particle around the GKP string with the momentum $p$ in 
the mirror kinematics. The ellipsis correspond to higher L\"uscher corrections associated with multiparticle exchanges and $n_{\star}$ is the number of particles 
of a given $\star$-flavor, e.g., $n_{\textrm{h}} = 6$ for scalar, etc. We shall find that the double Wick rotation symmetry is valid for all excitations and thus makes the 
analysis akin to the one typically performed in relativistic theories, 
\be
\label{FS}
Y^{\textrm{mirror}}_{\star}(p) \sim {\rm e}^{- R E_\star (p)}\, \, \,  \longrightarrow \, \, \, E^{\rm FS} \sim \sum_{\star} \int d p \, {\rm e}^{- R E_\star (p)} \, ,
\ee 
where $E_\star (p)$ is given by the dispersion relation of a real excitation and $R \sim 2\log{S}$. This result immediately implies that the contribution of a 
$\star$-particle to the energy~(\ref{LuscherIntro}) scales as
\be
E_{\star}^{\rm FS}  \sim 1/S^{2 m_\star}
\, ,
\ee
with $m_{\star} = E_{\star}(p=0)$ being the mass of the excitation. As a consequence, within our accuracy at weak coupling, i.e., $\sim 1/S^2$, we can 
restrict the summation in Eq.\ (\ref{LuscherIntro}) to the set of twist-one excitations only, i.e., $\star = $ h (hole = scalar), f (fermion), gf (gauge field). This 
also explains why at strong coupling only the scalar sector, which is controlled by the O(6) sigma model, remains relevant: while all other excitations 
maintain their masses of order one~\cite{FT02, AM07}, the holes' mass becomes exponentially small. These conclusions survive a more careful analysis 
given below, but it is interesting to point out that this consideration requires a modification of~(\ref{FS}) by an extra $\sim R^0 \sim \log^0{S}$ contribution,
\be
R E_\star (p) \to 2E_\star (p)\log{\bar{S}} + 2 \delta E_\star\, .
\ee 
This seems to reflects a curious feature of the propagation of excitations in the GKP background, probably related to the loss of its homogeneity, observed 
explicitly at strong coupling~\cite{FT02, GRRT}. At weak coupling, this new factor determines the dependence of the finite-size corrections on the coupling 
constant, which according to our analysis conducted below is $\delta E_\star \sim - \log g^2$ resulting in $E^{\rm FS}  \sim g^4$.  Supersymmetry somewhat 
delays the onset of the finite size corrections by one order of the coupling, however, and as a consequence one finds $E^{\rm FS}  \sim g^6$.%
\footnote{The cancellation between bosonic and fermionic contributions may not be as effective if one considers $\sim 1/S^4$ and higher suppressed 
corrections. Bound states that contribute at this order do not have super-partners, for instance. It makes us believe that the onset of the finite-size corrections 
starts at two loops, though not apparent at the level of the $\sim 1/S^2$ contribution.} Hence, compared to the wrapping effects introduced in the magnon 
picture the finite size corrections arise starting already from three loop order for twist-two operators. This immediately exhibits the fundamental difference of 
our finite size correction compared to the conventional L\"uscher formula \re{GammaWrapLeading} that one add to the ABA energy. Therefore, blindly adding 
our \re{FS} to the ABA energy would not produce the correct energy. This is not totally surprising since after all, the L\"uscher formula \re{FS} that we are 
proposing does not quite sum up effects due to propagation of virtual magnons, as it is the case for the wrapping energy \re{GammaWrapLeading}, but instead 
of holes which, though somewhat dual to them, are nevertheless distinct. 

The L\"uscher formula was given schematically in~(\ref{LuscherIntro}). An important part of our analysis will be therefore to make it work at a quantitative level. 
To this end we shall analyze the transformation to the mirror kinematics separately for all our excitations and construct the relevant $Y$ functions. We shall see 
that they can all be expressed in terms of the solution to the BES equation and its siblings~\cite{FZ09} at the given accuracy in the spin. These can be though 
as fixing the asymptotic data, directly constructible from the ABA equations. This has a major benefit that the formula~(\ref{LuscherIntro}) can be univocally 
determined at any value of the coupling.  It will then be possible to construct a weak coupling expansion of~(\ref{LuscherIntro}) to high orders of perturbation
theory and find, in particular, that it admits the expansion
\be\label{FSpol}
E^{\textrm{FS}} = {(-1)^S \over S^2}\sum_{n\ge 2}\mathcal{P}_{n} g^{2n}\, + o(1/S^2),
\ee
where $\mathcal{P}_{2n}$ is a coupling independent polynomial in the variable $\log{\bar{S}} \equiv \log{S} + \gamma_{\textrm{E}}$ or degree $n-2$. It thus predicts 
that the higher-loop corrections are dressed by higher powers of $\log{\bar{S}}$ with no bound as the loop order $n$ becomes infinite. This can be seen as the first 
evidence that the formula~(\ref{LuscherIntro}) is capable to reconcile the gauge and string theory predictions through its intricate dependence both on the 
coupling $g$ and the length $\sim \log{S}$.

Our proposal for the twist-two energy would remain incomplete without full control on the second halve of the equation~(\ref{Ansatz}) given by the bulk energy. 
It is much more difficult to advocate its general expression beyond the fact that it should incorporate the leading asymptotics~(\ref{ABE}). Our approach will 
remain unfortunately empirical and is advocated by the comparison to the ABA predictions as follows. The ABA equations, or equivalently the asymptotic Baxter 
equation, can be solved at any coupling and to high accuracy at large spin. A thorough analysis demonstrates that the energy stemming from ABA can be 
decomposed in terms of sign preserving (regular) $E^{\rm reg}$ and alternating $E^{\rm alt}$ contributions
\be\label{RegAlt}
E^{\rm ABA} = E^{\rm reg} + E^{\rm alt}\, ,
\ee
with $E^{\rm alt} \propto {\rm e}^{i \vartheta}$, which depends on the momentum $\vartheta$ of the state. The latter expression evaluates to $(-1)^S$ on twist-two 
states with integer spin. This requires including into our consideration states with nonzero momentum $\vartheta = \pi$, realized for odd (unphysical) spins $S$. 
This continuation was surreptitiously performed in the finite-size energy above, see Eq.~(\ref{FSpol}), where the factor ${\rm e}^{i \vartheta} = (-1)^S$ can be thought 
as twisting the boundary conditions for excitations around the GKP string.%
\footnote{The linearity in ${\rm e}^{i \vartheta}$ of the first L\"uscher correction originates directly from the linearity in the $Y$ functions.} 
The physical configurations corresponding to even values of the spin are then associated with the periodic boundary conditions. A direct analysis shows that the 
quantity $E^{\rm alt}$ is suppressed by two powers of the spin at large $S$, to any order at weak coupling. The general structure of its expansion is furthermore 
akin to the finite-size energy: both are given by a series of terms with increasing logarithmic enhancement in $S$ as one passes to higher and higher orders in the 
coupling. Though their general forms appear to be similar, the accompanying expansion coefficients are different. Nevertheless, the former smilarity, together with the 
expectation that the L\"uscher formula~(\ref{Luscher}) should complement in some way the ABA prediction, inspires us to promote the substitution rule
\be\label{SubsIntro}
E^{\textrm{alt}} \rightarrow E^{\textrm{FS}}\, ,
\ee
as a mean of exacting the ABA prediction. It is equivalent to the statement
\be
E^{\textrm{bulk}} = E^{\textrm{reg}}\, ,
\ee
which together with~(\ref{LuscherIntro}) determines the (full) energy~(\ref{Ansatz}). This, of course, should be understood as an empirical rule at the currently
considered accuracy in large spin. Some physical evidence for it will be presented shortly. However, before turning to it, we would like to stress that by means 
of Eq.~(\ref{SubsIntro}) our proposal for exacting of the ABA prediction with the help of the L\"uscher formula becomes complete: at the given accuracy in the 
spin, the bulk energy can be obtained from the ABA equations and these same equations can be used to construct all ingredients involved in the L\"uscher 
formula~(\ref{LuscherIntro}). It is found, for instance, that in distinction with the alternating or finite-size energy the large spin expansion of the bulk energy is 
much more rigid. With the help of an appropriate expansion parameter $\eta = \eta(S) = S+\ldots$ alluded to before, it takes the form
\be
E^{\textrm{bulk}} =  2A\left(\log{\bar{\eta}}-{1\over 12\eta^2}\right) + B + C/\eta^2 + o(1/\eta^2)\, ,
\ee
where $C$ is a new coefficient function of the coupling only. After re-expansion in the spin, the bulk energy displays a logarithmic enhancement at order $\sim 1/S^2$ 
but this one is bounded by $\log^2{S}$ for any coupling, contrary to the alternating or finite-size energy.

Perhaps more importantly, the rule~(\ref{SubsIntro}) makes it possible to compare the content of~(\ref{LuscherIntro}) with existent results for the large spin 
expansion of twist-two anomalous dimension up to five loops, including wrapping corrections. These results based on the L\"uscher formula `on the BMN 
vacuum' are built on a much firmer basis than ours and it is thus more than a requirement to successfully compare the two. To this end, one merely observes that 
\be\label{WrapVsFS}
E^{\textrm{FS}} = E^{\textrm{alt}} + E^{\textrm{wrap}}
\ee
holds as a corollary of our previous discussion. This should be the case at order $O(1/S^2)$ at large spin and to any order at weak coupling. After extracting the 
relevant large spin expression for $E^{\textrm{wrap}}$ from~\cite{BJL, BF09, LRV}, we will be able to verify Eq.~(\ref{WrapVsFS}) up to five loops.%
\footnote{This agreement is actually observed only for the physical configuration, that is, for even spin.} We notice finally that, through the identity~(\ref{WrapVsFS}) 
and the common structure of the finite-size and alternating ABA energy at large spin, our proposal allows us to make immediate qualitative predictions at higher 
loops, notably, suggesting that
\be
E^{\textrm{wrap}}_{n-\textrm{loop}} \sim {\log^{n-2}{S} \over S^2}\, .
\ee

This discussion sets up the stage of our analysis. Before plunging into detailed computations of the bulk and finite-size energy, let us come back to our problem 
and let us see how our proposal helps us uncover the weak-to-strong transition in coupling.

To some extent, the explanation of the weak-to-strong transition is already embodied in the ABA prediction. To our accuracy in the spin, the main role is played by 
the alternating component of the energy that displays the enhancement of the logarithmic growth as the loop order increases. This behavior is shared by the 
L\"uscher formula~(\ref{LuscherIntro}), which also motivated the substitution rule~(\ref{SubsIntro}). What happens next is that the weak coupling expansion breaks 
down as soon as $g^2 \sim 1/\log{S}$, calling for a resummation. All of these corrections can be properly encoded into the formula
\be\label{ExAltEn}
E^{\textrm{alt}} = {1\over \pi}\oint dp_{\textrm{h}}(u-i/2)Y_{\textrm{h}}^{\textrm{mirror}}(u-\ft{i}{2}) + \ldots\, ,
\ee
with ellipsis standing for corrections suppressed stronger than $\sim 1/S^2$ at weak coupling. This identity looks pretty similar to the L\"uscher formula
(\ref{LuscherIntro}) if not for a different choice of integration contour and the absence of contributions from other GKP excitations than the holes. The similarity 
is enough however to understand the behavior of $E^{\textrm{alt}}$ at strong coupling. Namely, when the coupling gets large, the energy~(\ref{ExAltEn}) gets 
higher due to the reduction of the mass $m = m_{\textrm{h}}$ of the scalar, since $Y_{\textrm{h}}^{\textrm{mirror}} \sim 1/S^{2m_{\textrm{h}}}$. It then takes a
very suggestive form  
\be
E^{\textrm{alt}} = -{2m\over \pi}(-1)^S K_{1}(mR) + \ldots\, ,
\ee
where $K_{1}(z) \sim \exp{(-z)}$ for $z\gg1$ is the modified Bessel function, $m \sim \exp{(-\pi g)}$ the mass gap~\cite{AM07} and $R \sim 2\log{S}$ the length. 
For even spin, this equation gives the contribution to the vacuum energy of two massive bosons subject to periodic boundary conditions on a cylinder of size 
$R \gg 1/m$. For comparison, the finite-size energy~(\ref{LuscherIntro}), which is also dominated by the contribution from the lightest modes, i.e., $6$ scalars, 
at strong coupling, reads
\be\label{EFSSC}
E^{\textrm{FS}} = -{6m\over \pi}(-1)^S K_{1}(mR) + \ldots\, .
\ee
It suggests that the ABA equation accounts only for 2 out of 6 scalars excitations. The former may be viewed as the two scalar fields $Z$, $\bar{Z}$ which are not 
treated as independent dynamical excitations in the BMN picture, since $Z$ is part of the background and $\bar{Z}$ is a BMN mass $2$ excitation that decays into 
two BMN scalars\footnote{This is the traditional interpretation of the mixing $Z\bar{Z} \sim Y\bar{Y} \sim X\bar{X}$ between the various scalars in the BMN background.}. 
In light of it, the proposed substitution $E^{\rm alt} \to E^{\rm FS}$ can be understood as correcting for the proper number of scalar excitations not captured by ABA.
Adding further contributions of fermions and other bosons, the formula~(\ref{LuscherIntro}) together with~(\ref{SubsIntro}) appears as a natural attempt for exacting 
the quntitative predictions for the energy of the GKP string.

The results~(\ref{EFSSC}) is transparent enough to allow one to guess the final step towards the string perturbative result. It reveals simultaneously the limitation 
of our expression at strong coupling. Namely, as was already said earlier, the result~(\ref{LuscherIntro}) accounts only for the first L\"uscher correction to the 
vacuum energy computed within the O(6) sigma model. The latter however predicts an infinite number of higher L\"uscher corrections, with all of them encrypted
into a system of TBA equations, see, e.g., \cite{BH} for a recent discussion. From this more established approach, it becomes obvious that the asymptotics
(\ref{EFSSC}) can only prevail provided $mR \gg 1$. This requires an extraordinarily large length that the perturbative string analysis cannot afford. The latter 
always operates in a regime where the Compton wavelength $\sim 1/m$ of the O(6) particles is much bigger than the length $R$, i.e., $mR \ll 1$. In this
regime, the natural excitations appear to be $5$ massless Goldstone bosons, instead. The stringy perturbative prediction necessarily resums an infinite tower  
of L\"uscher corrections, which from the weak-coupling standpoint all originate from highly subleading contributions suppressed as $\sim 1/S^{2n}$ with $n \ge 1$. 
Tracing back these corrections would constitute a considerable improvement of our approach and would entail working out a system of TBA equations for the 
dynamics of excitations on top of the GKP string. This is however beyond the scope of our present analysis.

A reader would notice that our previous discussion is, of course, a well understood phenomenon in relativistic quantum field theory, which can be studied in great 
details corresponding system becomes integrable. The result for the finite size energy can be encapsulated into the formula~\cite{Cardy, Ztba, KM}
\be
E^{\rm FS} = - \frac{\pi c(mR)}{6R} + \ldots\, ,
\ee
where the omitted terms stand for corrections $\sim 1/R^{d}$ with $d\ge 2$, associated with irrelevant deformations of the O(6) sigma model, and `UV' contributions 
$\sim 1/S^2$ from fermions and gauge fields. Here $c(mR)$ is the effective central charge of the O(6) sigma model. It  interpolates between $c(mR) \sim \exp{(-mR)} $ 
when $mR \gg 1$ (massive regime) and $c(mR) \sim 5$ when $mR \ll 1$ (massless regime). In between, it is controlled by the full set of TBA equations for the vacuum 
state of the O(6) sigma model. Our expression~(\ref{LuscherIntro}) is only applicable in the massive regime while the perturbative string analysis works in 
the UV regime. In this latter domain, which is far beyond our reach, the O(6) sigma model can be used to shed some light on the structure of the large spin expansion. 
It is known, for instance, that in the O($n$) sigma model the effective central charge can be found to take the form~\cite{BH,Warringa}%
\footnote{This result was derived from the leading~\cite{BH} and subleading~\cite{Warringa} large $n$ expression by assuming that the two-loop coefficient, obtained 
after re-expansion in the renormalized coupling $e^{2}$ of the O($n$) model, is a polynomial in $n$ of degree 2 vanishing for $n=2$ (i.e., for free theory). The latter re-expansion is 
found by means of $1/\log{(1/mR)} \sim \beta_{0} e^2 \sim (n-2) e^2$.}
\be
c(mR \ll 1) = (n-1) - {3(n-1) \over 2\log{(1/mR)}} + O\left({\log \log{(1/mR)} \over \log^2{(1/mR)}}\right)\, .
\ee
This renormalization-group improved expansion resums the standard (renormalized) perturbative expansion taken at fixed renormalization scale $\sim 1/R$ in the 
O($n$) model. Specializing to $n=6$ and re-expanding in terms of the string tension $\sqrt{\lambda} \equiv 4\pi g \gg 1$ leads to
\be
c(mR) = 5 - {30 \over \sqrt{\lambda}} + O\left({\log{R} \over \lambda}\right)\, ,
\ee
indicating that the stringy loop expansion will receive enhancement by powers of $\log{R} \sim \log{\log{S}}$ at higher loops. This can only make sense 
at large spin if $\log{R} \ll \sqrt{\lambda}$, which eventually is just a rephrasing of the condition $mR \ll 1$, since $m \sim \exp{(-\sqrt{\lambda}/4)}\, $.

Our above discussion illustrates how rich and complicated is the structure of the subleading large spin corrections to the twist-two anomalous dimension 
and how subtle the interpolation between the gauge and string predictions becomes. 

Our subsequent consideration is organized as follows. In the next section we will formulate the long-range Baxter equation and construct its solution at large spin 
for the twist-two operator, up to the relevant accuracy. As a follow up, we shall derive the all-loop expression for the bulk (i.e., regular) and alternating contribution 
to the energy, both encapsulated in the solution to a system of integral equations. Then, in Section \ref{FSsection}, we advocate and postulate the L\"uscher formula 
for excitations propagating on the GKP background. We discuss there the rotation to the mirror kinematics for all flavors of excitations and provide all-order 
representation for their associated $Y$ functions. Later on, we perform the analysis up to five loops and demonstrate the equivalence of our formalism to the one 
available in the literature. Several appendices summarize technical details.

\section{ABA solution}
\label{ABA}

In this section, we shall derive the ABA expression for the twist-two anomalous dimension in the large-spin expansion up 
to the order $O(1/S^2)$. We will make extensive use of the long-range Baxter equation~\cite{Bel06, Bel09}, whose content 
is equivalent to the Asymptotic Bethe Ansatz equations of Ref.\ \cite{BS05}, and apply the strategy developed in Ref.\ \cite{BGK06} 
for construction of corresponding large-spin solutions. After exposing the main ingredients of the Baxter equation in the next
section, we engineer two non-polynomial solutions, --- the main building blocks of our framework, --- out of which the sought 
polynomial solution \re{PolSol} can be obtained. Then we address the next step of our formalism and derive a system of 
equations, similar to the ones worked out for the cusp anomalous dimension~\cite{BES}, that allow us to extract appropriate 
scaling dimensions to all orders in 't Hooft coupling.

Before we move on to solving the Baxter equation, let us comment on the parameter we shall use to carry the large spin analysis. As we will establish later 
in this section, the large-spin expansion is naturally carried out in terms of the parameter $\eta$ that is determined by the conserved charge 
$\mathfrak{q}_2 =  \eta^2$ entering the transfer matrix $t (u)$,
\be\label{etaS}
\eta^2 = \mathbb{C}_{2} + \delta
\, .
\ee
The latter is essentially defined by the renormalized quadratic Casimir
\be\label{Casimir}
\mathbb{C}_{2} = (S+\ft{1}{2}\gamma+1)(S+\ft{1}{2}\gamma)
\ee
of the underlying $\mathfrak{sl}(2)$ algebra. This is the case up to the additive contribution $\delta$ which scales logarithmically with the 
spin $\delta = x \log{\bar{\eta}} + y + o(\eta^0)$, with $\log{\bar{\eta}} \equiv \log{\eta} - \psi(1) = \log{\eta}+\gamma_{\textrm{E}}$, with 
$\gamma_{\textrm{E}}$ being the Euler-Mascheroni constant, and where $x$ and $y$ are functions of the coupling only. Since our
analysis relies on the ABA equation, the anomalous dimension $\gamma \equiv \Delta-\Delta_{0}$ in~(\ref{Casimir}) stands for its ABA 
prediction, $\gamma = \gamma^{\textrm{ABA}}$. The ABA equation, or equivalently the Baxter equation, also provides us with the 
coefficient $\delta$, related to some higher spin-chain charges.

At the level of the ABA equation, the relation~(\ref{etaS}) between $\eta$ and $S$ can be unambiguously computed for any $S$ and/or 
expanded at large spin to any accuracy. For many reasons, however, we might prefer to restrict ourself to this part of the relation~(\ref{etaS}) 
that is not contaminated by wrapping effects. At weak coupling, this happens if the relation $\eta = \eta(S) = S + \ldots$ is worked out at large 
spin up to (and excluding) $1/S^2$ corrections. This stands for the fact that the main ingredients entering~(\ref{etaS}), namely $\gamma$ 
and $\delta$, are expected to be uniquely determined at large spin up to corrections $\sim 1/\eta^2$ which is the typical scale for wrapping 
effects~\cite{BJL, BF09}. In our discussion of contributions $\sim 1/S^2$ to the anomalous dimension, the suggested accuracy for determination 
of $\eta = \eta(S)$ is good enough.%
\footnote{This statement follows from the logarithmic scaling of the anomalous dimension, $\gamma \sim \log{\eta} = \log{(S+\delta S)} \sim \log{S} 
+\delta S/S + \ldots$, such that contributions $\sim 1/S^2$ to the parameter $\delta S$ cannot affect $\gamma$ at the working accuracy.} We stress 
that these remarks apply to any order at weak coupling (only). At strong coupling, wrapping effects enter earlier~\cite{GSSV} with the scale 
$\sim 1/\log{S}$, truncating its expansion at large $S$ to the first asymptotic term, $\eta(S) = S+\ldots $, i.e., right before this scale enters its 
right-hand side. This limitation will not affect however our discussion of the strong coupling regime.

\subsection{Long-range Baxter equation}

The Baxter equation that we will be solving here was given above in Eq.\ \re{LRBaxterEq}. Below, we will discuss in turn 
two important ingredients determining it, the dressing factors and the transfer matrix. The notations that we will 
be using presently will be however different from the ones adopted in Refs.\ \cite{Bel06,Bel09}

\subsubsection{Dressing factors}

To start with, let us notice that the ratio of the dressing factors entering the left-hand side of the Baxter equation
\re{LRBaxterEq} could serve as a generating function for the spin-chain conserved charges,
\be\label{GenFunc}
\left[ {\Delta_{-}(u) \over \Delta_{+}(u)} \right]^{1/2} 
=  
\prod_{j=1}^{S}{1-g^2/xx^{-}_{j} \over 1-g^2/xx^{+}_{j}}
= 
\exp{\left[-i\sum_{n\ge 1}q_{n+1}\left({g^2 \over x}\right)^{n}\right]} 
\, ,
\ee
which are defined as \cite{BDS},
\be
\label{ConsrvdChargs}
q_{n+1} ={i \over n}\sum_{j=1}^{S}
\left( (x^+_j)^{-n}- (x^-_j)^{-n} \right)
= {1\over 2\pi}\oint {dx \over x^{n+1}}\log{{Q(u+\ft{i}{2}) \over Q(u-\ft{i}{2})}}\, .
\ee
These are defined in terms of the renormalized rapidity $x = (u+\sqrt{u^2-(2g)^2})/2$ which is the large solution, i.e.,
$x(u) \sim u$ as $u \to \infty$, of the Zhukowski map $u = x +g^2/x$. Here and elsewhere in the paper we will be using 
the notations $x^{\pm} \equiv x \left(u \pm \ft{i}{2} \right)$. The contour of integration in~(\ref{ConsrvdChargs}) encloses 
the cut along $u^2 <(2g)^2$ present in the $x$ and goes in the anticlockwise direction. The lowest charge, with $n=1$, 
determines the ABA anomalous dimension or equivalently the energy,
\be
E^{\rm ABA} = L+ 2 g^2 q_{2}
\, .
\ee
Recall that the energy is determined by the scaling dimension $\Delta$ minus the spin $S$. The latter stands for the 
number of magnons propagating on the spin chain of length $L$, where according to the spin-chain/operator dictionary the 
integer $L$ reflects the twist of the operator, or the number of scalar fields $Z$ building it up.

On the other hand, the individual dressing factors contain more information than their ratio. They can be written as
\be\label{DefCoeffBis}
\begin{aligned}
\log{\Delta_{+}(u)} &=  L\log{x} +2\sum_{n=1}^{\infty}\gamma_{n}\left({i g \over x}\right)^{n}\, ,  \\
\log{\Delta_{-}(u)} &=  L\log{x} +2\sum_{n=1}^{\infty}\bar{\gamma}_{n}\, \bigg({g \over ix}\bigg)^{n}\, ,
\end{aligned}
\ee
where the expansion coefficients $\gamma_{n}$, $\bar{\gamma}_{n},$ are complex conjugate of each other. The latter quantities 
are dynamical:  they depend on the state under study, that is on the solution $Q(u)$ we aim at constructing, and thus on 
the coupling constant. Their precise form shall be given shortly.

In view of the representation~(\ref{DefCoeffBis}), the dressing factors $\Delta_{\pm}(u)$ do not define analytic functions 
of $u$. At leading order in weak coupling, the expression~(\ref{DefCoeffBis}) reduces to
\be\label{wcdress}
\Delta_{+}(u) = \Delta_{-}(u) = u^L + \mathcal{O} (g^2) \, ,
\ee
and, after plugging it back to~(\ref{LRBaxterEq}), one recovers the Baxter equation for the short-range $\mathfrak{sl}(2)$ 
Heisenberg spin chain of conformal spin $s=1/2$. However beyond certain order in the weak coupling expansion, the 
dressing factors develop poles at $u\sim 0$ which resum into cuts with branch points at $u = \pm 2g$ at finite value of 
the coupling. For this reason one needs to be careful about the meaning of the expressions $\Delta_{\pm}(u\pm \ft{i}{2})$ 
entering ~(\ref{LRBaxterEq}): namely, the long-range Baxter equation is understood to be evaluated on the sheet where
(\ref{wcdress}) hold true at large rapidity, $u \sim \infty$.

An equivalent representation for the dressing factors~(\ref{DefCoeffBis}) can be found after introducing the generating 
function $\gamma(t)$ as an infinite series
\be\label{BessDec}
\gamma(t) = \sum_{n \ge 1}(2n)\gamma_{n}J_{n}(t)\, ,
\ee
in Bessel functions $J_{n}(t)$, with well-known asymptotic behavior $J_{n}(t) \sim t^{n}$ for $t \sim 0$. Similar equation 
holds for $\bar{\gamma}(t)$ via the obvious replacement $\gamma_{n} \rightarrow \bar{\gamma}_{n}$ above. We note that since 
for $t$ real the Bessel functions are real, then $\bar{\gamma}(t) = (\gamma(t))^{*}$. In terms of these functions we find
\be
\begin{aligned}
\log{\Delta_{+}(u)} &= L\log{x} + \int_{0}^{\infty}{dt \over t}\gamma(\pm 2gt)\e^{\pm iut}\, , \\
\log{\Delta_{-}(u)} & = L\log{x} + \int_{0}^{\infty}{dt \over t}\bar{\gamma}(\mp 2gt)\e^{\pm iut}\, ,
\end{aligned}
\ee
where the upper (lower) sign refers to $u$ in the upper (lower) half plane, respectively. This immediately 
follows from the identity
\be
\label{BesselRep1overX}
\left({ig \over x}\right)^{n} = n\int_{0}^{\infty}{dt \over t}J_{n}(\pm 2gt)\e^{\pm iut}\, , \qquad n=1, 2, \ldots ,
\ee
where the $\pm$ corresponds to $\Im{\rm m} (u) \gtrless 0$.

Now, we can write the coefficients $\gamma_{n}$ as
\be
\begin{aligned}\label{StrEq1}
\gamma_{n} &= -{1\over 2i\pi}\oint {dx \over x}\bigg({g \over ix}\bigg)^{n}\log{Q(u+\ft{i}{2})} + d_{n} \, 
= -{1\over 2i\pi}\oint {dx \over x}\bigg({x \over ig}\bigg)^{n}\log{Q(u+\ft{i}{2})} + d_{n}
\, , \\
\bar{\gamma}_{n} &= -{1\over 2i\pi}\oint {dx \over x}\left({ig \over x}\right)^{n}\log{Q(u-\ft{i}{2})} + \bar{d}_{n} 
= -{1\over 2i\pi}\oint {dx \over x}\left({ix \over g}\right)^{n}\log{Q(u-\ft{i}{2})} + \bar{d}_{n}
\, .
\end{aligned}
\ee
Here the contour of integration goes anti-clockwise and surrounds the interval $u^2 < (2g)^2$ in all cases. The 
coefficients $d_{n}, \bar{d}_{n},$ are complex conjugate of one another and account for the modification stemming 
from the BES/BHL dressing phase~\cite{BES, BHL}. It is important to realize that $d_{2n-1}$ is purely imaginary while 
$d_{2n}$ is real as it follows from their definitions. This implies that dressing phase only affects the imaginary part of
 the coefficients $\gamma_{2n-1}$ and the real part of the coefficients $\gamma_{2n}$. Therefore, in the representation 
 for $d$'s, that will appear to be convenient in our subsequent discussion, 
\be\label{StrEq2}
d_{2n-1} = \int_{0}^{\infty}{dt \over t}J_{2n-1}(2gt){\gamma_{+}(2gt)-\bar{\gamma}_{+}(2gt) \over \e^{t}-1}
\, , \qquad 
d_{2n} = \int_{0}^{\infty}{dt \over t}J_{2n}(2gt){\gamma_{-}(2gt)+\bar{\gamma}_{-}(2gt) \over \e^{t}-1}\, ,
\ee
only the first term in Eq.\ \re{StrEq1} involving the Baxter polynomial will contribute to the right-hand sides of Eqs.\
\re{StrEq2}. This property becomes obvious since the integrand of Eqs.\ \re{StrEq2} involves the combinations
$\gamma_{\pm}(t) \equiv [\gamma(t)\pm \gamma(-t)]/2$ (and similarly for $\bar{\gamma}_{\pm}(t)$) possessing 
specific parity under the reflection $t \to - t$, which project out the $d$-dependence on the right-hand side of Eqs.\ \re{StrEq1}.
The representation of the dressing phase given in~(\ref{StrEq2}) is just one among many. The reason to select it 
here is that it becomes convenient in large-spin analysis when one aims to cast the Baxter equation into the form of an infinite 
system of equations for the defining coefficients $\gamma_{n}, \bar{\gamma}_{n}$.

We note that for a polynomial solution~(\ref{PolSol}), the contour integrals in~(\ref{StrEq1}) are just a fancy way of
rewriting of the finite series representation for $\gamma_n$
\be
\gamma_{n} = {1 \over n}\sum_{j=1}^{S} \bigg( \frac{g}{ix^-_j} \bigg)^{n} + d_{n}\, ,
\ee
and similarly for $\bar{\gamma}_{n}$ by complex conjugation.  Making use of the reality properties of $d_n$'s for even/odd
values of its index, one finds that we have the following relations between the conserved charges \re{ConsrvdChargs} and $\gamma$'s
\be\label{ConsChg}
q_{2n} = (-1)^{n+1}g^{1-2n}\left(\gamma_{2n-1}+\bar{\gamma}_{2n-1}\right)
\, , \qquad 
q_{2n+1} = (-1)^{n+1}g^{-2n}i\left(\gamma_{2n}-\bar{\gamma}_{2n}\right)
\, .
\ee
The relations~(\ref{ConsChg}) give us a clue on the typical scaling of the coefficients $\gamma_{n}, \bar{\gamma}_{n}$ 
at weak coupling. Namely, assuming the charges $q_{n}$ are non-vanishing at one loop, i.e. $q_{n} = O(g^0)$, we get that 
$\gamma_{n} \sim \bar{\gamma}_{n} \sim g^{n}$. This could have been inferred directly from the representation~(\ref{StrEq1}) 
as well. As a particular application of~(\ref{ConsChg}), we can find the ABA energy as
\be\label{deltaABA}
E^{\rm ABA} = L + 2g\left(\gamma_{1}+\bar{\gamma}_{1}\right)\, .
\ee
We see that the same quantities $\gamma_{n}, \bar{\gamma}_{n},$ parameterizing the dressing factors of the Baxter equation, 
immediately determine the energy we are interested at constructing. A fruitful strategy, which is easily implementable at large spin, 
is then the following: first, obtain a representation of the (large-spin) solution $Q(u)$ to the Baxter equation with unspecified 
$\gamma_{n}, \bar{\gamma}_{n},$ and then derive a system of equations for these coefficients by imposing the relations~(\ref{StrEq1}). 
The last step shall be to compute the energy via~(\ref{deltaABA}).

\subsubsection{Transfer matrix $t(u)$}

The second ingredient in the Baxter equation is the function $t(u)$. It can be decomposed into two contributions
\be
t(u) = p(u) + s(u)\, ,
\ee
where $p$ is a polynomial of degree $L$ in $u$ and $s(u)$ is a non-polynomial function of $u$, whose presence 
is needed to match the analytical structure of the left-hand side of~(\ref{LRBaxterEq}).

The polynomial component, $p(u)$, can be parameterized as
\be\label{Poltu}
p(u) = 2u^{L} - \mathfrak{q}_{1}u^{L-1} - \mathfrak{q}_{2}u^{L-2} - \ldots - \mathfrak{q}_{L}\, ,
\ee
where the coefficients $\mathfrak{q}_{n}$ generalize the auxiliary conserved charges of the short-range Heisenberg 
spin chain alluded to above, to which they reduce when $g=0$. The singular part, $s(u)$, admits another type of 
expansion, which can be chosen as
\be\label{Sintu}
s(u) = 2\sum_{n\ge 1}s_{n} \bigg({ig \over x^{+}}\bigg)^{n} + 2\sum_{n\ge 1}\bar{s}_{n} \bigg({g \over ix^{-}}\bigg)^{n}\, ,
\ee
with $s_{n}, \bar{s}_{n},$ being complex-conjugate coefficients, both independent on $u$. An alternative representation, 
that holds for $-1/2 < \Im{\rm m}(u) < 1/2$, is given by
\be
\label{SuToSt}
s(u) = \int_{0}^{\infty}{dt \over t}s(2gt)\e^{iut-t/2} +  \int_{0}^{\infty}{dt \over t}\bar{s}(2gt)\e^{-iut-t/2}\, ,
\ee
where $s(t) = \sum_{n \ge 1}(2n) s_{n} J_{n}(t)$, and similarly for $\bar{s}(t)$ with $s_{n} \rightarrow \bar{s}_{n}$. 

For a polynomial solution, $Q(u) \sim u^{S}$, the two first coefficients in $p(u)$ can be written as
\be\label{qtm}
\mathfrak{q}_{1} = -2ig(\gamma_{1}-\bar{\gamma}_{1})
\, , \qquad 
\mathfrak{q}_{2} = j(j-1) + \ft{1}{4}L +2g^2L+2g^2(\gamma_{2}+\bar{\gamma}_{2})+g^2(\gamma_{1}-\bar{\gamma}_{1})^2\, .
\ee
where $j$ is the renormalized (total) $\mathfrak{sl}(2)$ spin of the state,
\be
j = S+g(\gamma_{1} + \bar{\gamma}_{1}) +\ft{1}{2}L\, ,
\ee
since $g(\gamma_{1} + \bar{\gamma}_{1})$ is exactly half of the ABA anomalous dimension, see Eq.~(\ref{deltaABA}). 
The relations~(\ref{qtm}) are a direct consequence of the asymptotics $Q(u) \sim u^{S}$ for relevant solution to the Baxter 
equation~(\ref{LRBaxterEq}).

At weak coupling, the polynomial part $p(u)$ gives the leading contribution to $t(u)$. The non-polynomial component 
$s(u)$ appears suppressed, with the typical perturbative scaling for the expansion coefficients $s_{n}$ being  $s_{n} = O(g^{n})$. 
Their precise scaling actually depends on the state and on the spin-chain length under consideration. A general expression for 
these coefficients, in terms of the Baxter polynomial and dressing factors, reads
\be\label{sn}
\begin{aligned}
s_{n} = {1\over 2(2i\pi)}\oint {dx \over x}\left[\left({x \over ig}\right)^{n} - \bigg({g \over ix}\bigg)^{n}\right] 
\Delta_{+}(u){Q(u+\ft{i}{2}) \over Q(u-\ft{i}{2})}
\, , \\
\bar{s}_{n} = {1\over 2(2i\pi)}\oint {dx \over x}\left[\left({ix \over g}\right)^{n} -\bigg({ig \over x}\bigg)^{n}\right]
\Delta_{-}(u){Q(u-\ft{i}{2}) \over Q(u+\ft{i}{2})}
\, ,
\end{aligned}
\ee
where the contour of integration goes counterclockwise and encompasses the cut along the interval $u^2 < (2g)^2$ in the $u$-plane as 
we discuss in Appendix \ref{s-coefficients}. The latter cut is hidden in the relation $x=x(u)$. The contour of integration in~(\ref{sn}) 
should be chosen in such a manner that it encircles all singularities emanating from $u= \pm \ft{i}{2}$ as the coupling increases. This 
includes the previously addressed cut, for finite values of the coupling. It is moreover not excluded that some of the singularities in 
$\Delta_{\pm}(u)$, that originally resided on the second sheet of the Riemann surface $|x|^2 < g^2$, will migrate to the first sheet. It 
appears that this is not the case in our problem, as could be inferred from the analysis of the analytical properties of the solution to the 
BES equation offered in~\cite{KSV08}. It is therefore safe to assume that the contour of integration in~(\ref{sn}) surrounds closely the 
cut in $u$ plane, that it equivalent to the circle of radius $g$ for the variable $x$.

Note finally that we can introduce a set of secondary, i.e., complementary to Eqs.\ \re{ConsChg}, charges as%
\footnote{Note that the complementary charges are real, by their definition. They do not seem to have a simple spin-chain interpretation, however.}
\be
\tilde{q}_{2n} = (-1)^{n+1}g^{1-2n}i\left(\gamma_{2n-1}-\bar{\gamma}_{2n-1}\right)
\, , \qquad 
\tilde{q}_{2n+1} = (-1)^{n}g^{-2n}\left(\gamma_{2n}+\bar{\gamma}_{2n}\right)
\, ,
\ee
which allow us to derive an alternative representation for $s_n$'s
\be\label{altsn}
s_{n} = {1\over 2(2i\pi)}\oint {dx \over x}\left[\left({x \over ig}\right)^{n} - \bigg({g \over ix}\bigg)^{n}\right] x^{L}
\exp\left[-i\vartheta -i\sum_{n\ge 1}q_{n+1}x^n + \sum_{n\ge 1}\tilde{q}_{n+1}\left({g^2 \over x}\right)^{n}\right]\, ,
\ee
where $\vartheta$ is the spin-chain total momentum,
\be
\label{QuasiMomentum}
\e^{i\vartheta} = \prod_{j=1}^{S}{x^{+}_{j} \over x^{-}_{j}}\, .
\ee
The expression~(\ref{altsn}) is convenient for the weak coupling expansion. Indeed, the contour integral in~(\ref{altsn}) 
is easily performed then, after using the known behavior of $q_{n} \sim \tilde{q}_{n} \sim g^0$ and expanding the integrand 
at weak coupling (and fixed $x$).

\subsubsection{Comments on twist-two operators}

Since we are considering a length-two chain only in the current work, let us indicate the simplification that apply to this particular case.

The polynomial twist-two solution at a given (integer) spin $S$ is unique and is described by a (real) symmetric distribution of roots, 
such that $Q(- u)=(-1)^S Q(u)$, with the change of sign for odd spins reflecting the presence of the Bethe root at $u=0$. This implies 
that the coefficients $\gamma_{n}$ are real, $\gamma_{n} = \bar{\gamma}_{n}$, and, in particular, the twist-two anomalous dimension 
is merely determined by one of them,
\be\label{ttwoABA}
E^{\rm ABA} = 2 + 4g\gamma_{1}\, .
\ee
It also follows that the polynomial part of $t(u)$ is completely fixed, with in particular $\mathfrak{q}_{1} =0$. Therefore we can write
\be
p(u) = 2u^2 -\eta^2
\, ,
\ee
where
\be
\label{EtaToS}
\eta^{2} \equiv \mathfrak{q}_{2} = j(j-1) + \ft{1}{2} +4g^2+4g^2\gamma_{2}\, , \qquad j =  S+2g\gamma_{1}+1\, .
\ee
This is the definition of the quantity $\eta = S + \ldots$ that we shall employ as an expansion parameter at large spin, in place of the spin 
$S$ itself. It is of type~(\ref{etaS}) suggested before, since $\mathbb{C}_{2} = j(j-1)$, with $\delta = \ft{1}{2} +4g^2+4g^2\gamma_{2}$. 
The latter coefficient scales logarithmically at large spin, $\gamma_{2} \sim \log{\eta}$, as do all the quantities $\gamma_{n}$, including 
in particular $\gamma_{1}$ governing the energy~(\ref{ttwoABA}).

In the same vein, the coefficients $s_n$ that control the non-polynomial part of the transfer matrix are all real, again as a consequence of 
the parity of the twist-two Baxter polynomial. We stress furthermore that $s(u)$ vanishes identically in the first two orders in perturbation 
series and starts from three-loop order for twist two operators. With regards to the large-spin limit, it is important to realize that the coefficients 
$s_{n}$ are of order $\sim \eta^{0} \sim S^{0}$ at weak coupling, up to logarithmically enhanced factors whose maximal power increases with 
the loop order. Finally, another distinguished feature of these coefficients that shows up at the leading $\sim \eta^0$ order at large spin is 
that they acquire the $(-1)^S$ spin-alternating character. This is due to the presence of the overall phase $\exp{(i\vartheta)}$ in their 
expression, see Eq.~(\ref{altsn}), which, as is well known, evaluates to $\exp{(i\vartheta)} = (-1)^S$ for twist two operators.

Keeping these remarks in mind, we are now equipped with all required machinery to attack the construction of the large-spin solution to the 
Baxter equation that will eventually determine all the coefficients $\gamma_{n}$.

\subsection{Non-polynomial solutions and large-spin expansion}

The Baxter equation \re{LRBaxterEq} is a second order finite-difference equation which admits two linearly independent solutions. When 
solving the equation order by order in the coupling constant for finite values of the spin $S$, one of the solutions is chosen as polynomial 
\re{PolSol} while the other one is not. Only the first one corresponds to the eigenstate of the eigenvalue problem for the anomalous dimension 
mixing matrix. However, when considering the large-spin expansion, it appears fruitful to look for a pair of solutions $Q_{\pm}(u)$ both of 
which are non-polynomial. They enter in our analysis through the ansatz
\be\label{DecQ}
Q(u) = Q_{+}(u) + Q_{-}(u)
\, ,
\ee
where the left-hand side is of course the sought polynomial solution. We shall also demand in the following that the relative normalization 
in~(\ref{DecQ}) is fixed by the extra condition $Q_{-}(u) = (Q_{+}(u^{*}))^{*}$.

The rationale behind the decomposition~(\ref{DecQ}) is the observation \cite{BGK06} that the sought solution to the Baxter equation, $Q(u)$, 
admits two different asymptotic expansions at large spin depending on the sign of the imaginary part of $u$. Namely, to arbitrary order in 't 
Hooft coupling the two functions $Q_\pm$ possess the behavior 
\be\label{ScQpm}
Q_{\pm}(u) \sim \eta^{\mp 2iu}\, ,
\ee
for $\eta \sim S \gg 1$ and fixed $u \sim O (\eta^0)$. As a consequence, the function $Q_{+}(u)$ dominates in~(\ref{DecQ}) for $u$ in the 
upper-half plane, and analogously for $Q_{-}(u)$ when $u$ is in the lower-half $u$-plane.%
\footnote{Notice however that for real values of $u$ both $Q_{-}(u)$ and $Q_{+}(u)$ are of the same order.} Moreover, due to their specific 
scaling (\ref{ScQpm}) the functions $Q_{\pm}$ satisfy simplified equations at large spin. These are obtained by neglecting one of the two 
terms in the left-hand side of the Baxter equation \re{LRBaxterEq} and read~\cite{BGK06}
\be\label{HBEq}
\Delta_{\pm}(u\pm\ft{i}{2})Q_{\pm}(u\pm i) = t(u)Q_{\pm}(u)\, .
\ee
In fact, since $Q_{\pm}(u\mp i)/Q_{\pm}(u) \sim 1/\eta^2$ while $Q_{\pm}(u\pm i)/Q_{\pm}(u) \sim \eta^{2}$ and $t(u) \sim - \eta^2$, not only 
do the simplified equations~(\ref{HBEq}) apply at large spin, but they are also valid up to corrections of order $O (1/\eta^4)$. For our purpose 
of constructing the large-spin solution up to order $O(1/\eta^2)$ it is therefore sufficient to consider~(\ref{HBEq}).

The construction of the large-spin solution along these lines was proposed in~\cite{BGK06} at one loop, to leading order at large spin. It was
further extended to higher loops in~\cite{BKP09} and to all loops in~\cite{V08} where it was shown to be equivalent to the BES equation~\cite{BES}, 
which is based on the Bethe-root density approach~\cite{ES}. To two-loop order, but more accurately in the spin, the analysis of the large spin 
Baxter equation was refined in~\cite{BC07} and proved to reproduce correctly the large spin expansion of the two-loop, twist-two anomalous 
dimension of~\cite{KLV}. Our analysis below essentially consists in a generalization to higher order, either in the loop or spin parameter, of all 
previous considerations. It reproduces, in particular, the BES equation and its siblings~\cite{FZ09} for the leading order solution and consistently 
reduces at two loops to the solution found in~\cite{BC07} for the subleading $\sim 1/\eta^2$ contributions.

We put forward this agreement with established analysis because the equation~(\ref{DecQ}) on which it is based is to some extent ambiguous. 
Though, in practice, the functions $Q_{\pm}(u)$ are easily constructed in a iterative fashion, following~\cite{BGK06, BC07}, the decomposition~(\ref{DecQ}) 
and asymptotics~(\ref{ScQpm}) at large spin does not seem to fix them unequivocally. It is always possible, for instance, to find new solutions, given a 
pair of solutions $Q_{\pm}(u)$, as $Q_{\pm}(u)+\alpha Q(u)$. We did not identify a convenient criterium that would pick out uniquely two such solutions 
among all possible pairs. Nevertheless, since we shall explicitly construct such a pair of solutions at large spin, it is always possible at the end to test if 
they sum up correctly to a polynomial solution, which is in principle unique. In this regard, we shall verify that our functions define a solution via~(\ref{DecQ}) 
that is holomorphic at the working accuracy in the spin. To ensure that it is the correct one would require testing that it possesses the expected asymptotics 
$\sim u^S$ at large $u$. This would turn out to be difficult because it would entail considering the regime $u \sim \eta \sim S$, and ultimately $u\gg \eta$, 
which are not covered by our analysis. We shall therefore content ourselves with partial agreement alluded to above, which together with our ability to 
reproduce all known results up to four loops, make us confident about the validity of the above solutions. It would be interesting, however, and perhaps 
crucial for analysis at higher order at large spin, to address more seriously the problem of uniqueness that we emphasized before.

Prior to presenting our results for the functions $Q_{\pm}$, let us comment a bit further on the magnitude of corrections stemming from dropped terms of 
in the Baxter equation. The above estimate $\sim 1/\eta^2$ was based on the weak coupling expansion. However, due to the logarithmic enhancement 
displayed by the power suppressed contributions, the hierarchy observed between various corrections within the context of the perturbative expansion, 
which follows from the behavior~(\ref{ScQpm}), does not survive at strong coupling. For instance, corrections of the type $1/\eta^2$ and $1/\eta^4$ both 
receive some log-enhanced contributions $g^{2n}\log^{n}{\eta}$ that change the analytical properties of the result when resummed to all orders in 't Hooft 
coupling. In fact, it is expected that all corrections of the form $\sim 1/\eta^{2k}$ shall become of the same order when $g \to \infty$. Then, as one 
undergoes a transition from weak to strong coupling, the typical scale could change from $1/\eta$ to $1/\log{\eta}$, the latter standing for the size of 
corrections found at strong coupling within the ABA framework~\cite{GSSV}. For this reason, when referring to $\sim 1/\eta^2$ corrections in the 
following discussion, we shall always mean contributions which are of this order to all loops, i.e., we shall use the `weak coupling scale' for the ordering 
of the large-spin corrections.

Now we are in a position to give the solutions to the truncated Baxter equations \re{HBEq}. Namely, up to $\sim 1/\eta^2$ corrections,
we find
\be\label{logQp}
\begin{aligned}
\log{Q_{+}(u^+)} &= \log{c_{+}}-2iu\left(\log{\bar{\eta}}-{1\over 12\eta^2}\right) 
+ 
{2i u^3 \over 3\eta^2} -2\int_{0}^{\infty}{dt \over t}{\e^{iut}J_{0}(2gt)-iut-1 \over  \e^{t}-1} \\
& + \int_{0}^{\infty}{dt \over t} {\e^{iut}\gamma(2gt)-2g\gamma_{1} t \over \e^{t}-1} 
- 
{2\over \eta^2}\int_{0}^{\infty}{dt \over t}{\e^{iut}s_{+}(2gt) \over \e^{t}-1} 
- 
{1\over \eta^2}\int_{0}^{\infty}{dt \over t}\e^{iut}s(-2gt)\, ,
\end{aligned}
\ee
where $u^{+} \equiv u+i/2$ and $c_{+}$ is an arbitrary constant. Due to the last term in the right-hand side of this equation, the representation~(\ref{logQp}) 
holds for $\Im{\rm m}(u) > 0$. Notice that, in distinction, the other integrals in~(\ref{logQp}) only assume that $\Im{\rm m}(u) > -1$. Analogously,
\be\label{logQm}
\begin{aligned}
&\log{Q_{-}(u^{-})} 
= \log{c_{-}} + 2iu\left(\log{\bar{\eta}}-{1\over 12\eta^2}\right) - {2i u^3 \over 3\eta^2} 
-2\int_{0}^{\infty}{dt \over t}{\e^{-iut}J_{0}(2gt)+iut-1 \over  \e^{t}-1} \\
&\, \, \, \, \, \, \, \, \, \, \, \, \, \,  \qquad  + \int_{0}^{\infty}{dt \over t} {\e^{-iut}\gamma(2gt)-2g\gamma_{1} t \over \e^{t}-1} 
- {2\over \eta^2}\int_{0}^{\infty}{dt \over t}{\e^{-iut}s_{+}(2gt) \over \e^{t}-1} 
- {1\over \eta^2}\int_{0}^{\infty}{dt \over t}\e^{-iut}s(-2gt)\, ,
\end{aligned}
\ee
where $u^{-} \equiv u-i/2$ and with now $\Im{\rm m}(u) < 0$ and $c_{-} = c_{+}^{*}$. In both equations, we used the definition $\log{\bar{\eta}} \equiv \log{\eta} 
+ \gamma_{E}$, with $\psi(z)$ being the digamma function and $\gamma_{\rm E}$ the Euler-Mascheroni constant, 
$\gamma_{\rm E} = - \psi (1)$. Also we introduced a parity-even combination $s_{+}(t) = [s(t)+s(-t)]/2$ of $s(t)$ defined 
in Eq.\ (\ref{SuToSt}). We notice finally that the functions~(\ref{logQp}, \ref{logQm}) have the property that $c_{-}Q_{+}(u) = c_{+}
Q_{-}(-u)$, wherefrom we find
\be\label{cpcm}
{c_{-} \over c_{+}} = {Q_{-}(-u) \over Q_{+}(+u)} = (-1)^{S}\, ,
\ee
after making use of the condition $Q(-u) = (-1)^{S}Q(u)$ obeyed by the twist-two polynomial $Q(u)=Q_{+}(u)+Q_{-}(u)$. This identity will be useful later on 
in our discussion.

\begin{figure}[t]
\begin{center}
\mbox{
\begin{picture}(0,260)(230,0)
\put(0,0){\insertfig{9}{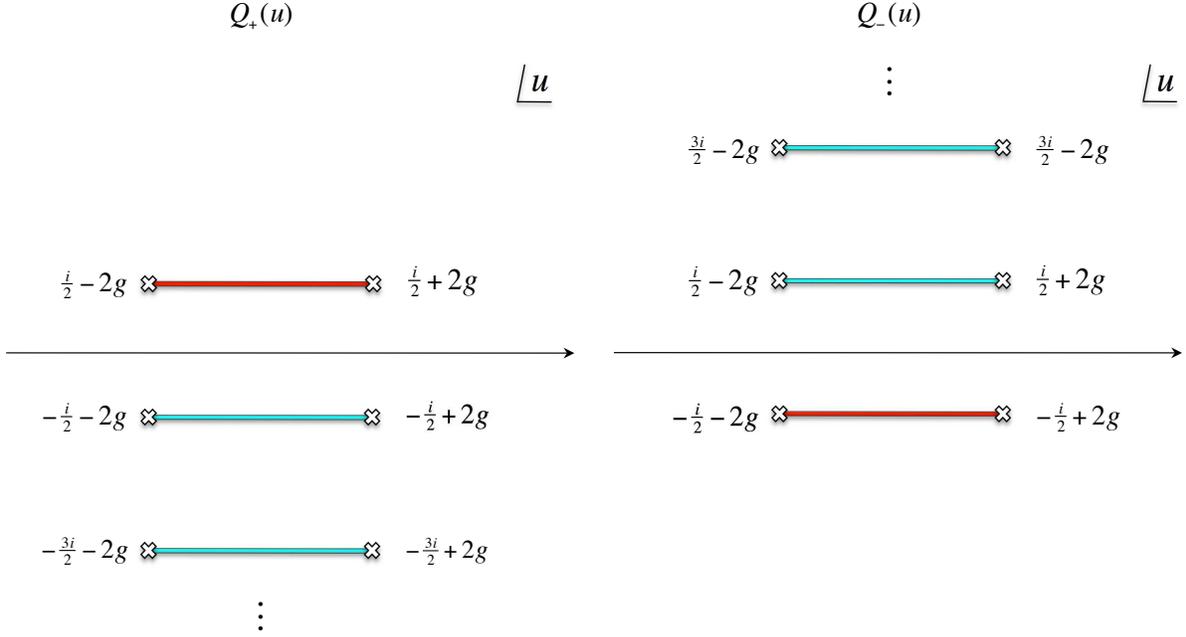}}
\end{picture}
}
\end{center}
\caption{ \label{BaxterSingularities} Singularity structure of solutions $Q_\pm (u)$ of the truncated Baxter equation.
The $Q_\pm (u)$ are holomorphic functions in the upper/lower half of the complex plane at leading order of the large
spin expansion and possess an infinite number of (blue) cuts for $\Im{\rm m} (u) \lessgtr 0$, respectively. However, they
develop (red) cuts in physical region once subleading, i.e., $1/\eta^2$, effects are taken into account.}
\end{figure}

It would not be difficult to verify that~(\ref{logQp}) and (\ref{logQm}) are solutions to the equations~(\ref{HBEq}) at the given accuracy in inverse spin 
expansion. We shall see later that they also obey the scaling~(\ref{ScQpm}) at weak coupling. Before that however, we would like to emphasize that 
even though the sum $Q(u) = Q_{+}(u) + Q_{-}(u)$ should be regular over the entire complex $u$-plane the individual functions $Q_{\pm}(u)$ are not. 
Indeed, it turns out that that these functions develop branch point singularities lying inside the strip $-2g < \Re{\rm e}(u) < 2g$, see 
Fig.~\ref{BaxterSingularities}. This is not evident in the realm of perturbation theory where, at any given order of the weak coupling expansion, 
both $Q_{\pm}(u)$ appear as meromorphic functions, though with increasing order of the poles. The presence of the cuts, however, implies that the 
equation~(\ref{HBEq}) for the functions $Q_{\pm}(u)$ apply only upon the additional requirement that the expressions $Q_{\pm}(u\pm i)$ are 
understood as argument shifts performed outside of the strip alluded to above. This is only under this restriction that the functions given below can be 
observed to solve~(\ref{HBEq}).

Let us look at the weak coupling limit. In this case, we can neglect the second lines in Eqs.\ (\ref{logQp}) and (\ref{logQm}) which enter starting from two 
loops only. This is because the functions $\gamma$ and $s$ are small at weak coupling. Then, after simplifying the first lines by keeping only the first 
term in the Taylor expansion for $J_{0}(2gt) = 1 +O(g^2)$ and applying the integral
\be
\int_{0}^{\infty}{dt \over t}{\e^{iut} - iut -1\over \e^{t}-1} = \log{\Gamma(1- iu)} + iu\psi(1)\, ,
\ee
one immediately finds that
\be
Q_{\pm}(u^{\pm}) 
= 
{c_{\pm} \over \Gamma(1\mp iu )^2} \exp{\left[\mp 2iu\left(\log{\eta}-{1 \over 12\eta^2}\right )\pm {2iu^3 \over 3\eta^2} 
+ 
o(1/\eta^2)\right]}\, .
\ee
We explicitely verify the behavior~(\ref{ScQpm}) which holds as long as $u \ll \eta$.%
\footnote{Higher-loop contributions would not affect this behavior at any given order, though they will come with a logarithmic enhancement.} After 
expansion at large values of the spin $S$, by using the one-loop relation $\eta^2 = S(S+1)+\ft{1}{2}$, this expression is find to be in complete 
agreement with the earlier results of Refs.\ \cite{BGK06, BC07} for the leading and $O(1/\eta^2)$ corrections, respectively. As one can see explicitly 
from this result both $Q_{\pm}(u)$ appear regular over the entire complex $u$-plane, and so is their sum $Q (u)$. This property persists at two loop 
order, but breaks down starting from three loop and is reflected in the appearance of poles at $u = \pm (i/2-in)$, with $n=0, 1, 2, \ldots$, in $Q_\pm (u)$, 
respectively. The regularity of the solution is restored however in the linear combination $Q (u) = Q_+ (u) + Q_- (u)$ within the given $\sim 1/\eta^2$ 
accuracy. This important property can be verified, for any value of the 't Hooft coupling, as follows.

Already at leading order of large-spin expansion, the functions $Q_{+/-} (u)$ develop an infinite number cuts in the 
lower/upper half-plane, respectively, with branch points at $-i/2-in\pm 2g$ and $i/2+in\pm 2g$, with $n= 0, 1, 2, \ldots\,$,  
as demonstrated by the blue cuts in Fig.\ \ref{BaxterSingularities}. These singularities do not compensate each another 
but instead add up such that $Q(u)$ appears to have an infinite number of cuts in the strip $-2g < \Re \textrm{e}{(u)} 
< 2g$ rather than being regular. However due to the fact that $Q_{\pm}(u) \sim \eta^{\mp 2iu}$, it is easy to convince
oneself that the singularities at $\Im\textrm{m}(u) = \pm i/2$ are actually of order $1/\eta^2$ in accord with the conclusion 
arrived from the weak coupling analysis sketched in the previous paragraph. Namely, if we consider the vicinity of 
$\Im\textrm{m}(u) \sim i/2$ we find immediately that $Q(u)$ is dominated by $Q_{+}(u) \sim \eta$ which is obviously 
regular to leading order of the large spin expansion while $Q_{-}(u) \sim 1/\eta$, which is singular there, is actually 
suppressed by $1/\eta^2$ compared to the former contribution. So in order to confirm the regularity of $Q(u)$ for 
$\Im\textrm{m}(u) \sim i/2$ we should observe that $Q_{+}(u)$ has a cut of order $O(1/\eta^2)$ (see the red cut in
Fig.\ \ref{BaxterSingularities}) which compensates the singularities introduced by $Q_{-}(u)$. This is indeed the case, 
thanks to the presence of the last contribution in Eq.\  (\ref{logQp}) stemming from the non-polynomial effects in the 
transfer matrix. Let us see how the compensation mechanism works in more details.

To order $O(1/\eta^2)$, which is the accuracy of our considerations, it suffices to consider the $\Im\textrm{m}(u) \sim i/2$
domain more carefully to define $Q(u)$ in the upper-half plane. Analogous analysis will be applicable for the lower 
semi-plane with obvious modifications. We start by rewritting $Q(u)$ making use of the Baxter equation obeyed by 
$Q_- (u)$ as follows
\be
Q(u) = Q_{+}(u) + Q_{-}(u) 
= Q_{+}(u)\left(1+{Q_{-}(u) \over Q_{+}(u)}\right) 
\simeq Q_{+}(u)\left(1-{\Delta_{-}(u-\ft{i}{2})Q_{-}(u-i) \over \eta^2 Q_{+}(u)}\right) \, ,
\ee
where the last equality holds only up to $o(1/\eta^2)$ corrections. To the same accuracy, we can replace
$Q_\pm (u)$ by $Q (u)$ in the second term
\be\label{logQcheck}
\log{Q(u+\ft{i}{2})} \simeq \log{Q_{+}(u+\ft{i}{2})} - {\Delta_{-}(u)Q_{-}(u-\ft{i}{2}) \over \eta^2 Q_{+}(u+\ft{i}{2})} 
\simeq  \log{Q_{+}(u+\ft{i}{2})} - {\Delta_{-}(u)Q(u-\ft{i}{2}) \over \eta^2 Q(u+\ft{i}{2})} \, ,
\ee
which is valid for $-1/2 < \Im{\rm m}(u) < 1/2$. Then from the condition of $\log{Q(u+\ft{i}{2})}$ to be regular around 
$u=0$ we should require that (see Appendix \ref{s-coefficients} for clarification of the origin of this equation)
\be\label{Regularityfirst}
{1\over 2i\pi}\oint {dx \over x} \left[\left({ig \over x}\right)^{n}-\left({ix \over g}\right)^{n}\right] \log{Q(u+\ft{i}{2})} = 0\, ,
\ee
for any $n =1, 2, \ldots\, $. We can check the fulfillment of this condition by plugging~(\ref{logQcheck}) into it. Then we 
immediately find that
\be\label{Regularity}
\begin{aligned}
{\eta^2\over 2i\pi}\oint {dx \over x} \left[\left({ig \over x}\right)^{n}-\left({ix \over g}\right)^{n}\right] \log{Q_{+}(u+\ft{i}{2})} 
&= {1\over 2i\pi}\oint {dx \over x}\left[\left({ig \over x}\right)^{n}-\left({ix \over g}\right)^{n}\right] {\Delta_{-}(u)Q(u-\ft{i}{2}) 
\over Q(u+\ft{i}{2})} \\
&= 2s_{n}\, ,
\end{aligned}
\ee
where in the last equality we used Eq.\ (\ref{sn}) defining the expansion coefficient $\bar{s}_n$ together with the reality 
condition$s_{n} = \bar{s}_{n}$ valid for the twist two sector. Now substituting the explicit solution (\ref{logQp}) in the left-hand 
side of Eq.~(\ref{Regularity}), we immediately find that the only non-vanishing contribution comes from the last term in Eq.\ (\ref{logQp})
\be
-{1\over \eta^2}\int_{0}^{\infty}{dt \over t}\e^{iut}s(-2gt) 
= 
-{2\over \eta^2}\sum_{n\ge 1}s_{n}\bigg({g \over ix}\bigg)^{n} \subset \log{Q_{+}(u+\ft{i}{2})}\, .
\ee
It leads to the identity $s_n = s_n$ in Eq.\ (\ref{Regularity}), after evaluation of the contour integral in the 
left-hand side of~(\ref{Regularity}) by taking the residues. This completes the verification of regularity of the complete
solution $Q (u)$ up to $O(1/\eta^2)$ accuracy of our solution for $\Im{\rm m}(u) \sim 1/2$. A similar reasoning applies 
in the lower-half plane, for $\Im{\rm m}(u) \sim -1/2$. 

It is clear however that we cannot compensate singularities that are further away from the real axis in the complex
plane (see Fig.\ \ref{BaxterSingularities}) if we do not go beyond the $1/\eta^2$ approximation of our framework. For 
$\Im{\rm m}(u) \sim \pm (\ft12 + n)$, the singularities produce effects of order $O(1/\eta^{2+2n})$ such that for $n \geq 1$ 
they fall outside of the accuracy of our consideration. In conclusion our solution $Q(u)$ is indeed regular everywhere in 
the complex plane up to $o(1/\eta^2)$ corrections. The proof of polynomiality however is beyond reach due to the
order-of-limits problem, alluded to before, as this would require sending $u$ to infinity prior to performing the large $\eta$-expansion.
Our construction was done in the opposite order and holds for $u$ fixed and for $\eta \gg 1$.

\subsection{Bulk and ABA energy}

The final step towards the construction of the ABA energy $E^{\rm ABA}$ is to derive a closed system of equations for the 
coefficients $\gamma_{n}$ introduced in Eq.\ \re{DefCoeffBis}. This is easily done with the help of the representation
(\ref{logQp}) for the Baxter solution. In the analysis that follows, it will be instructive to decompose all quantities into the 
so-called regular, or bulk, and alternating components as we advocated in the introduction. This separation is echoed 
by the expansion for $\gamma_n$'s
\be
\label{GammaToBulkAlt}
\gamma_{n} = \gamma^{\textrm{bulk}}_{n} + \gamma^{\textrm{alt}}_{n}\, ,
\ee
with each individual contribution obeying an independent set of equations. In the above decomposition, the first and 
second terms admit the following scaling with large $\eta$
\be\label{gamm-bulk-alt}
\gamma^{\textrm{bulk}}_{n} = \gamma^{\o}_{n} \log{\bar{\eta}} + \delta \gamma_{n}^{\o} + O(1/\eta^2)
\, , \qquad 
\gamma^{\textrm{alt}}_{n} = O((-1)^S/\eta^2)\, ,
\ee
respectively. In both cases, we will find it possible to cast the system of equations for $\gamma^{\textrm{.}}_{n}$ in the 
form
\be\label{GenSyst}
\gamma^{\textrm{.}}_{n} + \int_{0}^{\infty}{dt \over t}J_{n}(2gt){\gamma^{\textrm{.}}_{+}(2gt) 
- 
(-1)^{n} \gamma^{\textrm{.}}_{-}(2gt) \over \e^{t}-1} = \kappa^{\textrm{.}}_{n}\, ,
\ee
where $\kappa^{\textrm{.}}_{n}$ is a given set of coupling- and spin-dependent sources which possess asymptotics
behavior with $\eta$ analogous to one for $\gamma_n$'s as will be shown in the subsequent sections. As should be 
obvious, Eq.\  (\ref{GenSyst}) generalizes the system of equations for the solution $\gamma^{\o}_n$ determining the cusp 
anomalous dimension. The latter controls the leading large-spin asymptotics of the bulk coefficients, see Eq.\ 
(\ref{gamm-bulk-alt}), and solves (\ref{GenSyst}) for 
\be\label{kBES}
\kappa^{\o}_{n} = 2g\, \delta_{n, 1}\, .
\ee
Here (and below) $\delta_{n, m}$ is the Kronecker symbol with well-known properties, $\delta_{n, m} = \{ 1, n= m ; 0, n 
\neq m \}$. On the other hand, the subleading function $\delta\gamma^{\o}_{n}$ solves Eq.\ \re{GenSyst} with the 
following inhomogeneous term
\be\label{kdelta}
\delta \kappa_{n}^{\o} = 2\int_{0}^{\infty}{dt \over t}{J_{n}(2gt)J_{0}(2gt) -gt \delta_{n, 1}\over \e^{t}-1}\, .
\ee
It turns out that the solution $\gamma^{\o}_n$ for the leading logarithmic asymptotics plays a distinguished role in the 
construction of energy eigenvalues for more general $\kappa^{\textrm{.}}_{n}$'s. Indeed, it allows one to project out
the $n=1$ component $\gamma_{1}^{\textrm{.}}$ from the general-$n$ solution $\gamma^{\textrm{.}}_{n}$ to Eq.\ 
(\ref{kBES}) with a generic set of source coefficients $\kappa_{n}^{\textrm{.}}$ via the identity \cite{BK08}
\be\label{Wi}
4g\gamma^{\textrm{.}}_{1} = -\sum_{n\ge 1}(-1)^{n}2n\, \gamma^{\o}_{n}\, \kappa^{\textrm{.}}_{n}\, .
\ee
The demonstration of this identity can be obtained from~\cite{BK08}. So from the knowledge of the coefficients 
$\kappa_{n}^{\textrm{.}}$ and $\gamma^{\o}_{n}$, where the latter is a function of the coupling only, one can 
determine $\gamma^{\textrm{.}}_{1} $ which controls the energy $E^{\rm ABA}$. Before we move on to the construction 
of explicit results for $\gamma_{1}^{\textrm{bulk/alt}}$ obtained by means of the representation~(\ref{Wi}), we will explain
in the following few paragraphs how one achieves the separation between the bulk and alternating components and
derives the system of equations \re{GenSyst}. Here are the main steps.

To begin with, let us decompose the solution $Q_+ (u)$ in the upper half-plane as
\be
Q_{+}(u) = Q_{+}^{\textrm{bulk}}(u)Q_{+}^{\textrm{alt}}(u)\, ,
\ee
where the two factors have different scaling with $\eta$, namely $Q_{+}^{\textrm{bulk}}(u) \sim \eta^{-2iu}$ and 
$Q_{+}^{\textrm{alt}}(u) = 1 + O((-1)^S/\eta^2)$, respectively. Substituting the decomposition \re{GammaToBulkAlt}
to (\ref{logQp}) and attributing non-po\-ly\-nomial parts of the transfer matrix to $Q_{+}^{\textrm{alt}}(u)$, one deduces
\be\label{logQpbulk}
\begin{aligned}
\log{Q^{\textrm{bulk}}_{+}(u+\ft{i}{2})} 
=& \, -2iu\left(\log{\bar{\eta}}-{1\over 12\eta^2}\right) + {2i u^3 \over 3\eta^2} 
-2\int_{0}^{\infty}{dt \over t}{\e^{iut}J_{0}(2gt)-iut-1 \over  \e^{t}-1}  \\
& + \int_{0}^{\infty}{dt \over t} {\e^{iut}\gamma^{\textrm{bulk}}(2gt)-2g\gamma^{\textrm{bulk}}_{1} t \over \e^{t}-1}\, ,
\end{aligned}
\ee
and
\be\label{logQpalt}
\begin{aligned}
\log{Q^{\textrm{alt}}_{+}(u+\ft{i}{2})} 
=& \, \, \, \int_{0}^{\infty}{dt \over t} {\e^{iut}\gamma^{\textrm{alt}}(2gt)-2g\gamma^{\textrm{alt}}_{1} t \over \e^{t}-1} 
- {2\over \eta^2}\int_{0}^{\infty}{dt \over t}{\e^{iut}s_{+}(2gt) \over \e^{t}-1} 
-{2\over \eta^2}\sum_{n\ge 1}s_{n}\left({g \over ix}\right)^{n}\, ,
\end{aligned}
\ee
where in the last term we made use of the equality
\be
-{1\over \eta^2}\int_{0}^{\infty}{dt \over t}\e^{iut}s(-2gt) = -{2\over \eta^2}\sum_{n\ge 1}s_{n}\left({g \over ix}\right)^{n}\, ,
\ee
to get an alternative representation for $\log{Q^{\textrm{alt}}_{+}(u+\ft{i}{2})}$ which turns out more convenient for our 
purposes. At this point the following remark is in order on the analytical properties of the above two components.
While the alternating contribution $Q^{\textrm{alt}}_{+}(u)$ inherits the cut in the upper half plane with the branch points 
at $u=i/2\pm 2g$ from the one of the full $Q_+ (u)$, the $Q^{\textrm{bulk}}_{+}(u)$ does not and is thus holomorphic in
the upper half-plane. However, both $Q^{\textrm{bulk}}_{+}(u)$ and $Q^{\textrm{alt}}_{+}(u)$ have infinitely many cuts in 
lower-half plane, with branch points at $u=-in/2 \pm 2g$, $n=1, 2, \ldots$.

Note that in order to evaluate $\gamma_{n}^{\textrm{alt}}$ from Eq.\ \re{StrEq1}, we need to account for the presence 
of $Q_{-}(u)$ in $Q(u)$ to stay within the accuracy of our approximation. To this end we first rewrite
\be
\log{Q}(u) = \log{Q_{+}}(u) + \log{(1+Y_{\textrm{h}}(u))}
\, , \qquad 
Y_{\textrm{h}}(u) = {Q_{+}(u) \over Q_{-}(u)} \sim (-1)^S\eta^{4iu}\, ,
\ee
and observe that in the upper-half plane, $\Im{\rm m}(u) > 0$, and particularly in the vicinity of $u=i/2$,
\be
Y_{\textrm{h}}(u+\ft{i}{2}) \sim (-1)^S/\eta^2 
\, .
\ee
So $Y_{\textrm{h}}(u+\ft{i}{2})$ is a small correction to $\log{Q}(u+\ft{i}{2})$ in the upper-half plane that contributes
at this order in the large-$\eta$ expansion to the alternating part only. This counting allows one to write
\be\label{logQ-Y}
\log{Q}(u+\ft{i}{2}) = \log{Q_{+}}(u+\ft{i}{2}) + Y_{\textrm{h}}(u+\ft{i}{2}) + \ldots\, , 
\ee
where the ellipsis stand for $O(1/\eta^4)$ corrections. The second term $Y_{\textrm{h}}(u+\ft{i}{2})$ contributes 
to the final expression for $\gamma^{\textrm{alt}}_{n}$ after plugging~(\ref{logQ-Y}) into~(\ref{StrEq1}) at 
power-suppressed level. So in summary, the two formulas we use for the extraction of $\gamma_{n}^{\textrm{bulk/alt}}$ 
can be written as
\be\label{gam-n}
\begin{aligned}
\gamma^{\textrm{bulk}}_{n} 
&= -{1\over 2i\pi}\oint {dx \over x}\bigg({g \over ix}\bigg)^{n}\log{Q_{+}^{\textrm{bulk}}(u+\ft{i}{2})} 
+ d_{n}^{\textrm{bulk}}\, ,\\
\gamma^{\textrm{alt}}_{n}\, \, \,  \,  
&= -{1\over 2i\pi}\oint {dx \over x}\bigg({g \over ix}\bigg)^{n}\log{Q_{+}^{\textrm{alt}}(u+\ft{i}{2})}  
-{1\over 2i\pi}\oint {dx \over x}\bigg({g \over ix}\bigg)^{n}Y_{\textrm{h}}(u+\ft{i}{2})+ d_{n}^{\textrm{alt}}\, ,
\end{aligned}
\ee
where the separation of the dressing phase contribution into the bulk and alternating terms merely follows from 
the one for the coefficients $\gamma_{n}$ in Eq.~(\ref{GammaToBulkAlt}).

The final step then boils down to evaluation of Eq.\ (\ref{gam-n}) with the help of~(\ref{logQpbulk}) and (\ref{logQpalt}). 
This together with the expression for the coefficients $d_{n}$ defining the dressing phase and the identity
\be
J_{n}(2gt) = {1 \over 2i\pi}\oint {dx \over x} \bigg({g \over ix}\bigg)^{n}\e^{iut} 
= 
{1 \over 2i\pi}\oint {dx \over x} \left({x \over ig}\right)^{n}\e^{iut}\, , \qquad u=x+g^2/x
\, ,
\ee
gives the system of equations~(\ref{GenSyst}) with prescribed $\kappa_{n}^{\textrm{.}}$

\subsubsection{Bulk energy}

Now we turn to finding the bulk energy making use of the representation \re{Wi}. The source term defining it is rather
simple and reads
\be\label{kbulk}
\kappa^{\textrm{bulk}}_{n} 
= \left(\log{\bar{\eta}}-{1\over 12\eta^2}\right)\kappa^{\o}_{n} + \delta \kappa_{n}^{\o} 
+ \kappa_{n}^{\textrm{sub}} + o(1/\eta^2)\, ,
\ee
where $\kappa^{\o}_{n}, \delta \kappa^{\o}_{n},$ were previously given in Eqs.\ (\ref{kBES}) and (\ref{kdelta}), respectively,
while the last contribution is
\be\label{ksub}
\kappa_{n}^{\textrm{sub}}  = -{2g^3 \over \eta^2}\delta_{n, 1} + {2g^3\over 3\eta^2}\delta_{n, 3}\, .
\ee
One could obviously absorb the term $\sim \delta_{n, 1}$ present in $\kappa_{n}^{\textrm{sub}}$ into the coefficient 
in front of $\kappa^{\o}_n$ in~(\ref{kbulk}) since the latter is localized at $n=1$ as well, and then deal separately with 
the component $\sim \delta_{n, 3}$ in $\kappa_{n}^{\textrm{sub}}$, which to an extent defines the only new problem. 
The decomposition~(\ref{kbulk}, \ref{ksub}) is nevertheless convenient since it systematically separates contributions
in the $1/\eta^2$ expansion that appear with different transcendentalities at a given order in weak coupling.

If we are interested in the (bulk) energy only, $E^{\textrm{bulk}} = 2 + 4g\gamma_{1}^{\textrm{bulk}}$, then only the 
coefficient $\gamma_{1}^{\textrm{bulk}}$ matters, and, by means of Eq.\ (\ref{Wi}), we know that the former admits the 
representation
\be\label{bulkenform}
E^{\textrm{bulk}} 
= 
4g\gamma_{1}^{\o}\left(\log{\bar{\eta}}-{1\over 12\eta^2}\right) 
+ 
2\int_{0}^{\infty}{dt \over t}{\gamma^{\o}(-2gt)J_{0}(2gt) -2g\gamma_{1}^{\o}t\over \e^{t}-1} 
-
{4g^3 \over \eta^2}\gamma^{\o}_{1} + {4g^3\over \eta^2}\gamma^{\o}_{3}\, ,
\ee
up to $o(1/\eta^2)$ corrections. Therefore, from the knowledge of the solution to the BES equation, that is, from the 
coefficients $\gamma^{\o}_{n}$ or equivalently their generating function $\gamma^{\o}(t) = \sum_{n \ge 1}(2n) 
\gamma^{\o}_{n}J_{n}(t)$, one obtains the expression for the bulk energy up to order $O(1/\eta^2)$. As we can see 
it takes the form
\be\label{Ebulk-C}
E^{\textrm{bulk}} = 2A\left(\log{\bar{\eta}}-{1\over 12\eta^2}\right) + B + C/\eta^2 + \ldots\, ,
\ee
where $A \equiv 2g\gamma_{1}^{\o}$ is the cusp anomalous dimension and $B$ the twist-two virtual scaling function~\cite{FZ09},
\be
B \equiv 
2+4g\delta\gamma_{1}^{\o} = 2+ 2\int_{0}^{\infty}{dt \over t}{\gamma^{\o}(-2gt)J_{0}(2gt) -2g\gamma_{1}^{\o}t\over \e^{t}-1}\, ,
\ee
where we cast it in the representation for the function $B$ equivalent to the one given in~\cite{FZ09}. The weak coupling 
expansion of the new coefficient $C$ is given up to five-loops by
\be
C  \equiv -4g^3 \gamma^{\o}_{1} + 4g^3\gamma^{\o}_{3} 
= -
8g^4 + {8\pi^2 \over 3}g^6 -{32\pi^4 \over 15}g^8 +\left({2048\pi^6 \over 945} + 64\zeta_{3}^2\right)g^{10} + \ldots\, . 
\ee

Re-expanding the bulk energy in terms of the spin $S$ we induce odd powers in the inverse spin in addition to the
general structure in terms of the $\eta$-variable,
\be
E^{\textrm{bulk}} = 2A\log{\bar{S}} + B + \sum_{n = 1, 2}C_{n}/S^n + \ldots\, .
\ee
Here $C_{n}$ is a polynomial of degree $n$ in $\log{\bar{S}} = \log{S} + \gamma_{\rm E}$ with coupling constant-depending
coefficients. For the reason that the degree of the polynomials $C_{n}$ remains bounded at any value of the coupling, the 
bulk energy is  quite gentle from the point of view of the interpolation between the regimes of weak and strong coupling.
The explicit expressions for the polynomials $C_{n}$ are not quite illuminating and will not be quoted here. They immediately 
follow from~(\ref{Ebulk-C}) and the $\eta = \eta(S)$ relation for twist two case, see Eq.~(\ref{EtaToS}). Up to five loops, all what is needed 
are the expressions for $\gamma^{\o}_n$ up to $n=4$,
\be\label{gammn}
\begin{aligned}
\gamma_{1}^{\o} &= 2g-{2\pi^2 \over 3}g^3 +{22\pi^4 \over 45}g^5 -\left({146 \pi^6 \over 315}+16\zeta_{3}^2\right)g^7 
+ 
\left({7096 \pi^8 \over 14175}+{32 \pi^2\zeta_{3}^2\over 3}+320\zeta_{3}\zeta_{5}\right)g^9 + \ldots\, , \\
\gamma_{2}^{\o} &= 4\zeta_{3}g^4 -\left({4\pi^2\zeta_{3} \over 3}+40\zeta_{5}\right)g^6 
+ 
\left({32\pi^4\zeta_{3} \over 45}+{40\pi^2\zeta_{5} \over 3}+420\zeta_{5}\right)g^8 + \ldots\, ,\\
\gamma_{3}^{\o} &= -{2\pi^4 \over 45}g^5+{74\pi^6 \over 945}g^7-\left({326 \pi^8 \over 2835}
+
32\zeta_{3}\zeta_{5}\right)g^9 + \ldots\, ,\\
\gamma_{4}^{\o} &= 4\zeta_{5}g^6-\left({4\pi^2\zeta_{5} \over 3}+84\zeta_{7}\right)g^8 + \ldots\, ,\\
\end{aligned}
\ee
and similarly for $\delta\gamma_{n}^{\o}$,
\be\label{dgammn}
\begin{aligned}
\delta\gamma_{1}^{\o} &= -6\zeta_{3}g^3 + \left({4\pi^2 \over 3}\zeta_{3}
+
40\zeta_{5}\right)g^5 - \left({14\pi^4\zeta_{3} \over 15}+{20\pi^2\zeta_{5} \over 3}+350\zeta_{7}\right)g^7 \\
& \qqqquad \qqqquad \qqqquad\qquad + \left({32\pi^6\zeta_{3} \over 35}
+{208\pi^4\zeta_{5} \over 45}+{140\pi^2\zeta_{7} \over 3}+ 48\zeta_{3}^3+3528\zeta_{9}\right)g^9+ \ldots\, , \\
\delta\gamma_{2}^{\o} &= {\pi^2 \over 6}g^2-{4\pi^4 \over 45}g^4+\left({43\pi^6 \over 630}
-
12\zeta_{3}^2\right)g^6 - \left({892 \pi^8 \over 14175}-{8\pi^2\zeta_{3}^2 \over 3}-208\zeta_{3}\zeta_{5}\right)g^8 + \ldots\, ,\\
\delta\gamma_{3}^{\o} &= {2\zeta_{3} \over 3}g^3 -10\zeta_{5}g^5
+
\left({2\pi^4\zeta_{3} \over 15}-{4\pi^2\zeta_{5} \over 3}+126\zeta_{7}\right)g^7  \\
& \qqqquad \qqqquad \qqqquad\qquad
-\left({664\pi^6\zeta_{3} \over 2835}+{8\pi^4\zeta_{5} \over 45}-{70\pi^2\zeta_{7} \over 3}+1568\zeta_{9}\right)g^9 + \ldots\, ,\\
\delta\gamma_{4}^{\o} &= {\pi^4 \over 180}g^4 -{4\pi^6 \over 315}g^6 
+ 
\left({269\pi^8 \over 14175}-12\zeta_{3}\zeta_{5}\right)g^8 + \ldots\, .
\end{aligned}
\ee
Here we have also included coefficients that will be useful later.

\subsubsection{Alternating contribution}

Finally, let us address the alternating contribution to the ABA energy.  The source terms for this one is more involved but 
can be derived in a straightforward manner from~(\ref{logQpalt}) and \re{gam-n} to take the form
\be\label{kappalt}
\kappa_{n}^{\textrm{alt}} 
= {2\over \eta^2}\int_{0}^{\infty}{dt \over t}{J_{n}(2gt)s_{+}(2gt) \over \e^{t}-1} 
- {1\over 2i\pi}\oint {dx \over x} \bigg({g \over ix}\bigg)^{n}Y_{\textrm{h}}(u+\ft{i}{2})\, ,
\ee
where $s_+ (t)$ is found from the infinite series
\be
\label{sPlusinBessel}
s_{+}(t) = \sum_{n\ge 1}2(2n)s_{2n}J_{2n}(t)
\ee
with the expansion coefficients that read
\be
s_{2n} = {1\over 4i\pi \eta^2}\oint {dx \over x} \left[\left({ig \over x}\right)^{2n}-\left({ix \over g}\right)^{2n}\right]
Y_{\textrm{h}}(u+\ft{i}{2})\, ,
\ee
up to corrections smaller than $1/\eta^2$ at any order in weak coupling.

The system of equations~(\ref{GenSyst}) for $\gamma_{n}^{\textrm{alt}}$ appears closed once one realizes that to 
compute $\kappa_{n}^{\textrm{alt}}$, to arbitrary order in the coupling, one merely needs to know $Y_{\textrm{h}}(u+i/2)$ at 
leading order at large spin. The latter function can be considered as being known once the solutions $\gamma^{\o}_n$ 
to the BES equation together with its subleading components $\delta\gamma^{\o}_n$ have been constructed.%
\footnote{This remark becomes obvious after looking at the Eq.~(\ref{Yalt}) where $Y_{\textrm{h}}(u+\ft{i}{2})$ is given in terms 
of $\gamma_{n}$. Manifestly, to construct $Y_{\textrm{h}}(u+\ft{i}{2})$ at order $O(1/\eta^2)$ we only need to use that $\gamma_{n} 
= \gamma^{\o}_{n} \log{\bar{\eta}} + \delta \gamma_{n}^{\o} + O(1/\eta^2)$.} Then, after plugging the coefficients $\kappa_{n}^{\textrm{alt}}$ 
into the system of equations~(\ref{GenSyst}) and solving the latter, one gets the expression for $\gamma_{n}^{\textrm{alt}}$.

Instead of solving the system of equations~(\ref{GenSyst}) with the inhomogeneous terms~(\ref{kappalt}), we apply the 
formulae~(\ref{Wi}) in order to extract the coefficient $\gamma_{1}^{\textrm{alt}}$ which controls the energy. The expression 
so obtained is parameterised by the solution to the BES equation, i.e., $\gamma^{\o}$ solving~(\ref{GenSyst}) with
(\ref{kBES}), and the coefficients~(\ref{kappalt}). It simplifes after using the fact that $s_{+}(2gt)$ in~(\ref{kappalt}) can 
be decomposed over the even Bessel functions \re{sPlusinBessel} and then the equations obeyed by the solution  $\gamma^{\o}$ . 
Proceeding along these lines we derive the following representation for $E^{\textrm{alt}} = 4g\gamma_{1}^{\textrm{alt}}$,
\be\label{Ealt}
E^{\textrm{alt}}  
= {1\over 2i\pi} \oint{dx \over x}
\bigg[\sum_{n \ge 1}2(2n-1)\gamma_{2n-1}^{\o}\left({ig \over x}\right)^{2n-1}
+
\sum_{n \ge 1}2(2n)\gamma_{2n}^{\o}\left({ix \over g}\right)^{2n}
\bigg] Y_{\textrm{h}}(u+\ft{i}{2})\, ,
\ee
which holds again up to $o(1/\eta^2)$ corrections. In this formula, the contour of integration goes around the cut along
$u^2 <(2g)^2$ present in $x(u)$ and $Y_{\textrm{h}}(u+\ft{i}{2})$.

To evaluate the integral~(\ref{Ealt}) at weak coupling, we can use that
\be\label{Yalt}
Y_{\textrm{h}}(u+\ft{i}{2}) 
= -\left({x \over \eta}\right)^{2}\exp{\bigg[i\vartheta 
+ 2\sum_{n \ge 1}\gamma_{2n-1}\left({ix \over g}\right)^{2n-1}
+ 2\sum_{n \ge 1}\gamma_{2n}\left({ig \over x}\right)^{2n}\bigg]}\, ,
\ee
up to $o(1/\eta^2)$ corrections (at any order in weak coupling). This representation follows from the consideration analogous to the one which lead to 
the integrand in Eq.\  \re{altsn}. We see that due to the presence of the factor $\exp{(i\vartheta)} = (-1)^{S}$ we immediately
find that $Y_{\textrm{h}}(u+\ft{i}{2}) \sim (-1)^{S}/\eta^2$ and thus $E^{\textrm{alt}} \sim (-1)^{S}/\eta^2$ confirming 
and explaining the naming scheme we adopted for this contribution.

Now, recalling that $\gamma_{n} \sim g^{n}$ for $g\sim 0$, we observe that we can compute $E^{\textrm{alt}}$ at weak 
coupling by expanding those terms in~(\ref{Yalt}) that involve inverse powers of $x$, e.g., $(g/x)^n$, and then perform 
the contour integral in~(\ref{Ealt}) by the method of residues. For instance, the leading order contribution at weak coupling
arises from
\be
E^{\textrm{alt}} = -{(-1)^{S}\over \pi \eta^2} g \gamma_{1}^{\o}
\oint dx \exp{\bigg[2\sum_{n \ge 1}\gamma_{2n-1}\left({ix \over g}\right)^{2n-1}\bigg]} + \ldots = 0 + O(g^4)
\, ,
\ee
which in light of the dependence $\gamma_{1}^{\o} \sim g$ implies the $E^{\textrm{alt}}$ vanishes at one loop. Proceeding 
to higher loops is straightforward, though slightly more tedious, and after some algebra one could derive the following
expression
\be\label{Ealt5}
E^{\textrm{alt}} 
= -{(-1)^{S}\over \eta^2}\bigg[8g^2\gamma_{1}^{\o}\gamma_{1}\gamma_{2} 
+ 12g^2\gamma_{3}^{\o}\gamma_{1}+ 8g^2\gamma_{1}^{\o}(\gamma_{4}
+\gamma_{2}^2)(\gamma_{3}+\ft{2}{3}\gamma_{1}^3) 
-8g^2\gamma_{2}^{\o}(\gamma_{4}+\gamma_{2}^2) + \ldots\bigg]\, ,
\ee
which suffices for evaluation up to five-loop order.%
\footnote{Note that the validity of~(\ref{Ealt5}) up to five loops also relies on the fact that the coefficients 
$\gamma_{n}^{\o}$ are actually more suppressed at weak coupling than naively expected. They indeed 
satisfy $\gamma^{\o}_{n} \sim g^{n+2}$ in place of $\gamma^{\o}_{n} \sim g^{n}$, except for $n=1$ where 
the naive scaling applies, $\gamma^{\o}_{1} \sim g$. The latter properties allowed us to remove some terms 
in~(\ref{Ealt5}), as $\sim g^2\gamma^{\o}_{5}\gamma_{3}$ and $\sim g^2\gamma^{\o}_{3}\gamma_{2}\gamma_{3}$ 
for instance, that eventually enter at $O(g^{12})$ instead of $O(g^{10})$.} 

Finding the weak coupling expansion is now straightforward by substituting the large-$\eta$ expansion for the
coefficients $\gamma_{n}$, i.e., $\gamma_{n} = \gamma^{\o}_{n}\log{\bar{\eta}} + \delta\gamma^{\o}_{n} + O(1/\eta^2)$, 
together with the explicit weak coupling results~(\ref{gammn}) and (\ref{dgammn}). Eventually we find $E^{\textrm{alt}}$ to
admit the following generic form
\be\label{EaltPol}
E^{\textrm{alt}} = {(-1)^S \over \eta^2}\sum_{n\ge 2}g^{2n}P_{n} + o(1/\eta^2)\, ,
\ee
where $P_{n}$ is a polynomial of degree $(n-2)$ in the $\log{\bar{\eta}}$ variable. As we shall see in a later section, 
the structure~(\ref{EaltPol}) is actually correct to all loops, though this is not transparent from~(\ref{Ealt5}). This pattern 
contrasts with the simplicity of the bulk energy, which, even after re-expanding in the spin $S$, displays a logarithmic 
enhancement in $\log{\bar{S}}$ that is strictly bounded in power at any coupling.%
\footnote{The structure~(\ref{Ealt}) on the other hand reflects the fact that $E^{\textrm{alt}}$ is non-linear in the coefficients 
$\gamma_{n} \sim \log{\bar{\eta}}$, see Eq.~(\ref{Ealt5}).}
We want to emphasize again that the corrections we are neglecting in~(\ref{EaltPol}) are smaller than $1/\eta^2$ to 
any order at weak coupling only. These corrections are indeed expected to exhibit the same kind of logarithmic 
enhancement as the one accompanying the above $O(1/\eta^2)$ contributions, and their resummation at finite coupling 
may considerably complicate the picture. At strong coupling, the contributions controlled by~(\ref{Ealt}) could not capture 
the leading large-spin expression of $E^{\textrm{alt}}$ anymore.%
\footnote{Strictly speaking this phenomenon is not restricted to the alternating component of the ABA energy, 
because as we increase the accuracy in $1/\eta$ we will generate corrections quite similar to the ones controlled 
by the right-hand side of~(\ref{Ealt}) but with $Y_{\textrm{h}}(u+i/2)$ raised to some power. These corrections have 
no reason to alternate with the spin anymore and thus should contribute to the regular (i.e., non-alternating) part of 
the energy. One may expect however that there exists a more general decomposition of the ABA energy into a 
`simple' bulk contribution and a `complicated' remaining piece, with the latter part being identical to the alternating 
component of the energy to the leading $\sim 1/\eta^2$ order but otherwise receiving regular contributions beyond 
it. A possible example for such a remaining piece can be obtained from the right-hand side of~(\ref{Ealt}) with 
$Y_{\textrm{h}}(u+i/2) \rightarrow \log{(1+Y_{\textrm{h}}(u+i/2))}$. This would be suggestive of a TBA-like interpretation 
of the decomposition between the presumed all-order-in-$1/\eta$ bulk and remaining parts, which is however far 
beyond the scope of our analysis.} This result is anticipated to be the case at strong coupling in the regime where 
perturbative string theory is applicable and estimates for subleading large-spin corrections suggest that they scale 
as $\sim 1/\log{S}$, see~\cite{SNZ, BDFPT, GRRT}.

After this general remarks we come back to our calculation and present the weak coupling expansion for 
$E^{\textrm{alt}}$ up to five loop order. The polynomials entering Eq.\ (\ref{EaltPol}) are given by
\be
\begin{aligned} \label{ABAPol}
&P_{2} = 0\, , \\
&P_{3} = -{16\pi^2 \over 3}\log{\bar{\eta}}\, , \\
&P_{4} = -128\zeta_{3}\log^2{\bar{\eta}} +{112\pi^4 \over 15}\log{\bar{\eta}} +16\pi^2\zeta_{3}\, , \\
&P_{5} =-{128\pi^4 \over 45}\log^{3}{\bar{\eta}} + \left(1280\zeta_{5}+128 \pi^2 \zeta_{3}\right)\log^2{\bar{\eta}} 
+ 
\left(768\zeta_{3}^2 -{8992 \pi^6 \over 945}\right) \log{\bar{\eta}}-\left({320\pi^2\zeta_{5} \over 3}
+
{896\pi^4 \zeta_{3} \over 45}\right)\, .
\end{aligned}
\ee
According to our preliminary findings, we encoded the fact that $E^{\textrm{alt}}$ does not receive corrections till
two loops by starting the summation in Eq.\ (\ref{EaltPol}) with $n=2$ even though $P_{2}=0$. The reason for this
is that in distinction to the one-loop contribution $P_1$ which vanishes exactly at any spin, the two-loop alternating 
contribution is non-vanishing in general but turns out to be suppressed as $1/S^4$ at large spin. This can be easily 
verified starting from the exact (i.e., finite spin),  two-loop, twist-two anomalous dimension~\cite{KLV}.%
\footnote{Methods for separating regular and alternating contributions from exact representations usually involving 
nested harmonic sums can be found in or adapted from~\cite{KV}.}  We also checked along the same lines that our 
result (\ref{ABAPol}) is in agreement with the three-loop and four-loop-ABA anomalous dimension obtained in
Refs.\ \cite{KLOV, KLRSV}. The five-loop ABA anomalous dimension, though available~\cite{LRV}, is too cumbersome 
and difficult to handle at large spin to perform the comparison.

Before we close this subsection, let us give one more representation for the alternating energy~(\ref{Ealt}) which 
will be useful for our subsequent analysis at strong coupling. It reads
\be\label{RepReal}
E^{\textrm{alt}} = {i \over \pi}\oint dE_{\textrm{h}}(u+\ft{i}{2}) Y_{\textrm{h}}(u+\ft{i}{2}) + o(1/\eta^2)\, ,
\ee
where $E_{\textrm{h}}(u)$ is the energy of the scalar excitation (hole) computed in~\cite{B10, AGMSV10}.  The contour 
of integration in~(\ref{RepReal}) encircles the cut running from $u=-2g-i/2$ to $u=2g-i/2$ (as shown in Fig.\ \ref{Riemann}) 
present both in $Y_{\textrm{h}}(u)$ and $E_{\textrm{h}}(u)$. The proof of the representation~(\ref{RepReal}) goes as follows. 
First, we notice that the term in square brackets in the right-hand side of Eq.\ (\ref{Ealt}) can be written as
\be
2x{d\over dx}\bigg[\sum_{n \ge 1}\gamma_{2n}^{\o}\left({ix \over g}\right)^{2n}-\sum_{n \ge 1}\gamma_{2n-1}^{\o}\left({ig \over x}\right)^{2n-1} \bigg]\, ,
\ee
and then we use the formula
\be
E_{\textrm{h}}(u+\ft{i}{2}) = 1 + \sum_{n \ge 1}\gamma_{2n-1}^{\o}\left({ig \over x}\right)^{2n-1}-\sum_{n \ge 1}\gamma_{2n}^{\o}\left({ix \over g}\right)^{2n}\, .
\ee
The latter can be easily derived from the representation given for $E_{\textrm{h}}(u)$ given in~\cite{B10}. Plugging it back 
into Eq.\ (\ref{Ealt}) immediately yields (\ref{RepMirr}).

\section{Finite-size corrections}
\label{FSsection}

In this section, we will construct and analyze the expression for the finite-size corrections to the energy of the GKP string.

To understand and motivate our proposal it helps recalling some distinguished features of the spectrum and dynamics of excitations on the latter background. To start with, let us address the simplest set of excitations propagating on the GKP string, the holes. These naturally arise within the framework of
the asymptotic solution to the Baxter equation. It will suffice for our purposes to restrict the consideration to one loop order. What we are looking for 
is the solution to Eq.\ \re{LRBaxterEq} valid for the spectral parameter scaling as $u \sim O (S^0)$. The latter condition merely reflects the 
value of the argument of the Baxter function set to calculate the anomalous dimension, i.e., $u = \pm \ft{i}{2}$. The method is heavily based on the 
observation that the two terms entering the left-hand side of the Baxter equation \re{LRBaxterEq} possess different asymptotics as $S \to \infty$. The 
reason behind it becomes very transparent when one notices that the right-hand side behaves as $\sim S^2$ due to the appropriate scaling of the 
large-spin twist-$L$ transfer matrix
\be
t(u) = -\eta^2\prod_{j=1}^{L-2}(u-\delta_{j})\, ,
\ee
where $\eta = S+\ldots$. Introducing the ratio of the Baxter polynomials $\phi (u) = Q (u + i)/Q (u)$, one realizes that 
in order to match the scaling behavior of the left-hand side it should be either large or small. Therefore, it is legitimate
to neglect one of the terms. This procedure yields two first-order finite-different equations instead of one second order.
Then, the one-loop large-spin solution immediately reads
\be\label{BGKqp}
Q_{+}(u) = c_{+}\eta^{-2iu}{\prod_{j=1}^{L-2}\Gamma(-iu+i\delta_{j})\over \Gamma(\ft{1}{2}-iu)^L}
\, ,
\ee
where $c_+$ is an overall normalization constant. From the regularity of the general solution $Q(u) = Q_{+}(u) + Q_{-}(u)$, where $Q_{-}(u^*) = Q_{+}(u)^*$ 
and the real-hole assumption, i.e., $\Im{\rm m} (\delta_k) = 0$, we get the condition
\be\label{BYhole}
1+Y_{\textrm{h}}(\delta_{k}) = 0\, ,
\ee
for the ratio
\be
Y_{\textrm{h}}(u) = {Q_{-}(u) \over Q_{+}(u)} \, .
\ee
Equation~(\ref{BYhole}) is the requirement for the absence of poles in $Q(u)$ and yields the following quantization conditions
when one substitutes the explicit solution \re{BGKqp} in it,
\be\label{CC}
1+\eta^{4i\delta_{k}}{c_{-} \over c_{+}}
{\Gamma(\ft{1}{2}-i\delta_{k})^L\over \Gamma(\ft{1}{2}+i\delta_{k})^L}\prod_{j=1}^{L-2}{\Gamma(i\delta_{k}-i\delta_{j}) \over \Gamma(i\delta_{j}-i\delta_{k})} = 0\, ,
\ee
where the condition on the ratio of the unknown constants emerges from the total momentum condition \re{QuasiMomentum}
\be\label{eCC}
{c_{-} \over c_{+}} = \e^{i\vartheta}\prod_{j=1}^{L-2}{\Gamma(\ft{1}{2}+i\delta_{j}) \over \Gamma(\ft{1}{2}-i\delta_{j})}\, .
\ee

The equations~(\ref{BYhole}, \ref{CC}, \ref{eCC}) can be interpreted as the Bethe-Yang equations for the factorized scattering on top of the GKP string. 
Indeed the $Y$ function apparently factorizes as
\be\label{StrY}
Y_{\textrm{h}}(u) = \e^{i\vartheta}\e^{iP_{\textrm{h}}(u)}\prod_{j=1}^{L-2}S(u, \delta_{j}) \, ,
\ee
with the two-to-two $S$ matrix
\be
S(u, \delta_{j}) 
= 
{\Gamma(iu-i\delta_{j}) \over \Gamma(i\delta_{j}-iu)}{\Gamma(\ft{1}{2}-iu) \Gamma(\ft{1}{2}+i\delta_{j})  \over \Gamma(\ft{1}{2}+iu) \Gamma(\ft{1}{2}-i\delta_{j}) }\, ,
\ee
and the propagating phase
\be\label{OLPu}
\e^{iP_{\textrm{h}}(u)} = \eta^{4iu}{\Gamma(\ft{1}{2}-iu)^2 \over \Gamma(\ft{1}{2}+iu)^2}\, .
\ee

The interpretation of the quantity $2\log{\eta}$ as defining the physical length then suggests to introduce the reciprocal momentum $p(u) = 2u$ as it appears 
conjugated to the former in the progagating phase~(\ref{OLPu}). The form of the latter remains problematic however since it bears an extra phase in~(\ref{OLPu}) 
that does not seem to admit a clear interpretation. The S-matrix also is a bit surprising in form since it is not a function of the rapidity difference. Both 
phenomena can be traced back to the presence of the dressing factors in the Baxter equation, which are typical to spin-chain problems. In a relative set-up for 
the periodic Toda chain, these factors are absent and the quantization conditions have the standard Bethe form where the S-matrix is a function of the rapidity
difference only and the propagating phase is the exponential of the momentum times a parameter that plays the role of an effective length, see~\cite{GP,NS,KT} 
and references therein. These two features of the propagation and scattering on top of the GKP string are possibly related to the fact that this background is not 
quite homogeneous~\cite{FT02}, as observed  at strong coupling in explicit perturbative computations on the string theory side~\cite{BDFPT}. Though puzzling, 
these do not prevent us from trying to apply standard recipes to our problem. Note finally the presence of the total spin-chain momentum in~(\ref{StrY}). It endows
the current consideration with an extra lever arm allowing us to vary possible boundary conditions for the boson, and deviate from the conventional case of zero 
quasi-momentum for periodic boundary conditions.

The structure~(\ref{StrY}) is quite generic and holds to given accuracy in the large-spin expansion for any value of the 't Hooft coupling, the factorization being 
guaranteed by the approximation and/or the fact that the problem we solve is equivalent to an open spin-chain problem. This way one can define an all-loop 
momentum and all-loop S-matrix by considering the all-loop twist-$L$ problem at large spin. The so-obtained S-matrix was analyzed recently at both weak and 
strong coupling in Ref.\ \cite{DZ} and matched successfully against the finite-gap theory string predictions.%
\footnote{The result of~\cite{DZ} at strong coupling applies to the giant-hole kinematics, i.e., for two holes with momenta $\sim g\gg 1$. In the opposite regime 
where the momenta are small, the holes scatter as the asymptotics excitations in the O(6) sigma model~\cite{AM07}.} Here we are mainly interested in the 
propagating part of the $Y$ function, as this is precisely to this expression that the $Y$-function reduces when evaluated in the vacuum, i.e., in the absence of 
small holes, and which corresponds to the twist-two configuration.

The all-loop expression of the $Y$ function for a scalar (hole) in the vacuum can thus be parameterized as
\be\label{Yhu}
Y_{\textrm{h}}(u) = (-1)^S\e^{iP_{\textrm{h}}(u)} \, ,
\ee
where we used the fact that $\exp{(i\vartheta)} = (-1)^S$ holds for the twist two sector (and integer spin $S$). Here
\be\label{Phu}
P_{\textrm{h}}(u) = 2p_{\textrm{h}}(u)\log{\bar{\eta}} + 2\delta p_{\textrm{h}}(u)\, ,
\ee
where $p_{\textrm{h}}(u) = 2u +O(g^2)$ is the all-loop momentum of the hole excitation constructed in~\cite{B10} and $\delta p_{\textrm{h}}(u)$ is a new quantity 
which should account for, and deform at higher-loop, the $O(\log^0{\eta})$ correction present in~(\ref{OLPu}). Note that $P_{\textrm{h}}(u)$ is entirely determined 
by the vacuum (i.e., twist-two) solution. It accounts for the effective propagation of the scalar in the background of covariant derivatives. A concise expression for it 
will be given later.

To define the L\"uscher corrections we would like to continue $Y_{\textrm{h}}$ to its mirror kinematics and then write down the finite-size correction to the energy in
the form
\be\label{EFSh}
E^{\textrm{FS}}_{\textrm{h}} = -{n_{\textrm{h}}\over 2\pi}\int dp^{\textrm{mirror}}_{\textrm{h}}(u) Y^{\textrm{mirror}}_{\textrm{h}}(u)\, ,
\ee
where $p^{\textrm{mirror}}_{\textrm{h}}(u)$ is the momentum of the hole with rapidity $u$ in the mirror kinematics and $n_{\textrm{h}}$ is the total number of 
scalar flavors, i.e., $n_{\textrm{h}}=6$ for the case at hand. As we shall find from our subsequent analysis, in analogy to a relativistic theory, the dispersion 
relation in the mirror kinematics of the scalar is identical to the real one. This is the double Wick rotation symmetry advocated in~\cite{AGMSV10}, with $p^{\textrm{mirror}}_{\textrm{h}}(u) = p_{\textrm{h}}(u)$ and $E^{\textrm{mirror}}_{\textrm{h}}(u) = E_{\textrm{h}}(u)$. This shall immediately lead to the 
following form of the mirror $Y$-function
\be
Y^{\textrm{mirror}}_{\textrm{h}}(u) \equiv (-1)^S\e^{-\mathcal{E}_{\textrm{h}}(u)}\, ,
\ee
where $\mathcal{E}_{\textrm{h}}(u)$ is the mirror of $-iP_{\textrm{h}}(u)$ entering~(\ref{Yhu}). The former quantity will be given as
\be
\mathcal{E}_{\textrm{h}}(u) = 2E_{\textrm{h}}(u)\log{\bar{\eta}}+2\delta E_{\textrm{h}}(u)\, ,
\ee
where $E_{\textrm{h}}(u)$ is just the energy of the hole with rapidity $u$ in the real kinematics and $\delta E_{\textrm{h}}(u)$ is the mirror of 
$-i\delta p_{\textrm{h}}(u)$, see Eq.~(\ref{Phu}), that we need to construct. The latter is an extra contribution that reflects this curious feature of the large spin background
that we already mentioned above. It plays an important role in the explicit computation of~(\ref{EFSh}) as it indeed provides the integration measure  
in Eq.\ (\ref{EFSh}) that makes the integral convergent to any order in the weak coupling expansion.

Of course the above $6$ scalars are not the only excitations on top of the GKP string and the L\"uscher formula would not be complete if these new degrees of
freedom were not included into the analysis. Of particular importance are the lightest excitations, with their twist equal to one, as they turn out to provide the 
leading contribution, $\sim 1/S^2$ at weak coupling, to the vacuum energy. This class of extra excitations includes $4+\bar{4}$ fermions and $1+1$ gauge fields. 
Introducing both of them into the game can be done by extending the previous considerations to higher-rank subsectors of the ABA system. The fermions can be 
unscrambled, for instance, from the $\mathfrak{sl}(2|1)$ subsector, while the gauge fields require a more thorough in-depth analysis~\cite{FRZ09}. The complete 
list of excitations was studied in~\cite{B10, GMSV}. In Appendix \ref{ExcitationsABA}, we provide for reader's convenience a brief outline explaining how these 
emerge from the ABA. This consideration serves as a preamble to the consideration that follows below where we explain how to construct the propagating phases 
for these excitations in the real kinematics. Then on a case-by-case  basis we pass to the mirror kinematics and eventually be able to evaluate the complete 
L\"uscher formula for leading finite-size corrections at large spin%
\footnote{Our notations below are slightly abusive, since in conventional treatments the degeneracy factor $n_{\star}$ associated to each contribution would 
actually be part of the quantity $Y_{\star}^{\textrm{mirror}}(u)$.}
\be\label{EfsFirst}
E^{\textrm{FS}}  = -\sum_{\star}{n_{\star}\over 2\pi}\int dp^{\textrm{mirror}}_{\star}(u) Y^{\textrm{mirror}}_{\star}(u) + \ldots \, ,
\ee
where the summation runs over all species of elementary (i.e., twist-one) excitations, $\star = $h, f, gf, with $n_{\star}$ the number of particles $\star$, and with the 
dots standing for higher L\"uscher corrrections as well as for corrections from composite (higher-twist) excitations.  For all excitations, we shall find that the double 
Wick rotation symmetry applies, with $p^{\textrm{mirror}}_{\star}(u) = p_{\star}(u)$ and $E^{\textrm{mirror}}_{\star}(u) = E_{\star}(u)$.

\subsection{Excitations in the real kinematics}

In this subsection, we shall generalize our previous discussion about the effective Bethe Ansatz equations to the case of arbitrary excitations. These equations 
can all be cast into the form
\be\label{RealYform}
1+Y_{\star}(u_{\star}) = 0\, ,
\ee
where $u_{\star}$ is the rapidity of the  $\star$-excitation. Our goal is to provide explicit expressions for all Y-functions $Y_{\star}(u)$.

For a collection of $K_{\star}$ excitations (of a given flavor), with $\star \neq \textrm{h}$, we have
\be\label{GenFormY}
Y_{\star}(u) = \sigma_{\star} \prod_{j=1}^{K_{4}}S^{\textrm{bare}}_{\star 4}(u, u_{4, j})\prod_{j=1}^{K_{\star}}S^{\textrm{bare}}_{\star \star}(u, u_{\star, j})\, ,
\ee
where
\be
\sigma_{\star} \equiv -S_{\star \star}^{\textrm{bare}}(u, u)\, .
\ee
It turns out that $\sigma_{\star} = -1$ for the fermionic excitation and $\sigma_{\star} = 1$ otherwise. Here $S_{\star \star'}^{\textrm{bare}}(u, v)$ is the bare 
S-matrix for scattering of the excitations $\star$ and $\star'$ carrying rapidities $u$ and $v$, respectively. It is different from the effective S-matrix controlling 
the quantizations conditions because the latter also receives contributions from the induced scattering, originating from the background of covariant derivatives 
parameterized by the distribution of main roots $u_{4, j}$, as reviewed in the Appendix \ref{ExcitationsABA}. The latter distribution gets deformed in the presence 
of impurities that generate the induced scattering. The explicit form of the $S_{\star \star}$ is not really relevant for our present discussion however, because at 
the end all we want is to evaluate~(\ref{GenFormY}) in the vacuum, that is, with the help of the twist-two distribution of roots obtained in the absence of impurities,
\be
Y_{\star}(u)\big|_{\textrm{vacuum}} = \sigma_{\star} \prod_{j=1}^{S}S^{\textrm{bare}}_{\star 4}(u, u_{4, j})\, ,
\ee
where the distribution of roots $u_{4, j}$ solves the twist-two ABA equations. These roots are encoded in the Baxter polynomial $Q(u) \equiv \prod_{j=1}^{S}(u-u_{4, j})$ 
that solves the twist-two Baxter equation \re{LRBaxterEq}  with $L=2$ and $K_{4} = S$. The resulting $Y$-functions (in the vacuum) are always parametrized as a pure 
phase
\be\label{Ystarreal}
Y_{\star}(u) = \sigma_{\star}(-1)^S\e^{iP_{\star}(u)}\, , 
\ee
where $P_{\star}(u) = 2p_{\star}(u)\log{\bar{\eta}} + \delta p_{\star}(u) + O(1/\eta^2)$, with $p_{\star}(u)$ being the all-loop momentum of the $\star$-excitation 
constructed in~\cite{B10}.

For the reader's convenience, let us quote the expressions for scattering of excitations on themselves. The corresponding bare S-matrix is trivial for fermions
\be
S^{\textrm{bare}}_{\textrm{f}\textrm{f}} = 1\, ,
\ee
meaning that there is no bare scattering between fermionic roots. For the gauge fields and their bound states we found that it takes the form
\be
S^{\textrm{bare}}_{\ell\ell}(u, v) = {u-v+2\ell i \over u-v-2\ell i}\prod_{n=1}^{\ell-1}\left(u-v+ni \over u-v-ni\right)^2\, .
\ee
However, none of these will be relevant below since in all expressions provided below we will not take into account the presence of extra excitations in the bath. 
They will only apply to the vacuum state.

\subsubsection{Scalar}

We start with the scalar (i.e., the hole) excitation that we have already extensively explored before. To recapitulate, it is associated with a dual solution of the 
equations for magnons. As such it is not surprising that
\be
Y_{\textrm{h}}(u) = Y_{\textrm{magnon}}(u)  = {Q(u-i) \over Q(u+i)}{\Delta_{-}(u-\ft{i}{2}) \over \Delta_{+}(u-\ft{i}{2})}\, .
\ee
Within the strip $-1 < \Im \textrm{m} (u) < 1$, which contains the real kinematics, we can use the fact that $Q(u\pm i) = Q_{\pm}(u\pm i)$ to leading order at large 
spin and then simplifies with the help of~(\ref{HBEq}). It leads to
\be\label{Yhagain}
Y_{\textrm{h}}(u) = {Q_{-}(u) \over Q_{+}(u)}\, ,
\ee
which we also encountered before. We want to evaluate~(\ref{Yhagain}) in the vacuum, i.e., without any excitations. This simply means that we should plug in
 the twist-two solution to the Baxter equation into this expression. Using Eqs.\ (\ref{logQp}) and (\ref{logQm}) we find it to admit the form
\be
Y_{\textrm{h}}(u) = (-1)^S\e^{iP_{\textrm{h}}(u)}\, ,
\ee
where
\be\label{Ph}
P_{\textrm{h}}(u) 
= 
4u\log{\bar{\eta}} + 4\int_{0}^{\infty}{dt \over t}{\sin{(ut)}\e^{t/2}J_{0}(2gt) - ut\over \e^{t}-1}
- 
2\int_{0}^{\infty}{dt \over t}{\sin{(ut)}\e^{t/2}\gamma(2gt) \over \e^{t}-1} \, ,
\ee
and we recall that $\log{\bar{\eta}} \equiv \log{\eta}-\psi(1)= \log{\eta}+\gamma_{\textrm{E}}$. Note that along the way we also used the condition $c_{-} / c_{+} 
= (-1)^S$, derived previously in~(\ref{cpcm}).

The one-loop estimate of~(\ref{Ph}) can be obtained  making use of the observation that the function $\gamma$ is small at weak coupling, thus leaving only the 
two first terms in the right-hand side of~(\ref{Ph}) relevant as $g \to 0$. The latter two simplify after noticing that $J_{0}(2gt) = 1+O(g^2)$ at leading order
which eventually leads to
\be
Y_{\textrm{h}}(u) = (-1)^S \eta^{4iu}{\Gamma(\ft{1}{2}-iu)^2 \over \Gamma(\ft{1}{2}+iu)^2}\, ,
\ee
after applying the identity
\be
2\int_{0}^{\infty}{dt \over t}{\sin{(ut)}\e^{t/2} -ut \over \e^{t}-1} = i\log{{\Gamma(\ft{1}{2}+iu) \over \Gamma(\ft{1}{2}-iu)}}+2\psi(1)u\, .
\ee
This confirms the result \re{OLPu} we reported in preamble to this section.

\subsubsection{Fermion}

We shall now turn to the fermionic case. As we adverted before, these excitations can be found by considering the $\mathfrak{sl}(2|1)$ subsector of the 
ABA system of equations~\cite{BS05}. Here we are mostly interested in the phase accumulated while a fermion propagates through the twist-two 
background, characterized by the distribution of roots $u_{4,j}$. The latter quantity is controlled, as demonstrated in Appendix~\ref{TO}, by
\be\label{Yf}
Y_{\textrm{f}}(x)  \equiv -(-1)^S \e^{iP_{\textrm{f}}(x)} = -\prod_{j=1}^{S}{x-x_{j}^{+} \over x-x_{j}^{-}} \, .
\ee
This expression illustrates one virtue specific to the fermion: its natural rapidity is the Zhukowsky variable $x$, and not $u$. To cover the whole $x$ plane 
we need two copies of the $u$ plane, to which we refer to as the small $|x| < g$ and large $|x| > g$ fermion domains. In all cases $u$ is related to $x$ via the 
Zhukowski map $u=x+g^2/x$, but its solution, $x(u)$, is chosen to be the large one for the fermion with $|x| > g$, such that $x \sim u$ at large $u$, and 
respectively the small one for the fermion with $|x| < g$, such that $x \sim g^2/u$ at large $u$. The fermion possesses a real value of the momentum $p$ 
when $x$ is real which implies that $u^2 > (2g)^2$, independent on whether $|x|$ is greater than $g$ or not. In particular, a fermion at rest with $p=0$ or 
equivalently $x=0$ is small with a rapidity $u\sim \infty$.

In practice, it is more convenient to present expressions for the $Y$-function parameterized by the rapidity $u$. It follows that we need to provide two 
representations, one for the large fermion with rapidity $u$ and another for the small fermion with the same rapidity. To distinguish between the two, we 
shall use the script `lf' and `sf', respectively. Of course these two expressions represent the same function of $x$, but in the two disjoint domains 
$|x| \gtrless g$. By continuity they should match on the circle $|x|  = g$ which is realized in $u$-space by the interval $u^2 < (2g)^2$. The two expressions 
below are such that they both have square root branch points at $u=\pm 2g$ and are glued together by an analytic continuation along their common cut 
along $u^2 < (2g)^2$.

Let us consider first the large fermion, i.e., carrying the rapidity $|x| > g$. In this case it is convenient to rewrite the ratios involved in~(\ref{Yf}) in the form
\be
\e^{iP_{\textrm{lf}}(u)} 
= (-1)^S\prod_{j=1}^{S}{u-u_{j}^{+} \over u-u_{j}^{-}}\prod_{j=1}^{S}{1-g^2/x x_{j}^{-} \over 1-g^2/x x_{j}^{+}} 
= (-1)^S{Q(u-\ft{i}{2}) \over Q(u+\ft{i}{2})}\left[{\Delta_{-}(u) \over \Delta_{+}(u)}\right]^{1/2}\, ,
\ee
where the last equality is a straightforward rephrasing done with the help of the definition of the Baxter polynomial and dressing factors. We want to 
evaluate the above expression at large spin with the rapidity variable $u$ belonging to the strip $-1/2 < \Im \textrm{m}(u) < 1/2$, since the latter contains 
the real kinematics $u^2 > (2g)^2$. In this strip, we can approximate $Q$ by $Q_{\pm}$ depending on the sign of the imaginary part of its argument and 
get
\be\label{EiPlf}
\e^{iP_{\textrm{lf}}(u)} = (-1)^S{Q_{-}(u-\ft{i}{2}) \over Q_{+}(u+\ft{i}{2})} \left[{\Delta_{-}(u) \over \Delta_{+}(u)}\right]^{1/2}\, .
\ee
Now it is an easy exercise to plug in the large-spin solutions (\ref{logQp}) and (\ref{logQm}) into Eq.\ (\ref{EiPlf}) and after a short derivation conclude that
\be\label{Plf}
P_{\textrm{lf}}(u) 
= 4u\log{\bar{\eta}} +4\int_{0}^{\infty}{dt \over t}{\sin{(ut)}J_{0}(2gt) -ut \over \e^{t}-1} 
-2\int_{0}^{\infty}{dt \over t}{\sin{(ut)}\gamma(2gt) \over \e^{t}-1} -\int_{0}^{\infty}{dt \over t} \sin{(ut)}\gamma_{-}(2gt)\, .
\ee
Here we also used the following integral representation
\be
\left[{\Delta_{-}(u) \over \Delta_{+}(u)}\right]^{1/2} 
= 
\exp{\left[2\sum_{n\ge 1}\gamma_{2n-1}\bigg({g \over ix}\bigg)^{2n-1}\right]} = \exp{\left[-i\int_{0}^{\infty}{dt \over t} \sin{(ut)}\gamma_{-}(2gt)\right]}\, ,
\ee
where $x$ is the large solution $u=x+g^2/x$. Here the equality after the second equality sign holds under the assumption that $u$ is real and such 
$u^2 > (2g)^2$, and thus the same restriction applies to Eq.\ (\ref{Plf}). The expression~(\ref{Plf}) is all we need for computing $P_{\textrm{lf}}(u)$, at 
any coupling, up to corrections explicitly inversely suppressed in spin.

The small fermion domain can be deduced by analytically continuing~(\ref{Plf}) through the central cut along $u^2 < (2g)^2$. Indeed this corresponds 
to the continuation to the domain $|x| < g$. It can be done making use of the equations obeyed by the function $\gamma(t)$. We do proceed differently, 
however, starting over from Eq.\ (\ref{Yf}) and then writing 
\be\label{EiPsf}
\e^{iP_{\textrm{sf}}(u)} =  \prod_{j=1}^{S}{1-x/x_{j}^{+} \over 1-x/x_{j}^{-}} = \left[{\Delta_{+}(u) \over \Delta_{-}(u)}\right]^{1/2} = \exp{\bigg[2\sum_{n\ge 1}\gamma_{2n-1}\bigg({ix \over g}\bigg)^{2n-1}\bigg]}\, ,
\ee
where $x$ is now the small solution of the map $u=x+g^2/x$, with $x \sim g^2/u$ at $u\sim \infty$. Here we simply factorized $\prod_{j} x^{+}_{j}/x^{-}_{j} = 
\exp{(i\vartheta)}$ in~(\ref{Yf}), which is equal to $(-1)^S$ for the twist two sector, and then applied several identities established earlier. For the real rapidity $u$, 
away from the cut $u^2 > (2g)^2$, we can also rewrite~(\ref{EiPsf}) as
\be\label{Psf}
P_{\textrm{sf}}(u) = \int_{0}^{\infty}{dt \over t} \sin{(ut)}\gamma_{-}(2gt)\, .
\ee

For illustration purposes and preparing ourselves to the analysis that follows, let us construct the weak coupling expansion for $Y_{\textrm{lf}}(u)$. To this end,
we neglect $\gamma$ in~(\ref{Plf}), approximate $J_{0}(2gt)$ by $1$, and find that the one-loop expression is simply given by
\be\label{YlfOL}
Y_{\textrm{lf}}(u) = -(-1)^S \eta^{4iu}{\Gamma(1-iu)^2 \over \Gamma(1+iu)^2}\, ,
\ee
where we used the result
\be
2\int_{0}^{\infty}{dt \over t}{\sin{(ut)} -ut \over \e^{t}-1} = i\log{{\Gamma(1+iu) \over \Gamma(1-iu)}}+2\psi(1)u\, .
\ee
This is the expected form for the fermion which has conformal spin $1$. It comes with an extra minus sign as compared to the  findings of~\cite{BGK06} which 
merely is due to the statistics. For a small fermion, we simply find $Y_{\textrm{sf}}(u) = (-1)^S$, which is essentially equivalent to the limit $u\rightarrow 0$ 
of~(\ref{YlfOL}).

\subsubsection{Gauge field and bound states}

Last but not least, the gauge field excitations and bound states thereof, are embedded into the ABA equations in the form of stacks of strings~\cite{FRZ09},
The details of this embedding are recalled in Appendix~\ref{TO}.  By applying the fusion method to these complex solutions, we obtain the following 
representation for the function $Y_{\ell}(u)$ associated to a bound state of $\ell$ gauge fields,
\be\label{YellGen}
Y_{\ell}(u) = \prod_{j=1}^{S}{x^{[- \ell]}-x^{+}_{j}\over x^{[+ \ell]}-x^{-}_{j}}{1-g^2/x^{[- \ell]}x^{-}_{j} \over 1-g^2/x^{[+ \ell]}x^{+}_{j}} 
= 
{Q(u-\ft{im}{2}) \over Q(u+\ft{im}{2})}\left[{\Delta_{-}(u+\ft{i\ell}{2})\Delta_{-}(u-\ft{i\ell}{2}) \over \Delta_{+}(u+\ft{i\ell}{2})\Delta_{+}(u-\ft{i\ell}{2})}\right]^{1/2}\, ,
\ee
where $x^{[\pm \ell]} = x(u^{[\pm \ell]})$ with $u^{[\pm \ell]} = u\pm \ft{i}{2} \ell$ and $m \equiv 1+\ell$.

At large spin and for $u$ residing within the strip $-\ft{m}{2}< \Im{\rm m}(u) < \ft{m}{2}$, we can evaluate~(\ref{YellGen}) with the help of $Q_{\pm}(u)$ determined
in Eqs.\ (\ref{logQp}) and (\ref{logQm}) neglecting corrections $\sim 1/\eta^2$ which are irrelevant here. Then $Y_{\ell}(u)$ if found to admit the expected generic 
form
\be\label{Yell}
Y_{\ell}(u) = (-1)^S\e^{iP_{\ell}(u)}\, ,
\ee
where
\be\label{Pl}
P_{\ell}(u) = 4u\log{\bar{\eta}} +4\int_{0}^{\infty}{dt \over t}{\sin{(ut)}\e^{-\ell t/2}J_{0}(2gt)-ut\over \e^{t}-1}
- 2\int_{0}^{\infty}{dt \over t}\sin{(ut)}\e^{-\ell t/2}\left[{\gamma_{-}(2gt) \over 1-\e^{-t}}+{\gamma_{+}(2gt) \over \e^{t}-1}\right]\, .
\ee
It is easily verified that to leading order in large spin, $P_{\ell}(u) = 2p_{\ell}(u)\log{\bar{\eta}} + O(\log^0{\bar{\eta}})$ with $p_{\ell}(u)$ being the momentum 
found in Ref.\ \cite{B10}.

The one-loop expression is obtained from~(\ref{Pl}) along the same lines as before. We get after a brief back-of-the-envelop calculation
\be
Y_{\ell}(u) = (-1)^S \eta^{4iu}{\Gamma(s_{\ell}-iu)^2 \over \Gamma(s_{\ell}+iu)^2}\, ,
\ee
where $s_{\ell} \equiv 1+\ell/2$. This is the anticipated result for excitations with conformal spin $s_{\ell}$.%
\footnote{Note however that these twist-$\ell$ excitations are not quite identical to the excitations (i.e., holes) of thel spin-$s_{\ell}$ Heisenberg spin chain 
at the one-loop level. This is correct for the dispersion relation or the vacuum contribution to $Y_{\ell}(u)$ but not for the two-to-two S-matrix of such 
excitations. There is no contradiction here with the findings of~\cite{FRZ09} where only the case of length $3$ was treated, implying a single excitation 
on top of the vacuum for us. The only exception to what have just been said is the gauge field, $\ell=1$, where the mapping to the conformal spin chain 
with $s=3/2$ is valid for any length (i.e., for any number of excitations for us).}

\subsection{Mirror kinematics and double Wick rotation symmetry}

For real physical kinematics the energy and momentum are both real. For bosons this holds provided the rapidity $u$ is real while for fermions it requires 
the reality of the corresponding Zhukowski variable $x$. The essence of the double Wick rotation is to look for a transformation $u \rightarrow u'$ such that
\be
E_{\star}(u') \equiv ip_{\star}^{\textrm{mirror}}(u)\, , \qquad p_{\star}(u') \equiv iE_{\star}^{\textrm{mirror}}(u)\, ,
\ee
become simultaneously purely imaginary. We will explicitly construct this transformation in the case of scalar, gauge field and bound states%
\footnote{Note that the mirror transformation depends on the flavor of the excitation. Moreover it is different from the one for magnons, which should not 
be too suprising since dispersion relations have different analytical properties in both cases.} and find that it leads to
\be\label{DWRS}
E_{\star}^{\textrm{mirror}}(u) = E_{\star}(u)\, , \qquad p_{\star}^{\textrm{mirror}}(u) = p_{\star}(u)\, .
\ee
These equalities are equivalent to the statement of the double Wick rotation symmetry of the dispersion relation proposed in~\cite{AGMSV10}. 

Then, from the knowledge of the transformation $u \rightarrow u'$ and the expression for the $Y$-function derived earlier, we shall construct the 
amplitude for the propagation of an excitation in the vacuum in the mirror kinematics. We shall see that it takes the form
\be\label{YmirrorType}
Y^{\textrm{mirror}}_{\star}(u) 
\equiv (-1)^{S}\e^{-\mathcal{E}_{\star}(u)} 
= (-1)^{S}\exp{\bigg[-2E_{\star}(u)\log{\bar{\eta}}-2\delta E_{\star}(u) + O(1/\eta^2)\bigg]}\, ,
\ee
where $\mathcal{E}_{\star}(u) \equiv -iP_{\star}(u')$, with $P_{\star}(u)$ introduced in~(\ref{Ystarreal}), and where $\eta = S$ to leading order at large spin. Both the energy $E_{\star}(u)$ and anomalous 
part $\delta E_{\star}(u)$ shall be found solely from the asymptotic solution to the Baxter equation. While the remaining $O(1/\eta^2)$ contributions 
in~(\ref{YmirrorType}) are beyond of our current accuracy and are expected to receive finite-size corrections. They will not be considered here. The integral 
representation for the energy and its weak coupling expansion were already derived in~\cite{B10, AGMSV10}, such that the new result here is for 
$\delta E_{\star}(u)$. This anomalous contribution parallels the one found in the real kinematics, as clear from the fact that
\be
\delta E_{\star}(u)  \equiv  -i\delta p_{\star}(u')\, .
\ee
One immediate consequence of~(\ref{YmirrorType}) at weak coupling is that
\be\label{ECest}
Y^{\textrm{mirror}}_{\star}(u) \sim 1/\eta^2 \sim 1/S^2\, ,
\ee
at large spin, for all twist one excitation. In other words, the finite-size corrections to the energy shall be suppressed with the spin, as expected. This is 
entirely due to the asympotics
\be\label{OLen}
E_{\star}(u) = 1 + O(g^2)\, .
\ee
Corrections to it lead to enhancement by powers of $\log{\bar{\eta}}$, with the maximal power increasing with the loop order.

Our treatment of the fermions is far less satisfactory than for the bosons. The reason being that we unable to unravel the mirror transformation 
$u\rightarrow u'$ in this case. Put differently, we did not realize how to achieve simultaneously purely imaginary values for $E_{\textrm{f}}(u')$ 
and $p_{\textrm{f}}(u')$ by transforming $u$ to $u'$ by a simple analytical continuation. The situation becomes however more transparent at 
strong coupling where the transformation $u \rightarrow u'$ can be accomplished, but only because the analytical properties in the $u$-plane 
of energy and momentum simplify substentially in this limit. It is found then that the fermion satisfies the double Wick rotation symmetry~(\ref{DWRS}) to all 
order at strong coupling. This strategy of considering the strong coupling regime as a governing principle also underlies our analysis of the the mirror 
transformation for bosons, as we shall illustrate for scalar below. Nevertheless, unlike the latter cases, the situation for fermions remains obscure  beyond 
perturbative analysis, whence the exact analytical structure is restored. Fortunately the strong coupling analysis is very suggestive and allows one to 
conjecture an expression for the mirror $Y$ function valid for any value of the coupling, perturbative or not. This will lead us to an educated guess for it 
which we shall employ afterwards for our weak coupling analysis. In this case, as well, we shall find that the representation~(\ref{YmirrorType}) holds 
true. Our guess has also a bestowal property that $E^{\textrm{mirror}}_{\textrm{f}}(u) = E_{\textrm{f}}(u)$ for any coupling. Then to put the fermion and 
boson on an equal footing we shall complement it with $p^{\textrm{mirror}}_{\textrm{f}}(u) = p_{\textrm{f}}(u)$, implying that the double Wick rotation 
symmetry holds for fermions at any value of the coupling. Altogether this will allow us to evaluate the fermionic contribution to the L\"uscher formula.

In the following, we shall make extensive analytic continuations of typical integrals defining $P_{\star}(u)$ as found in previous sections. These 
entail crossing certain cuts in the complex rapidity plane. This can be done efficiently by using integral equations for the functions $\gamma_{\pm}(t)$. 
With this application in mind, these equations are cast into a more convenient form
\be\label{EqCont}
\begin{aligned}
&\int_{0}^{\infty}{dt \over t}\sin{(ut)}\gamma_{-}(2gt) 
=- \int_{0}^{\infty}{dt \over t}{\sin{(ut)}\gamma(2gt) \over \e^{t}-1} + 2u\log{\bar{\eta}} + 2\int_{0}^{\infty}{dt \over t}{\sin{(ut)}J_{0}(2gt)-ut \over \e^{t}-1} \, ,\\
&\int_{0}^{\infty}{dt \over t}\cos{(ut)}\gamma_{+}(2gt) 
=- \int_{0}^{\infty}{dt \over t}{(\cos{(ut)}-J_{0}(2gt))\gamma(-2gt) \over \e^{t}-1} + 2\int_{0}^{\infty}{dt \over t}{(\cos{(ut)}-J_{0}(2gt)) J_{0}(2gt) \over \e^{t}-1} \, .
\end{aligned}
\ee
Both of these results are valid up to $O(1/\eta^2)$ corrections and, more importantly, for $u^2 < (2g)^2$. These equations~(\ref{EqCont}) can be deduced 
from the system (\ref{GenSyst}), with the known homogeneous terms $\kappa_{n} = \kappa_{n}^{\o}\log{\bar{\eta}} + \delta \kappa_{n}^{\o} + O(1/\eta^2)$, 
see Eqs.\ (\ref{kBES}, \ref{kdelta}), after observing that the coefficients $\gamma_{n}$ entering the Neumann expansion~(\ref{BessDec}) admit the integral 
representation
\be\label{CoeffInt}
\gamma_{2n} 
= 2(2n)\int_{0}^{\infty}{dt \over t} J_{2n}(2gt)\gamma_{+}(2gt)
\, , \qquad 
\gamma_{2n-1} 
= 2(2n-1)\int_{0}^{\infty}{dt \over t} J_{2n-1}(2gt)\gamma_{-}(2gt)\, ,
\ee
for $n=1, 2, \ldots\,$.%
\footnote{One has also
\be
\int_{0}^{\infty}{dt \over t} J_{0}(2gt)\gamma_{+}(2gt) = 0\, .
\ee} 
These integrals then follow straightforwardly from the Neumann expansion~(\ref{BessDec}) and the orthonormality of the Bessel's functions%
\footnote{We assume that we can commute summation and integration. This is not guaranteed by the convergence of the Neumann series~(\ref{BessDec}) 
alone, even if it converges over the whole complex $t$-plane, which is the case here. The important point is that the expansion coefficients $\gamma_{n}$ 
solving the system of equations~(\ref{GenSyst}) are well suppressed at large $n$, for any given value of the coupling.}
\be
\int_{0}^{\infty}{dt \over t}J_{2n}(t)J_{2m}(t) = {\delta_{n, m} \over 2(2n)}\, , \qquad \int_{0}^{\infty}{dt \over t}J_{2n-1}(t)J_{2m-1}(t) = {\delta_{n, m} \over 2(2n-1)}\, .
\ee
After plugging~(\ref{CoeffInt}) into~(\ref{GenSyst}), one easily arrives at the equations~(\ref{EqCont}) by performing suitable linear combinations implementing 
the change of basis $J_{n}(2gt) \rightarrow (\sin{(ut)}, \cos{(ut)})$ with $u^2 <(2g)^2$, see Ref.~\cite{BK08, BK09} for more details. 

Notice finally that Eqs.\ (\ref{EqCont}) imply that the two integrals
\be\label{TwoInt}
\int_{0}^{\infty}{dt \over t}\sin{(ut)}\gamma_{-}(2gt)\, , \qquad \int_{0}^{\infty}{dt \over t}\cos{(ut)}\gamma_{+}(2gt)\, ,
\ee
can be analytically continued into holomorphic functions of $u$ from the interval $u^2 < (2g)^2$ to the strip $-1 < \Im \textrm{m} (u) < 1$. This continuation 
is different from the one we would obtain starting from the interval $u^2 >(2g)^2$ and transforming the integrals to the form 
\be
\int_{0}^{\infty}{dt \over t}\sin{(ut)}\gamma_{-}(2gt) = \mp i\int_{0}^{\infty}{dt \over t}\e^{\pm iut}\gamma_{-}(2gt)
\, , \qquad 
\int_{0}^{\infty}{dt \over t}\cos{(ut)}\gamma_{+}(2gt) = \int_{0}^{\infty}{dt \over t}\e^{\pm iut}\gamma_{+}(2gt)\, ,
\ee
where the upper-/lower-script refers to $u$ in the upper-/lower-half plane. The latter continuation indeed comes with a cut along $u^2 < (2g)^2$ while the 
former does not. There is no contradiction however, since the two integrals in~(\ref{TwoInt}), which are both well defined for real $u$, are not analytical 
over the real $u$-axis. Though they define continuous functions of $u$, some of their derivatives jump at $u^2 = (2g)^2$ and thus the analytic continuation 
from inside and outside the interval $u^2 <(2g)^2$ can be inequivalent.

We are now in position to attack the transition to the mirror kinematics on an excitation-by-excitation basis.

\subsubsection{Scalar}

\begin{figure}[t]
\begin{center}
\mbox{
\begin{picture}(0,255)(160,0)
\put(0,0){\insertfig{9}{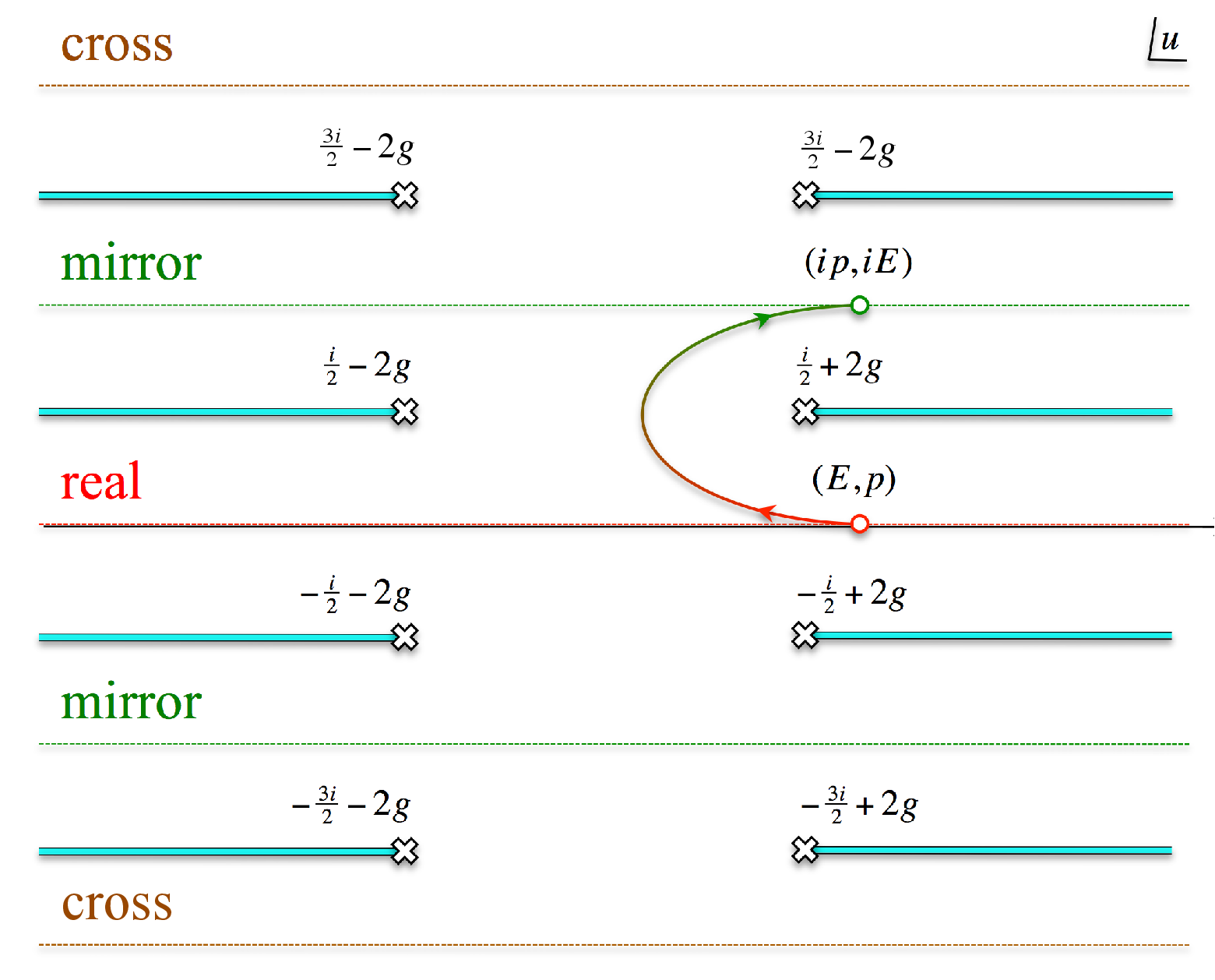}}
\end{picture}
}
\end{center}
\caption{\label{ScalarDWR} Rapidity plane for energy and momentum of a scalar excitation. The double Wick rotation is implemented by the shift $u\rightarrow u+i$ and depicted by the arrow. Composing two such rotations yields the particle to antiparticle transformation $(E, p) \rightarrow (-E, -p)$ related to the crossing symmetry. The 'mirror' line in the lower-half plane is associated to the transformation $(E, p) \rightarrow (-ip, -iE)$. Up to the presence of the cuts, this is the same pattern as the one observed for energy and momentum of a relativistic particle in the $\theta$-plane, with $\theta \equiv \pi u/2$.}
\end{figure}

We start our discussion with the scalar excitations. The transformation that supports the continuation to the mirror kinematics in this case is merely given by 
the shift $u\rightarrow u+i$. However this has to be done with due care since one encounters cuts in the rapidity plane along the way both in the energy and 
momentum of the scalar (see~\cite{B10}). To motivate the transformation alluded to above and understand how to deal with the issue of cuts, let us have a look 
at strong coupling where the dispersion relation becomes relativistic in a certain regime. The region in question is the low-momentum domain where 
a scalar excitation possesses the following energy and momentum
\be\label{EPcosh}
E_{\textrm{h}}(u) = m\cosh{\theta} + \ldots\, , \qquad p_{\textrm{h}}(u) = m\sinh{\theta} + \ldots\, ,
\ee
where the rapidity $\theta \equiv \pi u /2$ assumed to be fixed, while $g\gg 1$ and $m \sim \exp{(-\pi g)}$ is the mass gap~\cite{AM07} (see Eq.\ \ref{mass} 
below for explicit expression). The ellipsis in~(\ref{EPcosh}) stands for corrections suppressed as $m^3$ at fixed $\theta$, which incidentally break Lorentz 
invariance.%
\footnote{The Lorentz invariance of the problem arises at strong coupling and at low energy only~\cite{AM07}.} From the representation~(\ref{EPcosh}) it is 
clear that the double Wick rotation is obtained through the shift $\theta \rightarrow \theta +i\pi/2$, as is well known. It corresponds to $u\rightarrow u+i$ 
resulting in
\be\label{EipSc}
p^{\textrm{mirror}}_{\textrm{h}}(u) = -iE_{\textrm{h}}(u+i) = m\sinh{\theta} + \ldots = p_{\textrm{h}}(u)\, ,
\ee
and similarly for $E^{\textrm{mirror}}_{\textrm{h}}(u)$. A more general representation given in~\cite{B10} that includes neglected contributions in~(\ref{EPcosh}) 
shows that the same conclusions apply beyond the validity of relativistic approximation. Namely, the energy and momentum simultaneously achieve purely 
imaginary values under the shift $u\rightarrow u+i$ when starting from the mirror kinematics and, as it turns out, exhibit the double Wick rotation symmetry
(\ref{DWRS}). The latter representation, though still being limited to $u^2 < (2g)^2$, suggests a generalization for any $u$: the rapidity $u$ is carried over to 
the mirror kinematics by $u\rightarrow u+i$ assuming it is performed by going in between the first two branch points in the upper-half plane located at 
$u=i/2\pm 2g$, respectively. The situation is illustrated in Fig.~\ref{ScalarDWR}. Notice that the path to the mirror kinematics closes off at weak coupling when 
the two branch points collide. This explains why it is impossible to perform the double Wick rotation directly from the weak coupling expression for the energy and 
momentum. Not surprisingly, it opens up at strong coupling where this continuation is expected to make sense from the string theory perspective. This parallels, 
of course, the analysis for magnons.

It is not difficult, starting from the general representation proposed in Ref.\ \cite{B10}, to verify that the identities in~(\ref{DWRS}) are fulfilled for the hole
excitation with the energy and momentum defined in the mirror kinematics obtained via the procedure advocated above. The essential point here is to
use Eqs.\ (\ref{EqCont}) at intermediate steps for analytic continuation to higher imaginary values of the rapidity inside the strip $-2g < \Re \textrm{e} 
(u) < 2g$. Since we are interested in the transition to the mirror kinematics for the $Y$ function itself rather than only for the momentum $p_{\textrm{h}}(u)$, 
we shall explicitly illustrate the procedure for the quantity $P_{\textrm{h}}(u)$, which also includes the anomalous component $\delta p_{\textrm{h}}(u)$ in 
addition to $p_{\textrm{h}}(u)$. Thus we define
\be\label{Phmirr}
\mathcal{E}_{\textrm{h}}(u) \equiv -iP_{\textrm{h}}(u+i)\, ,
\ee
with the shift being performed through the strip $-2g < \Re \textrm{e} (u) < 2g$, see Fig.~\ref{ScalarDWR}.

To derive a convenient representation for $\mathcal{E}_{\textrm{h}}(u)$ we can start from $P_{\textrm{h}}(u)$ introduced in Eq.\ (\ref{Ph}) and perform 
the shift on the right-hand side of Eq.\ (\ref{Phmirr}) in two steps. First, we compute $P_{\textrm{h}}(u+i/2)$ for $u^2 <(2g)^2$. This is done by substituting 
$u\rightarrow u+i/2$ directly into~(\ref{Ph}), since the latter integral representation initially covers the strip $-1/2 < \Im \textrm{m}(u) < 1/2$. This procedure yields
\be\label{Interm}
P_{\textrm{h}}(u+i/2-i0) = 2i\log{(\eta/x)} + \pi-i\int_{0}^{\infty}{dt \over t}\e^{-iut}\gamma(2gt)+2\int_{0}^{\infty}{dt \over t}\sin{(ut)}\gamma_{-}(2gt)\, ,
\ee
where we used the first equation in~(\ref{EqCont}), which produces the last term in the right-hand side of~(\ref{Interm}), and also employed the explicit value 
of the integral
\be
\label{Int1}
\int_{0}^{\infty}{dt \over t}\bigg[\e^{-iut}J_{0}(2gt)-{t\over \e^{t}-1}\bigg] = -\log{x}+\psi(1)-{i\pi \over 2}\, ,
\ee
with $u=x + g^2/x$ and $|x| > g$. Note that $-i 0$ prescription on the left-hand side of Eq.\ (\ref{Interm}) remind us that this quantity is the boundary value 
(from above) of the original one given in~(\ref{Ph}). It is important to stress however that the representation~(\ref{Interm}) cannot be used readily if $u$ is 
given even a small positive imaginary part, as the first integral in the right-hand side of~(\ref{Interm}) (would possess an exponentially increasing integrand 
and thus) requires proper analytical continuation. To remedy to it, we can use the fact that
\be
-i\int_{0}^{\infty}{dt \over t}\e^{-iut}\gamma(2gt)+2\int_{0}^{\infty}{dt \over t}\sin{(ut)}\gamma_{-}(2gt) 
= i\int_{0}^{\infty}{dt \over t}\e^{iut}\gamma(-2gt)-2i\int_{0}^{\infty}{dt \over t}\cos{(ut)}\gamma_{+}(2gt)\, ,
\ee
since $\gamma(t) \equiv \gamma_{+}(t) + \gamma_{-}(t)$, and then cast (\ref{Interm}) into the form
\be\label{Intermbis}
i\mathcal{E}_{\textrm{h}}(u-i/2+i0) = 2i\log{(\eta/x)} + \pi +i\int_{0}^{\infty}{dt \over t}\e^{iut}\gamma(-2gt)-2i\int_{0}^{\infty}{dt \over t}\cos{(ut)}\gamma_{+}(2gt)\, ,
\ee
where by construction
\be
i\mathcal{E}_{\textrm{h}}(u-i/2) = P_{\textrm{h}}(u+i/2)\, ,
\ee
if $u^2 < (2g)^2$. Now we can go to second and last step and complete the transformation by shifting $u\rightarrow u+i/2$ in~(\ref{Intermbis}). This is easily
done at the level of the last term in the right-hand side of~(\ref{Intermbis}) by using the second equation in~(\ref{EqCont}), while for the first term we can use 
an identity similar to the one given before in Eq. \re{Int1}, which reads
\be
\int_{0}^{\infty}{dt \over t}\bigg[\e^{iut}J_{0}(2gt)-{t\over \e^{t}-1}\bigg] = -\log{\left({g^2 \over x}\right)}+\psi(1)+ {i\pi \over 2}\, ,
\ee
where $u= x +g^2/x$ and $|x| < g$. It is important to realize that this result holds for $|x| < g$ since in the process we have indeed crossed the boarder $|x| = g$ 
and thus can no longer consider $|x| > g$ as initially. Combining all results, we finally find that
\be\label{Mh}
\begin{aligned}
\mathcal{E}_{\textrm{h}}(u) & = -2\log{(g^2/\eta)} + 2\psi(1) - 4\int_{0}^{\infty}{dt \over t}{(\cos{(ut)}\e^{t/2}-J_{0}(2gt))J_{0}(2gt)-t/2 \over \e^{t}-1} \\
& \, \, \, \, \, \, + 2\int_{0}^{\infty}{dt \over t}{\cos{(ut)}\e^{t/2}-J_{0}(2gt) \over \e^{t}-1} \gamma(-2gt)\, .
\end{aligned}
\ee
This is the sought expression that determines $\mathcal{E}_{\textrm{h}}(u)$ to all loops, up to corrections vanishing at large spin. It is easily found in 
particular that
\be\label{PiE}
\mathcal{E}_{\textrm{h}}(u) = 2E_{\textrm{h}}(u)\log{\bar{\eta}} + \ldots\, ,
\ee 
after expanding~(\ref{Mh}) with $\gamma(t) = \gamma^{\o}(t)\log{\bar{\eta}} + \ldots$ and comparing the obtained expression with the general representation 
given in~\cite{B10} for $E_{\textrm{h}}(u)$.%
\footnote{The comparison also requires the use of the second equation in~(\ref{EqCont}) evaluated at $u=0$ and restricted to the leading component 
$\sim \log{\bar{\eta}}$.} Since, on the other hand, $\mathcal{E}_{\textrm{h}}(u) = 2E^{\textrm{mirror}}_{\textrm{h}}(u)\log{\bar{\eta}} + \ldots$ we verify our claim that 
$E_{\textrm{h}}^{\textrm{mirror}}(u) = E_{\textrm{h}}(u)$.

As we advocated before, the representation~(\ref{Mh}) is all we need for evaluation of the $Y$ function at large values of spin and any coupling. The weak 
coupling limit is by far the simplest one to derive. It follows from neglecting the second line in~(\ref{Mh}), since $\gamma$ is small, while replacing the Bessel
function $J_{0}(2gt)$ in the first line with its Taylor expansion $1 + O(g^2)$. This immediately yields
\be\label{olYh}
Y^{\textrm{mirror}}_{\textrm{h}}(u) = (-1)^S\exp{-\mathcal{E}_{\textrm{h}}(u)} = (-1)^{S}{\pi^2 g^4 \over \eta^{2}\cosh^2{(\pi u)}} + O(g^6)\, ,
\ee
where we applied the integral
\be
2\int_{0}^{\infty}{dt \over t}{\cos{(ut)}\e^{t/2}-1-t/2 \over \e^{t}-1} = \log{\left({\pi \over \cosh{(\pi u)}}\right)} +\psi(1) \, .
\ee
We note that the leading large-spin component of $Y$ due to the energy $E_{\textrm{h}}(u) = 1 + O(g^2)$ has produced, according to (\ref{olYh}), only
the overall $1/\eta^2$ dependence in it at a given order at weak coupling. Most of the expression comes therefore from the anomalous component
$\delta E_{\textrm{h}}(u)$. Notably, the latter is responsible for taming the large rapidity asymptotics in $Y^{\textrm{mirror}}_{\textrm{h}}(u)$ and guarantees,  
as we  shall see later, the convergence of corresponding integrals in the L\"uscher formula.%
\footnote{This contrasts it with a relativistic theory where the damping is produced by the large rapidity behavior of the energy.}
Finally, it also involves several factors of 't Hooft coupling, $Y^{\textrm{mirror}}_{\textrm{h}}(u)  \sim g^4$, immediately implying that the finite-size energy 
starts at two loop order.

\subsubsection{Gauge field and bound states}

The continuation to the mirror kinematics is slightly more involved for gauge fields and their bound states. Here, in complete similarity to the hole excitations
addressed in the previous section, it would also help to start the consideration with the strong coupling regime in order to figure out the correct continuation 
procedure. For the sake of generality, however, we shall present another suggestive path, which does no entail limitation to the strong coupling expansion. 
The transformation consists essentially of two steps. In the rapidity space, they combine to form a loop such that $u\rightarrow u$ at the end of the day. 
Nevertheless, this requires changing sheets of the Riemann surface where physical observables are defined, thus making the transformation quite non-trivial.

\begin{figure}[t]
\begin{center}
\mbox{
\begin{picture}(0,155)(155,0)
\put(0,0){\insertfig{5}{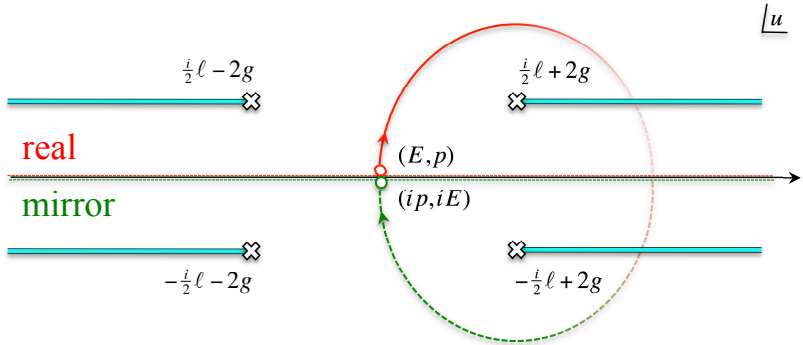}}
\end{picture}
}
\end{center}
\caption{\label{GluonDWR} Double Wick rotation for the gauge fields and bound states.}
\end{figure}

The first step is rather similar to the one adopted for the scalar. The energy $E_\ell$ and momentum $p_\ell$ for a bound state of $\ell$ gauge fields both 
have their first branch points in the upper-half plane at $u=\pm 2g + i\ell/2$. In order to reach the mirror kinematics we need first to go through them, see 
Fig.~\ref{GluonDWR}. Once this is done, we find ourselves on a new sheet where the energy and momentum both have a set of $\ell$ cuts centered 
around the origin and spaced from one another along the imaginary axis by $i$. The uppermost cut, i.e., the one from where we entered to this sheet, 
lies at $\Im \textrm{m}(u) = i\ell/2$, while the one at the bottom of the stack is at $\Im \textrm{m}(u) = -i\ell/2$. On this sheet, energy and momentum do not 
have more singular points than the ones associated with the cuts previously discussed, and, in particular, they are both analytic at infinity where they take 
the values
\be\label{Gth}
E_{\ell}(u=\infty) = \ell\, , \qquad p_{\ell}(u=\infty) = i\ell\, .
\ee
These expressions are exact at any value of the coupling%
\footnote{Note that have we decided at the very beginning to cross into the lower-half plane this would lead to the identity $E_{\ell}(p=-i\ell) = \ell$ which 
is complex conjugate of Eq.\ (\ref{Gth}).}. For the particular case of the gauge field, i.e., $\ell=1$, it is equivalent to the relation $E_{\textrm{gf}}(p=i) = 1$ 
that was first deduced in~\cite{AM07} as a consequence of the Goldstone theorem for broken symmetry. We stress that the points $u=\infty$ involved 
in~(\ref{Gth}) is completely different from infinity that could be reached by arriving to the real kinematics. The latter point indeed corresponds to the energy 
and momentum being real and taking arbitrarily large values%
\footnote{At this infinity, one finds $E_{\ell}(p=\infty) \sim \log{p}$, a result first derived at strong coupling in~\cite{DL10} and generalized to any coupling 
in~\cite{B10}.}.

However at the moment we are only half way through since we still did not strike our final destination of mirror kinematics. As can be seen from~(\ref{Gth}) 
the energy is essentially real here, while to perform the double Wick rotation we want it to be purely imaginary. Nevertheless, we made significant 
progress compared to our initial point. This can be illustrated by the following argument. We started with a particle at rest, in the real kinematics, which 
corresponds to $E_{\ell}(u= 0) = m_{\ell}$ and $p_{\ell}(u=0) = 0$, with $m_{\ell}$ being the mass of the excitation. Then we transported these along the 
imaginary line $u=i\xi$, with $\xi$ real, up to the point $\xi = \infty$ where~(\ref{Gth}) is fulfilled. Along the way, the energy was lowered from $m_{\ell}$ 
to $\ell$, since $m_{\ell} \geqslant \ell$.%
\footnote{This inequality is saturated at zero coupling as then $m_{\ell} = \ell$. At strong coupling we know that $m_{\ell} = \sqrt{2} \ell$, which is of course 
bigger than $\ell$. It is a bit of assumption that $m_{\ell}$ increases monotonically from $\ell$ to $\sqrt{2}\ell$ as the coupling grows. Though confirmed by 
the first correction at both weak and strong coupling, see~\cite{B10}, the latter claim has however not been proved for generic value of $g$.} So the next 
step naturally presents itself and suggests that what we would like to accomplish upon completing our path is to reduce the energy further, i.e., all the
way to $E_{\ell} = 0$. We can use the fact that the point $u=\infty$ is regular on the Riemann sheet where we currently reside and continue our continuation 
along $u=i\xi$ further with $\xi$ now ranging from $-\infty$ to $0$. This way we will get to the desired location $E_{\ell} = 0$.

In other words, in order to complete the transformation, the second, which is also our last, step consists in crossing through the cut lying at the bottom of the 
stack of $\ell$ cuts present on the sheet alluded to above. This way we pass yet to another Riemann sheet, which is pretty much the same as the original one 
that we started from. On this sheet we find the mirror kinematics precisely at the same position as the real kinematics. Hence, as announced before, the 
transformation is a loop from $u$ to $u$, up to the change of sheets. It is displayed in Fig.~\ref{GluonDWR}.

What is remarkable is that once all the above steps are performed making use of analytical expressions for the energy and momentum, one comes to the same 
conclusion as for the scalar excitation, namely
\be\label{DWRSgf}
E_{\ell}(u) \rightarrow ip^{\textrm{mirror}}_{\ell}(u) = ip_{\ell}(u)\, , \qquad p_{\ell}(u) \rightarrow iE^{\textrm{mirror}}_{\ell}(u) = iE_{\ell}(u)\, .
\ee
Hence we again recover the double Wick rotation symmetry~(\ref{DWRS}), in agreement with~\cite{AGMSV10}. The proof of~(\ref{DWRSgf}) is essentially the 
same as the one given earlier for scalars.

Since we established the precise form of the path in the complex plane that we have to follow to get from the real to mirror kinematics, we can start from the 
representation~(\ref{Yell}) for $Y_{\ell}(u)$ with (\ref{Pl}) and continue it to $Y_{\ell}^{\textrm{mirror}}(u)$. Going along these lines we find that the latter quantity 
is of the form~(\ref{YmirrorType}) with the exponent given by
\be\label{Mgf}
\begin{aligned}
\mathcal{E}_{\ell}(u) 
& = -2 \log{(g^2/\eta^{\ell})}-2(\ell-2)\psi(1) - 4\int_{0}^{\infty}{dt \over t}{(\cos{(ut)}\e^{-(\ell-2) t/2}-J_{0}(2gt))J_{0}(2gt)  + (\ell-2)t/2 \over \e^{t}-1} 
\\
& \, \, \, \, \, \, + 2\int_{0}^{\infty}{dt \over t}(\cos{(ut)}\e^{-\ell t/2}-J_{0}(2gt))\bigg[{\gamma_{+}(2gt) \over 1-\e^{-t}}-{\gamma_{-}(2gt) \over \e^{t}-1}\bigg]\, .
\end{aligned}
\ee
At large spin, Eq.\ (\ref{PiE}) derived for scalars is applicable here with obvious substitutions ${\rm h} \rightarrow \ell$.

The expression~(\ref{Mgf}) can be easily estimated at weak coupling. As was done earlier for scalars, we can indeed neglect the second line in~(\ref{Mgf}) and 
evaluate the first line using $J_{0}(2gt) = 1 + O(g^2)$. We get immediately that
\be
Y^{\textrm{mirror}}_{\ell}(u) = (-1)^{S}{g^4 \over \eta^{2\ell}}\Gamma({\ft{\ell}{2}+iu})^2\Gamma({\ft{\ell}{2}-iu})^2 + O(g^6)\, ,
\ee
where we used the following value for the integral
\be\label{UseIntGF}
2\int_{0}^{\infty}{dt \over t}{\cos{(ut)}\e^{-(\ell-2)t/2}-1+(\ell-2)t/2 \over \e^{t}-1} = \log{\left(\Gamma(\ft{\ell}{2}+iu)\Gamma(\ft{\ell}{2}-iu)\right)}-(\ell-2)\psi(1)\, .
\ee
As expected, the $Y$ function gets stronger suppression when the mass of the excitation increases, and for the twist-$\ell$ bound state, with mass $= \ell 
+O(g^2)$, it scales as $1/\eta^{2\ell} \sim 1/S^{2\ell}$. The dependence on the coupling, $Y^{\textrm{mirror}}_{\ell}(u) = O(g^4)$, is however the same for 
all excitations. Finally, we note that in the particular case of the gauge-field excitation, i.e., for $\ell=1$, we find
\be\label{olYgf}
Y^{\textrm{mirror}}_{\textrm{gf}}(u) = (-1)^{S}{\pi^2 g^4 \over \eta^{2}\cosh^2{(\pi u)}} + O(g^6)\, ,
\ee
which is thus identical to the one for scalar in Eq.\ (\ref{olYh}).

\subsubsection{Fermion}

Finally, we come to the discussion of fermions. In this last case to be studied, the mirror transformation is the most problematic and our understanding of the 
procedure does not really go beyond the realm of the strong coupling regime. Fortunately it is all we need in order to make an educated guess for the mirror of 
$Y_{\textrm{f}}$. So let us start with the strong coupling analysis.

We shall consider the case of a small fermion, i.e., the one carrying the rapidity $|x| < g$. At strong coupling, it implies that the momentum of the fermion is of 
order $ O(1)$, i.e., $p_{\textrm{sf}} \sim g^0$ for $g\gg 1$. This is the perturbative string regime where both the energy and momentum are found to admit an 
expansion in inverse powers of the 't Hooft coupling $1/g$ at fixed $\bar{u} \equiv u/(2g)$. The expansion is well-defined for any $\bar{u}$ in the complex plane 
except for the branch points $u=\pm 2g$ where the energy and momentum diverge. For illustration, we have~\cite{B10}
\be\label{SCR}
E_{\textrm{sf}} +p_{\textrm{sf}} = \left({\bar{u} + 1 \over \bar{u}-1}\right)^{1/4} + \ldots\, ,
\ee
with the ellipsis standing for terms being suppressed by powers of $1/g$ but enhanced by the inverse powers of $(\bar{u}^2-1)$. What happens as we get 
too close to the points $u=\pm 2g$ is that the strong coupling expansion needs to be resummed to all orders. If done properly, the singularities are resolved 
and the behavior around this points gets replaced by $E_{\textrm{sf}}(u\sim \pm 2g) \sim p_{\textrm{sf}}(u\sim \pm 2g) \sim g^{1/4}$. This is the transition to the 
near-flat-space regime that we shall not discuss here any further. In other words, we will avoid approaching too close to the branch points, which, as we recall, 
are the only singularities in the $u$-plane of the small fermion. 

\begin{figure}[t]
\begin{center}
\mbox{
\begin{picture}(0,135)(180,0)
\put(0,0){\insertfig{4.8}{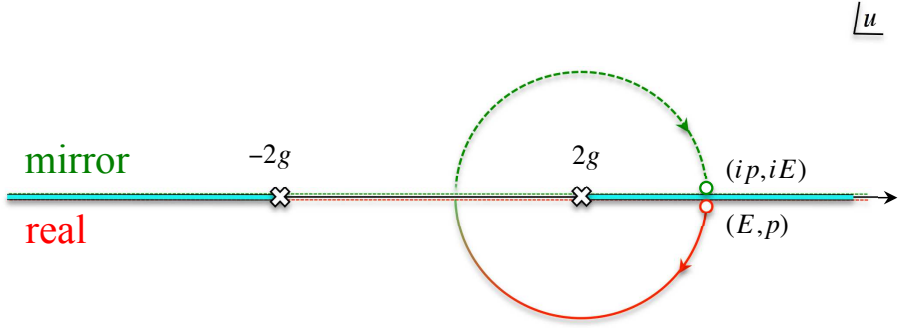}}
\end{picture}
}
\end{center}
\caption{\label{FermionStrongDWR} Double Wick rotation for the fermion in the strong coupling regime.}
\end{figure}

The mirror transformation is easy to elucidate for the perturbative string regime of the fermion. This corresponds to a clockwise rotation around $u=2g$,
see Fig.\ \ref{FermionStrongDWR}. So, that similarly to the gauge field and its bound states, the path that we sought for is a loop $u\rightarrow u$ and it 
turns out to yield
\be\label{EarrP}
E_{\textrm{sf}} \rightarrow ip_{\textrm{sf}}\, , \qquad p_{\textrm{sf}} \rightarrow iE_{\textrm{sf}}\, .
\ee
This is especially clear from~(\ref{SCR}) where such a rotation leaves the result invariant up to the phase $\exp{(i\pi/2)} =i$. More generally it would not be 
difficult to show that the symmetry~(\ref{EarrP}) is satisfied to any loop order in the $1/g$ expansion. We note finally that by passing four times around 
the point $u=2g$, we will go through the sequence of transformations
\be
E_{\textrm{sf}}(u)\rightarrow ip_{\textrm{sf}}(u) \rightarrow -E_{\textrm{sf}}(u) \rightarrow -ip_{\textrm{sf}}(u) \rightarrow E_{\textrm{sf}}(u)\, ,
\ee
in agreement with the fourth order nature of the branch point $u=2g$ apparent in Eq.\ (\ref{SCR}).

\begin{figure}[t]
\begin{center}
\mbox{
\begin{picture}(0,220)(160,0)
\put(0,0){\insertfig{7.7}{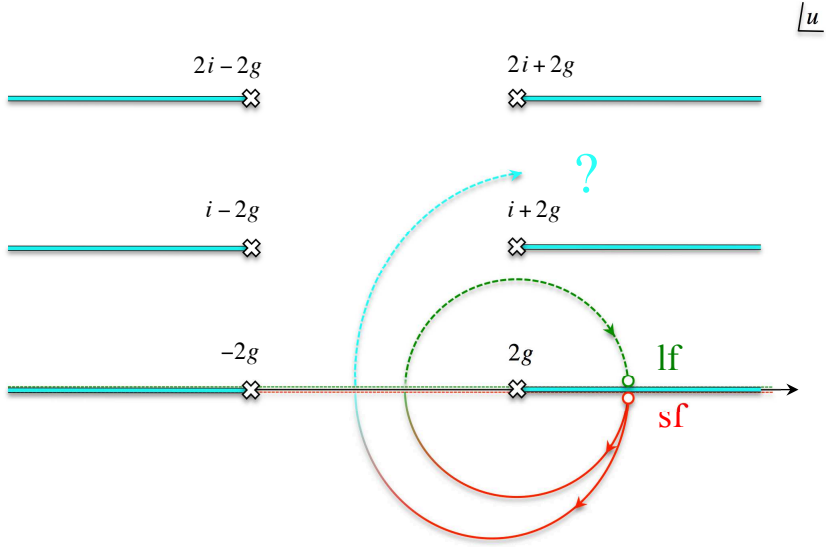}}
\end{picture}
}
\end{center}
\caption{\label{FermionWeakDWR} Continuation from the small to the large fermion domain is shown by the path going from the
lower to upper bank of the cut. On the other hand, the path to the mirror kinematics is not clear away from the strong coupling regime.}
\end{figure}

Something goes wrong at this moment however because we know from the exact expression for a small fermion that the point $u=2g$ is a genuine square-root 
branch point, see Ref.\ \cite{B10}. Moreover, rotating once around it simply amounts to changing from the small to large fermion domain,
\be\label{Outlet}
E_{\textrm{sf}}(u) \rightarrow E_{\textrm{lf}}(u)\, ,
\ee
as illustrated in Fig.\ \ref{FermionWeakDWR}. This cannot be a contender for the purported mirror transformation, since the outcome is not purely imaginary, 
and evidently contradicts~(\ref{EarrP}). There 
is no real contradiction here, rather it is a predicament which is easy to unravel. Indeed, everything we said above concerned $E_{\textrm{sf}}$ and 
$p_{\textrm{sf}}$ at strong coupling, for which the natural variable is the rescaled rapidity, $\bar{u} = u/(2g)$. The function $E_{\textrm{lf}}(u)$ that we should 
reach at the end point of the loop has an infinite number of cuts in the upper-half plane, as opposed to $E_{\textrm{sf}}(u)$. At strong coupling and in the 
variable $\bar{u}$, all these cuts condense into a single cut, the one with fourth order branch points apparent in~(\ref{SCR}). This was first elucidated 
in~\cite{KSV08} in the analysis of the analytical properties of the solution to the BES equation. In our case, it means that we never really get to the sheet 
where the energy becomes described by $E_{\textrm{lf}}(u)$. Note that if we were aiming at attempting this, we would have to resolve the singularities at 
$u=2g$ first, thus restoring the proper cut structure and then navigate between them. 

The next question is where exactly does the mirror transformation land us at the end? This is the question we did not answer. From our discussion above 
it seems clear that the mirror transformation requires a loop $u\rightarrow u$ crossing some of the aforementioned cuts. Unfortunately, all our attempts to 
deform the loop by crossing more cuts along the way have failed to produce a consistent mirror transformation. The fact that at strong coupling the infinite 
tower of cuts condense into one could imply that the exact mirror transformation cannot be obtained by crossing only a few of them.  If it is the case 
one would have to figure out how to implement the continuation through an infinite tower of cuts, which goes far beyond our current abilities, see Fig.\ 
\ref{FermionWeakDWR}. The situation 
is however not entirely hopeless, because, after all, we have control on the mirror transformation at strong coupling. This means that we should be able 
to construct the mirror of $Y_{\textrm{sf}}(u) \sim \exp{iP_{\textrm{sf}}(u)}$ at least in the latter regime. It is worth a try because, as we shall see, it produces 
an expression for $\mathcal{E}_{\textrm{sf}}(u)$ which is acceptable and meaningful at any coupling.

Our consideration to turn $P_{\textrm{sf}}(u)$ into the yet mysterious $\mathcal{E}_{\textrm{sf}}(u)$ goes as follows. The starting point is the identity
\be\label{RecRelA}
P_{\textrm{h}}(u-\ft{i}{2}+i0) = P_{\textrm{lf}}(u+i0) - \tilde{P}_{\textrm{sf}}(u+i0)\, ,
\ee
that relates $P_{\textrm{h}}$ and $P_{\textrm{lf}}$. In this equation everything is known except for the new quantity $\tilde{P}_{\textrm{sf}}(u+i0)$ that is 
defined as
\be\label{tildeP}
\tilde{P}_{\textrm{sf}}(u+i0) \equiv -2i\log{\left({g^2 \over \eta x}\right)}-\pi - i\int_{0}^{\infty}{dt \over t}\e^{iut}\gamma_{+}(2gt)\, ,
\ee
where $x$ is the small solution of the Zhukowski map $u=x+g^2/x$, i.e., such that $|x| < g$. The equation~(\ref{RecRelA}), with $\tilde{P}_{\textrm{sf}}$ as 
above, is a direct consequence of the expressions~(\ref{Ph}) and~(\ref{Plf}) for $P_{\textrm{h}}$ and $P_{\textrm{lf}}$, respectively. Now we can use the 
fact that the small and large fermion are connected for $u^2 < (2g)^2$, i.e., that $P_{\textrm{lf}}(u+i0) = P_{\textrm{sf}}(u-i0)$, to find that
\be\label{RecRelB}
P_{\textrm{h}}(u-\ft{i}{2}+i0) = P_{\textrm{sf}}(u-i0) - \tilde{P}_{\textrm{sf}}(u+i0)\, ,
\ee
when $u^2 < (2g)^2$. This introduces $P_{\textrm{sf}}(u)$ into the game. What is interesting in the equality~(\ref{RecRelB}) is that its left-hand side is 
exponentially small at strong coupling, as long as $u$ stays away from $u= \pm 2g$. This is a simple consequence of the fact that, as we discussed 
before, the regime $u^2 < 2g^2$ happens to be non-perturbative for the scalar excitation at strong coupling, thus resulting in the asymptotic behavior 
$P_{\textrm{h}} \sim \exp{(-\pi g)}$. Therefore, at strong coupling, the relation~(\ref{RecRelB}) simplifies to
\be\label{PassRule}
P_{\textrm{sf}}(u-i0) = \tilde{P}_{\textrm{sf}}(u+i0) \, , \qquad u^2 < (2g)^2\, ,
\ee
up to exponentially small corrections.%
\footnote{We emphasize again that these corrections grow however if $u$ gets close to $\pm 2g$, such that we should avoid these points, in agreement
with our previous discussion of the mirror transformation in the perturbative regime.} In other words, the two functions $P_{\textrm{sf}}(u)$ and 
$\tilde{P}_{\textrm{sf}}(u)$ are, at the perturbative level, analytically related to one another through the cut along $u^2 < (2g)^2$. It immediately follows 
that after going around the loop $u\rightarrow u$, describing the strong coupling mirror transformation, we get
\be
P_{\textrm{sf}}(u)  \rightarrow \tilde{P}_{\textrm{sf}}(u)\, .
\ee
Indeed, according to~(\ref{PassRule}), after passing in between the branch points $u=\pm 2g$, the function $P_{\textrm{sf}}(u)$ turns into 
$\tilde{P}_{\textrm{sf}}(u)$, to any order in the $1/g$ expansion. Thus the strong coupling mirror transformation for the small fermion leads to
\be
\mathcal{E}_{\textrm{sf}}(u) = -i\tilde{P}_{\textrm{sf}}(u)\, ,
\ee
with $\tilde{P}_{\textrm{sf}}(u)$ given in~(\ref{tildeP}). If we are only interested in the image in the mirror kinematics of a fermion with real momentum, 
then we can restrict ourselves to $u^2 > (2g)^2$ and write
\be\label{Msf}
\mathcal{E}_{\textrm{sf}}(u) = -2\log{\left({g^2 \over \eta x}\right)}+i\pi - \int_{0}^{\infty}{dt \over t}\cos{(ut)}\gamma_{+}(2gt)\, .
\ee
In a nutshell, what we have done by neglecting $P_{\textrm{h}}(u-\ft{i}{2}+i0)$ in~(\ref{RecRelB}) is precisely to remove an infinite tower of cuts depicted 
in Fig.~\ref{FermionWeakDWR} that was obstructing the mirror transformation. Thus we left the lower-half plane of the small fermion, which is free of singularities 
and controlled by $P_{\textrm{sf}}(u)$, and enter, after the above trick, into a `new' upper-half plane with similar analytical properties governed by 
$i\mathcal{E}_{\textrm{sf}}(u) =\tilde{P}_{\textrm{sf}}(u)$.

Now let us emphasize an important point, which also happens to become our working hypothesis, that the expression~(\ref{Msf}) is perfectly suitable 
for direct non-perturbative generalization. As it stands, it is perfectly well defined at any value of coupling and even obeys a very suggestive identity. 
Namely, when expanded at large spin, it leads to
\be\label{Consist}
\mathcal{E}_{\textrm{sf}}(u) = 2E_{\textrm{sf}}(u)\log{\bar{\eta}} + O(\log^0{\eta})\, ,
\ee
at any coupling, in perfect agreement with the putative double Wick rotation symmetry for the fermion. If we assume furthermore that 
$p^{\textrm{mirror}}_{\textrm{sf}}(u) = p_{\textrm{sf}}(u)$, which we know is satisfied at strong coupling, the situation becomes completely symmetric 
with the one observed for the bosonic excitations. Finally, it is interesting to note that a natural attempt to put fermions and bosons on an equal footing, 
as inspired by our findings for bosonic excitations, would allow us to conjecture that
\be
\mathcal{E}_{\textrm{sf}}(u)  = 2\log{\eta} - f(u) - \int_{0}^{\infty}{dt \over t}\cos{(ut)}\gamma_{+}(2gt)\, ,
\ee
with $f(u)$ being a spin-independent function of rapidity $u$. This ansatz would immediately warrant the validity of the large spin asymptotics~(\ref{Consist}). 
Our preceding analysis implies us that the strong-coupling mirror transformation yields
\be\label{Guessf}
f(u) = 2\log{\left({g^2 \over x}\right)}-i\pi\, ,
\ee
with $x$ being the small solution to $u=x+g^2/x$. Our proposal is to assume that~(\ref{Guessf}) is actually valid at any coupling, not only $g \to \infty$ where
it was rederived. We shall see that it leads to very sensible quantitative results at weak coupling in complete agreement with perturbative expectations.

Now being equipped with the expression for $\mathcal{E}_{\textrm{sf}}(u) $, we find $Y^{\textrm{mirror}}_{\textrm{sf}}(u)$ via the formula
\be\label{PtoY}
Y^{\textrm{mirror}}_{\textrm{sf}}(u) = -(-1)^S\e^{-\mathcal{E}_{\textrm{sf}}(u) }\, .
\ee
One peculiar feature of $Y^{\textrm{mirror}}_{\textrm{sf}}(u)$, as compared to the bosonic case, is that it displays a singular behavior at $u=\infty$, which 
corresponds to $x=0$, i.e., the fermion at rest. Indeed, one easily verifies from~(\ref{PtoY}) and~(\ref{Msf}) that $Y^{\textrm{mirror}}_{\textrm{f}}(x) \sim 1/x^2$ 
at small $x$. It originates from the term $2\log{x}$ in~(\ref{Msf}) and is therefore justified at any coupling. The coupling-dependent piece is indeed controlled 
by the integral involving the function $\gamma_{+}(t)$ in~(\ref{Msf}) which is always regular at $x=0$. This is apparent from the representation
\be
\int_{0}^{\infty}{dt \over t}\cos{(ut)}\gamma_{+}(2gt) =  2\sum_{n \ge 1}\gamma_{2n}\left({ix \over g}\right)^{2n} \sim x^2\, .
\ee
The above mentioned singular behavior of the $Y$-function requires a prescription for integration at small momenta in the L\"uscher formula, as we shall see
and elucidate in the following section. Finally, we notice the presence of $i\pi$ in~(\ref{Msf}) which after exponentiation compensates for the extra minus sign 
for the fermion present in~(\ref{PtoY}). This way the expression for $Y^{\textrm{mirror}}_{\textrm{sf}}(u)$ supports the form~(\ref{YmirrorType}) with 
$E_{\textrm{sf}}(u)$ and $\delta E_{\textrm{sf}}(u)$ being real functions.%
\footnote{These functions are real for $u^2 > (2g)^2$.}

The last step is to derive the expression $\mathcal{E}_{\textrm{lf}}(u)$ for the fermion in the large rapidity domain which is the pertinent domain at 
weak coupling. This is quite straightforward to do and follows from the analytical continuation of Eq.\ (\ref{Msf}) to $|x| > g$ region. Using the equations 
satisfied by $\gamma(t)$, we easily find that
\be\label{Mlf}
\begin{aligned}
\mathcal{E}_{\textrm{lf}}(u) &= -2\log{\left({g^2 \over \eta x}\right)}+i\pi - 4\int_{0}^{\infty}{dt \over t}{(\cos{(ut)}-J_{0}(2gt)) J_{0}(2gt) \over \e^{t}-1}\\
&\, \, \, \, \, \, \, + 2\int_{0}^{\infty}{dt \over t}{\cos{(ut)}-J_{0}(2gt) \over \e^{t}-1} \gamma(-2gt) + \int_{0}^{\infty}{dt \over t}\cos{(ut)}\gamma_{+}(2gt)\, ,
\end{aligned}
\ee
where $u^2 > (2g)^2$ and with now $x$ being the large solution of $u=x+g^2/x$. It has a form very similar to the ones derived for the bosons, which is 
very encouraging.

Finally, the evaluation of the mirror $Y$-function at weak coupling can be done along the same lines as for the bosons. Namely, neglecting the second line 
in~(\ref{Mlf}) and replacing $J_{0}(2gt)\rightarrow 1$, we obtain
\be\label{olYlf}
Y^{\textrm{mirror}}_{\textrm{lf}}(u) =  (-1)^{S}{\pi^2 g^4 \over \eta^{2}\sinh^2{(\pi u)}} + O(g^6)\, ,
\ee
after using the integral
\be
2\int_{0}^{\infty}{dt \over t}{\cos{(ut)}-1 \over \e^{t}-1} = \log{\left({\pi u\over \sinh{(\pi u)}}\right)}\, ,
\ee
together with $x = u +O(g^2)$. The previously observed small-$x$ singularity manifests itself in the divergent behavior $\sim 1/u^2$ of~(\ref{olYlf}) at 
vanishing $u$, at this order of the coupling. Except for that, we note a striking similarity with expressions deduced by a rigorous analysis for the bosons, 
both in the dependence on the coupling $\sim g^4$ and in suppression at large rapidity.

To conclude this section, let us add a comment on the mirror $Y$-function for the small fermion. The latter does not play a significant role in our analysis at 
weak coupling. Rather than performing the entire analysis anew in this domain, one could instead find the result directly from~(\ref{olYlf}). Namely, one
first performs the small momentum expansion by using the relation $p=2u+O(g^2)$ for the large fermion and then keeping the leading term at small $p$ 
from~(\ref{olYlf}) and finally reexpressing it in terms of the rapidity variable of the small fermion $p=2g^2/u + O(g^4)$. This would be observed to match the 
weak coupling expansion starting directly from~(\ref{PtoY}) and~(\ref{Msf}), yielding $Y^{\textrm{mirror}}_{\textrm{sf}}(u) = (-1)^S u^2/\eta^2 + O(g^2)$.

\subsection{L\"uscher formula}
\label{LF}

We are now in a position to evaluate the L\"uscher formula at weak coupling. We recall that it controls the finite-size corrections to the energy 
of the twist-two operators. Its form was previously motivated in~(\ref{EfsFirst}) and argued to be given by
\be\label{Luscher}
E^{\textrm{FS}}  
= 
-6\int{dp \over 2\pi}\, Y_{\textrm{h}}^{\textrm{mirror}}(p) 
-2\int{dp \over 2\pi}\, Y_{\textrm{gf}}^{\textrm{mirror}}(p) 
-8\int{dp \over 2\pi}\, Y_{\textrm{f}}^{\textrm{mirror}}(p) 
+ \ldots\, ,
\ee
where the ellipsis stand for contributions from bound states of gauge fields and/or from subleading order L\"uscher corrections, both suppressed 
by higher powers of the spin (at weak coupling).%
\footnote{Higher L\"uscher corrections coming from the scalar sector are no longer small at strong coupling because the mass of the scalar 
becomes vanishingly small as $g \to \infty$. But as long as the scalar mass is of order one, which is the case to any given order in the weak 
coupling expansion, the first L\"uscher correction is suppressed by $1/\eta^2 \sim 1/S^2$ and subleading ones by extra powers of the spin.}
The main ingredients entering this equation are the $Y$ functions in the mirror kinematics
\be\label{YLuscher}
Y_{\star}^{\textrm{mirror}}(p) \, ,
\ee
for the elementary twist-one excitations, $\star = \textrm{h}, \textrm{gf}, \textrm{f}$. These were constructed in the previous subsection to leading order 
at large spin and all orders of the 't Hooft coupling. The obtained asymptotic expressions are sufficient for our purposes and we do not need to worry 
about possible finite-size corrections to the $Y$'s themselves. These effects should enter at the next order of the large-spin expansion and would 
require us to develop systematic TBA equations for their determination.

Before we turn to the weak-coupling analysis of Eq.\ (\ref{Luscher}), some comments are in order concerning the choice of the integration contours 
there. When parametrized in terms of the momentum, it is clear that each integral in~(\ref{Luscher}) should be taken for real $p$ to run from $-\infty$ 
to $+\infty$. For the scalar and gauge fields contributions the same integration contour is maintained as one passes to the more convenient rapidity 
$u$ as the integration variable. However, the fermion case is a bit more tricky. The physical rapidity here is $x$ and the contribution from the fermions reads
\be\label{EFSf}
E^{\textrm{FS}}_{\textrm{f}}  = -8\int{dx \over 2\pi}{dp_{\textrm{f}} \over dx}\, Y_{\textrm{f}}^{\textrm{mirror}}(x) \, ,
\ee
with $x$ ranging from $-\infty$ to $+\infty$. But this integral should come with a prescription for integrating around the point $x=0$ on the real axis, 
corresponding to $p=0$. Indeed, as we already observed, the function $Y^{\textrm{mirror}}_{\textrm{f}}(x)$ is singular at small $x$ and behaves as
\be\label{Yx0}
Y_{\textrm{f}}^{\textrm{mirror}}(x) \sim 1/x^2 + O(x^2)\, .
\ee
We choose to avoid this singularity by integrating around the point $x=0$ either by going slightly above or below this point in the complex $x$-plane. 
The two choices are equivalent; their difference being proportional to
\be
\oint dx \,  {dp_{\textrm{f}} \over dx}Y_{\textrm{f}}^{\textrm{mirror}}(x)\, ,
\ee
evaluated around $x=0$, which vanishes since there is no residue at this point. The latter property also guarantees that the 
integral~(\ref{EFSf}) with the $\pm i0$ prescription around $x=0$ is real.

When written in terms of the rapidity $u$, the integral~(\ref{EFSf}) receives contribution both from the large and small fermion domains, corresponding 
to $|x| \gtrless g$, respectively. At weak coupling, it is however more efficient to evaluate integrals involving quantities computed for large fermions. The
reason behind it is that the large fermion domain covers the regime with $p =O(1)$ for $g \sim 0$ which is more important than the one implied by the 
small fermion domain, corresponding to $p\sim g^2$ for $g\sim 0$. To re-express~(\ref{EFSf}) in terms of $Y_{\textrm{lf}}^{\textrm{mirror}}(u)$ and 
$p_{\textrm{lf}}(u)$ solely, we can deform the contour of integration in Eq.\  (\ref{EFSf}) such that it resides entirely in the large fermion domain $|x| > g$. 
There are no obstructions encountered along the way since both $p_{\textrm{f}}(x)$ and $Y_{\textrm{f}}^{\textrm{mirror}}(x)$ are analytical inside the 
disk $|x| \leqslant g$, except for the double pole at $x=0$ in $Y_{\textrm{f}}^{\textrm{mirror}}(x)$ which is bypassed. Then one finds the representation
\be\label{EFSlf}
E^{\textrm{FS}}_{\textrm{f}}  = -8\int{du \over 2\pi}{dp_{\textrm{lf}} \over du}\, Y_{\textrm{lf}}^{\textrm{mirror}}(u) \, ,
\ee
where the contour of integration runs from $u=-\infty$ to $u=+\infty$ avoiding the cut along $u^2 <(2g)^2$ by going either slightly above or slightly below 
it. This is the representation we shall use for our computation at weak coupling.

Finally, let us note that the weak coupling expansion performs a kind of Laurent expansion in the integrand of Eq.\ (\ref{EFSlf}) which is valid away from 
the various cuts of $p_{\textrm{lf}}(u)$ and $Y_{\textrm{lf}}^{\textrm{mirror}}(u)$. In particular, the cut centered around $u=0$, which we evade in~(\ref{EFSlf}),
decomposes into contributions of the type $\sim g^{2n}/u^{2n}$. As a consequence, at weak coupling, all contributions obtained by expanding the integrand in Eq.\ (\ref{EFSlf}) in powers of $g^2$ are regularized by integrating with a $\pm i0$ prescription around $u=0$. This is how all integrals involving fermions entering the weak coupling expansion of~(\ref{EFSlf}) will be understood henceforth. A similar 
prescription was adopted in~\cite{GMSV, GMSVbis} when evaluating integrals relevant to the computation of gluon scattering amplitudes using the OPE for light-like Wilson loops~\cite{AGMSV10} and which display an intriguing resemblance with the ones considered presently.

\subsubsection{Weak coupling expansion}

The result~(\ref{Luscher}) for the finite-size energy correction admits an infinite series expansion at weak coupling
\be\label{FSEN}
E^{\textrm{FS}} = {(-1)^S \over \eta^2}\sum_{n\ge 2}\mathcal{P}_{n} \, g^{2n} \, ,
\ee
where $\mathcal{P}_{n}$ a polynomial in $\log{\bar{\eta}}$ of degree $n-2$ with $\log{\bar{\eta}} \equiv \log{\eta}+\gamma_{\rm E}$ as before. This structure is 
an immediate consequence of the general form for the $Y$ functions that we established earlier,
\be\label{Ystar}
Y_{\star}^{\textrm{mirror}}(u) = (-1)^S\exp{(-2E_{\star}(u) \log{\bar{\eta}} -2\delta E_{\star}(u))}\, ,
\ee
with the scaling $E_{\star}(u) = 1 +O(g^2)$, valid for all twist-one excitations, explaining the overall suppression factor in $E^{\textrm{FS}} \sim 1/\eta^2 \sim 
1/S^2$. The weak coupling expansion of the exponent (\ref{Ystar}) obviously generates powers of $\log{\bar{\eta}}$ increasing with the loop order and 
ultimately leads to the representation (\ref{FSEN}).

We will report in this section on the explicit five-loop calculation of the finite-size energy~(\ref{FSEN}), with details deferred to Appendix~\ref{LFWC}. In 
order to illustrate the problem at hand let us consider a particular set of contributions from the leading logarithmic effects. These are the coefficients in
Eq.\ (\ref{FSEN}) accompanying the maximal power of $\log{\bar{\eta}}$ at a given order in 't Hooft coupling, i.e., 
\be
\mathcal{P}_{n} = c_{n-2}\log^{n-2}{\bar{\eta}} + \ldots\, ,
\ee
where $n = 2, 3, \ldots\, $. For these, the expression derived from the L\"uscher formula~(\ref{Luscher}) is rather compact and the emerging integrals
are generic for its weak coupling expansion. The $c_n$ coefficients are generated by~(\ref{Luscher}) upon the substitution
\be\label{LLF}
\begin{aligned}
Y^{\textrm{mirror}}_{\textrm{h}}(u) &\rightarrow (-1)^{S}{\pi^2 g^4 \over \eta^2 \cosh^2{(\pi u)}}
\exp{\bigg[-4g^2(\psi(\ft{1}{2} +iu)+\psi(\ft{1}{2} -iu)-2\psi(1))\log{\bar{\eta}}\bigg]}\, , \\
Y^{\textrm{mirror}}_{\textrm{gf}}(u)  &\rightarrow (-1)^{S}{\pi^2 g^4 \over \eta^2\cosh^2{(\pi u)}}
\exp{\bigg[-4g^2(\psi(\ft{3}{2} +iu)+\psi(\ft{3}{2} -iu)-2\psi(1))\log{\bar{\eta}}\bigg]}\, , \\
Y^{\textrm{mirror}}_{\textrm{lf}}(u)  &\rightarrow (-1)^{S}{\pi^2 g^4 \over \eta^2 \sinh^2{(\pi u)}}
\exp{\, \bigg[-4g^2(\psi(1 +iu)+\psi(1-iu)-2\psi(1))\log{\bar{\eta}}\, \bigg]}\, ,
\end{aligned}
\ee
together with the fact that $p_{\star} = 2u+O(g^2)$ for all of these twist-one excitations. Of course the form~(\ref{LLF}) does not produce correct subleading-log
contributions beyond $O(g^4)$, however, this is irrelevant for the exploratory calculation we are after. Plugging~(\ref{LLF}) into Eq.\ (\ref{Luscher}) and 
comparing with~(\ref{FSEN}), we find
\be\label{cnn}
c_{n} = 6\,I_{\textrm{h}}^{(n)} + 2\, I_{\textrm{gf}}^{(n)} + 8\, I_{\textrm{f}}^{(n)}\, ,
\ee
where
\be\label{Ihgff}
\begin{aligned}
I_{\textrm{h}}^{(n)} &= - {(-4)^n \pi \over n!}\int du\,  {\left(\psi(\ft{1}{2}+iu) + \psi(\ft{1}{2}-iu)-2\psi(1)\right)^n  \over \cosh^2{(\pi u)}} \, ,\\
I_{\textrm{gf}}^{(n)} &= - {(-4)^n \pi \over n!}\int du\, {\left(\psi(\ft{3}{2}+iu) + \psi(\ft{3}{2}-iu)-2\psi(1)\right)^n  \over \cosh^2{(\pi u)}} \, ,\\
I_{\textrm{f}}^{(n)} &= - {(-4)^n \pi \over n!}\int du\, {\left(\psi(1+iu) + \psi(1-iu)-2\psi(1)\right)^n  \over \sinh^2{(\pi u)}} \, .
\end{aligned}
\ee
The contour of integration is chosen to be the real $u$-axis for the scalar and gauge fields, and similarly for the fermion but with a $\pm i0$ prescription
to go around the point $u=0$, as discussed before. For any $n = 0, 1, \ldots$ the integrals~(\ref{Ihgff}) are well-defined and real, yielding a definite 
answer for the expansion coefficients $c_{n}$. For low enough $n$, their evaluation is feasible but lengthy, see Appendix~\ref{LFWC} for details, and 
eventually gives
\be\label{c11dots}
c_{0} = 0
\, , \qquad 
c_{1} = -{16\pi^2 \over 3}
\, , \qquad 
c_{2} = -\left(256\zeta_{3} + {64\pi^2 \over 3}\right)
\, , \qquad 
c_{3} = -\left({448\pi^4 \over 45} +1024\zeta_{3}-{256\pi^2 \over 3} \right)
\, . 
\ee
This finding suggests that the degree of transcendentality of the coefficients $c_{n}$ increases with the loop order, in agreement with general expectation for twist-two anomalous dimension. It is also 
interesting to point out the vanishing of the constant $c_{0}$. This property is a consequence of the identity
\be
\int du{1 \over \sinh^2{(\pi (u\pm i0))}^2} = -\int du{1 \over \cosh^2{(\pi u)}} \, ,
\ee
which is obtained by shifting the contour of integration to the upper-/lower-half plane, $u\rightarrow u\pm \ft{i}{2}$. Then, due to the exact balance between 
the total numbers of fermions and bosons the coefficient $c_{0}$ vanishes. It is not entirely clear to us whether this cancellation should have been expected 
or if it merely reflects the enlarged symmetry algebra of the spectrum of excitations at weak coupling. A similar phenomenon was observed for the (BMN) 
L\"uscher formula in Refs.~\cite{BJ, BJL} where the cancellation occurs up to four loops. There, the effect relies on the supersymmetry of the BMN vacuum. 
In our case the cancellation takes place at the first order (i.e., two loops) only and is presumably restricted to the leading power corrections $\sim 1/\eta^2$ 
that we are considering in our study. For contributions suppressed stronger by the spin, it would indeed be rather surprising to find that the cancellation persists, 
since bound states entering into the game do not possess super-partners thus distorting the equality of fermionic and bosonic degrees of freedom. Nevertheless 
the phenomenon of cancellation turns out to be rather generic among the various integrals that contribute to the polynomials $\mathcal{P}_{n}$ in~(\ref{FSEN}).%
\footnote{Indeed, the weak coupling expansion of $Y^{\textrm{mirror}}_{\star}(u)$ generates many periodic factors that simplify among bosons and fermions 
after using the contour deformation trick adopted above. However, it also generates combinations of derivatives of $\psi$-functions that are not periodic and 
do not lead to a complete cancellation, see Appendix~\ref{MYWK} and~\ref{LFWC}.}

Summarizing, by taking into account omitted radiative corrections to Eq.\ (\ref{LLF}), we can immediately complete the polynomials in~(\ref{FSEN}) . The
analysis outlined in Appendix~\ref{LFWC} provides the polynomials up to five loops,
\be\label{FinalRes}
\begin{aligned}
&\mathcal{P}_{2} = 0\, , \\
&\mathcal{P}_{3} = -{16\pi^2 \over 3}\log{\bar{\eta}}\, , \\
&\mathcal{P}_{4} = -\left(256\zeta_{3}+{64\pi^2 \over 3}\right)\log^2{\bar{\eta}} + {112\pi^4 \over 15}\log{\bar{\eta}}+16\pi^2\zeta_{3}\, , \\
&\mathcal{P}_{5} = -\left({448\pi^4 \over 45}+1024\zeta_{3} -{256\pi^2 \over 3}\right)\log^{3}{\bar{\eta}} 
+ \left(2560\zeta_{5}+{896 \pi^2 \zeta_{3} \over 3}+ {1664\pi^4 \over 45}+256\zeta_{3}+{128\pi^2 \over 3} \right)\log^2{\bar{\eta}} \\
& \qquad \, \, \, 
+ \left(1536\zeta_{3}^2 -{8992 \pi^6 \over 945}+128\pi^2\zeta_{3}\right) \log{\bar{\eta}} -\left({320\pi^2\zeta_{5} \over 3}+{896\pi^4 \zeta_{3} \over 45}\right)\, .
\end{aligned}
\ee
An attentive reader would notice the absence of $\sim \log^0{\bar{\eta}}$ contribution in $\mathcal{P}_{3}$. Again, this stems from the cancellation between the
bosons and fermions in complete similarity to the identity $\mathcal{P}_{2} \equiv c_{0} = 0$ examplified before. 

\subsubsection{Finite-size vs wrapping corrections}

We are now in position to compare the finite-size corrections to the GKP vacuum, we are advocating here, and the wrapping corrections to states built on
the BMN ground state. We first recall what we have learnt and what we aim at proving.

On the one hand, we have shown that the twist-two ABA energy can be decomposed as
\be\label{Eaba}
E^{\rm ABA} = E^{\textrm{bulk}} + E^{\textrm{alt}}\, ,
\ee
where $E^{\textrm{alt}}$ is the alternating component of the ABA energy that scales as $E^{\textrm{alt}} \sim (-1)^S/\eta^2$ at large spin, up to powers of 
$\log{\eta}$ in the numerator, and so to any order at weak coupling, see Eqs.\ (\ref{EaltPol}) and (\ref{ABAPol}). The remaining component, $E^{\textrm{bulk}}$, 
does not alternate with the parity of the spin $S$ and has a much softer large-spin expansion, given in Eq.\ (\ref{Ebulk-C}) up to $O(1/\eta^2)$. On the 
other hand, motivated by the picture of the twist-two background as defining the ground state of the (long) GKP string, we have constructed the finite-size 
energy $E^{\rm FS}$ valid at leading order in large spin in the form of the L\"uscher formula~(\ref{Luscher}). We have seen that this quantity alternates with 
$S$ mimicking $E^{\textrm{alt}}$ and its evaluation at weak coupling have shown that it scales as $E^{\rm FS} \sim (-1)^S/\eta^2$, and, more generally, has 
the structure similar to the one observed for $E^{\textrm{alt}}$, see Eqs~(\ref{FSEN}) and (\ref{FinalRes}).

As alluded to in the Introduction, our prescription for correcting the ABA predictions $E^{\rm ABA}$ and recovering the complete twist-two energy $E$ is 
to perform the substitution
\be\label{SubtRule}
E^{\textrm{alt}} \rightarrow E^{\rm FS}\, ,
\ee
such that one replaces (\ref{Eaba}) by
\be\label{GenForbis}
E = E^{\textrm{bulk}} + E^{\rm FS}\, .
\ee
Though we do not have a compelling physical reason at weak coupling why this is the proper procedure, the similarity of the alternating ABA energy 
and finite-size corrections, together with the evidence that the ABA result has to be corrected, makes the substitution~(\ref{SubtRule}) as a working
hypothesis. We will later provide more physical evidence in favor of (\ref{SubtRule}) coming from the strong coupling analysis. However, first we will 
check that our substitution rule passes quantitative tests at weak coupling. At a qualitative level, one could regard the sum (\ref{GenForbis}) as a
decomposition of two different classes of contributions, shown in Fig.~\ref{Cylinder}. The bulk energy should resum those corrections associated to 
diagram with trivial topology as in Fig.~\ref{Cylinder} (d). The ones controlled 
by the finite-size energy are associated with graphs possessing non-trivial topology, with the first L\"uscher correction being associated to those 
with excitations wrapping once around the GKP string, see Fig.~\ref{Cylinder} (c). Since our understanding of the decomposition~(\ref{GenForbis}) 
is so far restricted to the leading contribution at large spin, it should hold true up to corrections suppressed stronger than $1/\eta^2 \sim 1/S^2$ to any 
order in weak coupling.

We notice that as a corollary of~(\ref{GenForbis}) we should recover the identity
\be\label{FSvsWrap}
E^{\rm FS} = E^{\textrm{alt}} + E^{\textrm{wrap}}\, ,
\ee
since by construction the wrapping corrections complete the ABA result, $E = E^{\rm ABA}+E^{\textrm{wrap}}$. One of the main results of this paper is to 
establish that the relation~(\ref{FSvsWrap}) is indeed correct up to five loops at weak coupling (allowing us to extrapolate this claim to any order in $g$). 
One might be concerned that though both $E^{\rm FS}$ and $E^{\textrm{alt}}$ alternate with spin this turns out not to be the case for $E^{\textrm{wrap}}$, 
as we shall see. (This is the case even for the leading order contribution to $E^{\textrm{wrap}}$ in the large-spin limit.) However, this fact is not surprising
at all if one recalls that only even values of the spin are physical, i.e, corresponds to state with zero quasimomentum consistently described by the long-range
ABA. The conclusion therefore is that~(\ref{FSvsWrap}) should be obeyed for even spins only. The fact that the extension to odd spins proposed here does 
not match the one realized in the construction of the wrapping corrections should not be so puzzling after all. The extension to odd spins of the (BMN) 
L\"uscher formula of Ref.\ \cite{BJL} was done under special requirements that do not seem to have a simple physical interpretation in our picture. Reciprocally, 
it appears that odd spin configurations are somehow associated, in our case, to a change in the boundary conditions imposed to excitations propagating 
around the GKP string; a property which bears no similarity with the steps adopted in~\cite{BJL}.

Keeping this caveat in mind, let us now come to the quantitative test. In to order to simplify the comparison, let us introduce a representation for the wrapping 
corrections similar to the ones found for the alternating and finite-size energies~(\ref{EaltPol}, \ref{FSEN}). Hence, we set
\be
E^{\textrm{wrap}} = {(-1)^S \over \eta^2}\sum_{n \ge 2} P^{\textrm{wrap}}_n g^{2n} + o(1/\eta^2)\, ,
\ee
where $P^{\textrm{wrap}}_n$ is a coefficient (weakly dependent on spin) emerging from the large-$S$ expansion of the $n$-loop wrapping energy 
that governs its $1/\eta^2$-scaling behavior. Testing Eq.\ (\ref{FSvsWrap}) is then equivalent to verifying
\be\label{PPwrap}
\mathcal{P}_{n} = P_{n} + P_{n}^{\textrm{wrap}}\, ,
\ee
with $\mathcal{P}_{n}$ and $P_{n}$ known explicitly from Eqs.\ (\ref{FinalRes}) and (\ref{ABAPol}) up to five loops. We first recall that since both 
$\mathcal{P}_{n}$ and $P_{n}$ are polynomials of degree $n-2$ in $\log{\bar{\eta}}$, the same should be the case for $P_{n}^{\textrm{wrap}}$, 
at least for even values of the spin. We will see that this is indeed the case up to five loops, independent of the parity of the spin, as was already 
elucidated at four loops in~\cite{BF09} by the large-spin expansion of available wrapping corrections obtained in Ref.\ \cite{BJL}.

The two particular cases of~(\ref{PPwrap}) yield $\mathcal{P}_{2} = P_{2}$ and $\mathcal{P}_{3} = P_{3}$, since the wrapping effects start at four loops
for twist-two operators, i.e., $P_{2, 3}^{\textrm{wrap}} = 0$. Comparing~(\ref{ABAPol}) and~(\ref{FinalRes}) we see that these relations are indeed 
satisfied, and what is more they are correct for both even and odd spins. To proceed further we need $P^{\textrm{wrap}}_4$ and $P^{\textrm{wrap}}_5$. 
Starting from the exact (finite-spin) expressions for the four-loop and five-loop wrapping corrections to twist-two anomalous dimension found in Refs.\ \cite{BJL} 
and \cite{LRV}, respectively, we derived the following expressions,
\be
\begin{aligned}\label{WrapPol}
P^{\textrm{wrap}}_4 
&= -\left(128\zeta_{3}+{64 \pi^2 \over 3}\sigma\right)\log^2{\bar{\eta}}\, , \\
P^{\textrm{wrap}}_5 
&= -\left({64\pi^4 \over 9}+\left(1024\zeta_{3}-{256\pi^2 \over 3}\right)\sigma\right)\log^3{\bar{\eta}} \\
&\, \, \, \, + \left(1280\zeta_{5} + {512\pi^2\zeta_{3} \over 3} + \left({1664\pi^4\over 45} + 256\zeta_{3}
+{128\pi^2 \over 3} \right)\sigma\right)\log^2{\bar{\eta}}+\left(768\zeta_{3}^2+128\pi^2\zeta_{3}\sigma\right)\log{\bar{\eta}}\, ,
\end{aligned}
\ee
with the signature factor $\sigma \equiv (-1)^S$. As expected they are polynomials with correct degrees of $\log{\bar{\eta}}$ at a given order in 't Hooft
coupling. For even spins, i.e., $\sigma=1$, we checked the agreement at four loops with the large spin result of~\cite{BF09}, since $\log{\bar{\eta}} = 
\log{\bar{S}} \equiv \log{S}+\gamma_{\rm E}$ at the given accuracy. The five-loop result seems to be new, however. What is important here is that, after 
plugging~(\ref{ABAPol}), (\ref{FinalRes}) and (\ref{WrapPol}) into Eq.\ (\ref{PPwrap}) for $n=4, 5,$ we immediately verify the validity of~(\ref{FSvsWrap}) 
up to five loops.%
\footnote{As a side remark, we notice the absence of $O(\log^{0}{\bar{\eta}})$ terms in both $P^{\textrm{wrap}}_4$ and $P^{\textrm{wrap}}_5$. It the 
latter observation extends to all loops, and if the equation~(\ref{FSvsWrap}) is also correct to all loops, it would turn into a non-trivial equality between 
$E^{\rm FS}$ and $E^{\textrm{alt}}$ for this kind of corrections.} We stress that the equality holds for $\sigma = 1$ only, indicating a discrepancy between 
our approach and the standard one~\cite{BJL} with regards to the extension to the (unphysical) odd values of the spin. It is interesting to note 
that this discrepancy does not seem to affect those contributions accompanying the maximal transcendentality at a given loop order and power of 
$\log{\bar{\eta}}$, since these are precisely the terms independent of $\sigma$ as can be verified by a simple inspection of Eq.\ (\ref{WrapPol}).

As demonstrated in this section, we succeeded in reproducing the exact twist-two energy at large spin, including the wrapping corrections not captured 
by the ABA equations. The five-loop agreement demonstrated above gives strong support to the correctness of our proposal for the L\"uscher formula
(\ref{Luscher}) and to the accompanying substitution rule~(\ref{SubtRule}). Though our result only captures the leading large-spin asymptotics, it provides 
us with a transparent physical understanding of the structure of the large spin expansion beyond wrapping order. This is of course a consequence of the
fact that all one needs in the description is the solution to the cusp equation and its siblings~\cite{FZ09} controlling the subleading $\sim \log^{0}{\bar{\eta}}$ 
contribution to the large-spin distribution of the Bethe roots. Both of these are easily constructed at weak coupling. One can also analyze these at strong 
coupling, and, in the next subsection, we will use them to suggest another check of the validity of the L\"uscher formula~(\ref{Luscher}). 

\subsubsection{Remarks on the strong coupling limit}

In this subsection, we will consider the L\"uscher formula~(\ref{Luscher}) at strong coupling. Its evaluation is slightly more demanding than at weak coupling, 
because each excitation develops several scaling regimes at strong coupling. Namely, the full kinematical range (assuming the reality of the momentum $p$) 
typically splits in various domains that chime in with those found for magnons at strong coupling, e.g., one can find a perturbative regime with $E, p\sim g^0$ 
which is typically probed by the perturbative string methods~\cite{FT02} and a semiclassical one with $E, p \sim g$ that can be analyzed using finite-gap
theory~\cite{DL10} or strong-coupling methods of Ref.\ \cite{B10}. 

For the present analysis, the domain of interest is the one that minimizes the energy. This is easily understood by noticing that at large spin each of the individual 
integrals in (\ref{Luscher}) can be evaluated by means of the saddle point approximation. The dominant contribution therefore comes from the regime where 
$E_{\star}(p)$ is minimal, i.e., with $p$ around $p=0$ such that $E_{\star}(p) = m_{\star} + O(p^2)$ with $m_{\star}$ being the mass of the  $\star$-excitation. Each 
integral in~(\ref{Luscher}) thus scales as
\be\label{EstiFS}
E^{\rm FS}_{\star} \sim 1/\eta^{2m_{\star}} \sim 1/S^{2m_{\star}}\, ,
\ee
up to some (fractional) powers of the length $\sim 2\log{\eta}$.

It follows from the estimate~(\ref{EstiFS}) that the leading contribution to the L\"uscher formula~(\ref{Luscher}) comes from the $6$ scalars. This reflects the 
fact that the scalar mass becomes vanishingly small at strong coupling being exponentially suppressed with $g$ \cite{AM07, FGR, BK08},
\be\label{mass}
m \equiv m_{\textrm{h}} = {2^{1/4}\lambda^{1/8} \over \Gamma(5/4)}\e^{-\sqrt{\lambda}/4}\bigg[1+O(1/\sqrt{\lambda})\bigg]\, .
\ee
with $\sqrt{\lambda} \equiv 4\pi g \gg 1$. This phenomenon was explained in~\cite{AM07} where the 6 scalars were identified at strong coupling with the 
multiplet of massive excitations of the O(6) sigma model, which encodes the low-energy effective theory of excitations on the long GKP string. In distinction, 
the masses of fermions and gauge fields remain of order $\sim g^0$ at strong coupling~\cite{FT02, AM07}, more precisely $m_{\textrm{f}}=1$ and 
$m_{\textrm{gf}}=\sqrt{2}$, yielding corrections to the energy suppressed at least as $1/S^2$ and thus negligible compared to the contribution of holes.

So let us focus on the scalar sector from now on and let us evaluate more accurately the L\"uscher formula, this time at strong coupling. For scalars, the 
domain connected to the point $p=0$ corresponds to the non-perturbative regime where these excitations become relativistic and their dispersion 
relation reads~\cite{BK08, B10}
\be\label{EPm}
E_{\textrm{h}}(u) = m\cosh{\theta}\, , \qquad p_{\textrm{h}}(u) = m\sinh{\theta}\, ,
\ee
where $\theta \equiv \pi u/2$ and the mass was introduced in Eq.\ (\ref{mass}). The result~(\ref{EPm}) is an approximation to the exact dispersion relation 
that holds at strong coupling for $u \sim 1$ up to corrections suppressed by higher powers of $m$. To compute $E^{\rm FS}$ we also need the expression for the scalar $Y$-function in the the mirror kinematics at strong 
coupling.  Actually we need the expression for $\delta E_{{\textrm{h}}}(u)$ since all other ingredients involved in $Y^{\textrm{mirror}}_{\textrm{h}}(u)$ are 
known. By using the method developed in Ref.\ \cite{BK08}, one can construct the latter and then to demonstrate that the complete expression 
reads
\be\label{Ymcosh}
Y_{\textrm{h}}^{\textrm{mirror}}(u) = (-1)^{S}\e^{-mR \cosh{\theta}} = (-1)^{S}\e^{-E_{\textrm{h}}(u) R}\, ,
\ee
where  $R$ is a parameter interpreted as the effective length of the O(6) sigma model. Up to corrections explicitly suppressed by the spin $S$, the effective 
length admits the strong coupling expansion with the dominant term behaving as
\be\label{efflength}
R = 2\log{\left({8\pi S \over \sqrt{\lambda}}\right)} + O(1/\sqrt{\lambda})\, .
\ee
What we immediately conclude from Eqs.\ (\ref{Ymcosh}) and (\ref{efflength}) is that both  $\delta E_{\textrm{h}}(u)$ and $E_{\textrm{h}}(u)$ are 
proportional to $\cosh{\theta}$. We should stress however that this behavior is restricted to this particular kinematical window where the O(6) description 
applies. It is lost once we include subleading non-perturbative corrections $\sim m^3$ which originate from irrelevant deformations of the O(6) 
sigma model.%
\footnote{So only after including these subleading effects do we recover that we are effectively dealing with a non-homogeneous background -- an 
intrinsic virtue of the GKP background otherwise observed in the perturbative string theory description~\cite{FT02, BDFPT}.}

Now, plugging the expressions~(\ref{Ymcosh}) and (\ref{EPm}) into the L\"uscher formula~(\ref{Luscher}), we find
\be\label{FSmcosh}
E^{\rm FS} = -{6m(-1)^{S} \over 2\pi}\int d\theta\,  \cosh{(\theta)}\e^{-m\cosh{(\theta)}R} =  -{6m(-1)^{S} \over \pi}K_{1}(mR) \, ,
\ee
where $K_{1}(z)$ is the modified Bessel function, with the well-known asymptotic behavior $K_{1}(z) \sim \exp{(-z)}$ as $z \to +\infty$. The equality
(\ref{FSmcosh}) holds up to non-perturbative corrections $\sim m^3$ coming from the scalars (and associated to irrelevant deformations of the O(6) model) and up to $\sim 1/S^2$, and higher suppressed, contributions due to the fermions, and gauge fields. Finally and more importantly, we also recall that it only takes into account the first L\"uscher correction, associated to the 
virtual propagation of a single excitation around the cylinder, see Fig.\ \ref{Cylinder} (c).

Summarizing, we find that the energy is given (for even spins) by
\be\label{Esc}
E = E^{\textrm{bulk}} -{6m \over \pi}K_{1}(mR) + \ldots\, .
\ee
This is the expected result for the energy of the vacuum in the O(6) sigma model defined on a cylinder of circumference $R$ with periodic boundary 
conditions~\cite{BH}. We note that from the low-energy point of view the bulk energy is a non-universal quantity that depends on the embedding 
of the O(6) model into the superstring sigma model. Were the latter model not UV finite,  the bulk energy would be associated to the UV divergent 
contributions to the vacuum energy that typically scale with the length to leading order at large length: $E^{\textrm{bulk}} \sim \Lambda^2R$ with 
$\Lambda$ some UV cut off. In most treatments, this quantity is discarded and the vacuum energy is substracted such that $E_{\textrm{vacuum}} 
= E^{\rm FS}$. This is the case if $E_{\textrm{vacuum}}$ is computed by means of the TBA equations for the O(6) sigma model, for instance, where the 
bulk energy is simply zero as it is the one used in the Bethe-Yang approximation. In our case, $ E^{\textrm{bulk}}$ is given as an expansion in the
inverse powers of $1/\eta$ with expansion coefficients being function of the coupling. Since we discarded in~(\ref{Esc}) corrections suppressed 
stronger than $1/\eta^2$ for $\eta \sim S \gg 1$ only the leading and subleading contributions to $E^{\textrm{bulk}}$ should be kept in (\ref{Esc}).  
One finds for it \cite{AABEK07, FZ09}
\be
E^{\textrm{bulk}} = {\sqrt{\lambda} \over \pi}\log{\left({8\pi S \over \sqrt{\lambda}}\right)} -{\sqrt{\lambda} \over \pi}  + O(\log{S}/S)\, ,
\ee
to leading order at strong coupling, in agreement with the string theory semiclassical result~\cite{GKP02, FT02}.

Since the finite-size energy is always suppressed at large spin, independently of the magnitude of the coupling, weak or large, the first 
two terms in the large spin expansion of the bulk energy, $E^{\textrm{bulk}} = 2A \log{\bar{S}} + B +O(\log{S}/S)$, always provide 
the leading contributions to the total energy $E$ at large spin. Both functions, $A$ and $B$, are well known for any value of the coupling 
\cite{BES,FZ09,AABEK07,BKK07,KSV08,BK09, Benna06, FT02,RTCusp}. The formula~(\ref{Esc}) predicts that the next term in the large spin expansion of $E$ is 
actually controlled by the finite-size energy at strong coupling and scales as $1/S^{2m}$ with $m\sim \exp{(-\pi g)}$. This conflicts at first sight with 
the perturbative string analyses~\cite{SNZ, BDFPT, GRRT} that detect a correction suppressed as $1/\log{S}$ at one-loop level --- a prediction that 
was confirmed using integrability in~\cite{GSSV}. We will come back later to this apparent discrepancy between our result and the one stemming 
from string theory, which as we shall see turns out to have a simple explanation, but first will compare the result~(\ref{Esc}) to the ABA prediction at 
strong coupling.

We have seen that our formula for the finite-size energy correctly reproduces the expectation from the O(6) sigma model. This agreement relies on 
the substitution~(\ref{SubtRule}) to reconstruct the exact energy. To shed some light on the physics behind this substitution let us compare the 
finite-size energy~(\ref{FSmcosh}) with the prediction stemming from the ABA equations. To do so, we replace $E^{\rm FS}$ by $E^{\textrm{alt}}$ 
in~(\ref{GenForbis}) in order to get back to the ABA energy~(\ref{Eaba}) and then evaluate the alternating ABA energy using the representation
(\ref{RepReal}). To compute the integral in~(\ref{RepReal}) at strong coupling we first deform the contour of integration in~(\ref{RepReal}) and derive 
an alternative representation which reads
\be\label{RepMirr}
E^{\textrm{alt}} = {1 \over \pi}\oint dp_{\textrm{h}}(u-\ft{i}{2}) Y^{\textrm{mirror}}_{\textrm{h}}(u-\ft{i}{2})\, .
\ee
Here the contour of integration goes anticlockwise along the cut $u-i/2 \in [-2g, 2g]$ present in both $Y^{\textrm{mirror}}_{\textrm{h}}(u)$ and 
$p_{\textrm{h}}(u)$. The fact that~(\ref{RepReal}) and~(\ref{RepMirr}) are equivalent is quite straighforward once we recall that the mirror 
kinematics is reached (for scalars) by analytically continuing through the cut involved in~(\ref{RepReal}). Putting it differently, one can pass from 
the representation (\ref{RepReal}) to~(\ref{RepMirr}) by going up and down through the cut. Crossing the latter, one finds
\be
Y_{\textrm{h}}(u+\ft{i}{2}) \rightarrow Y^{\textrm{mirror}}_{\textrm{h}}(u-\ft{i}{2})\, , \qquad  E_{\textrm{h}}(u+\ft{i}{2})\rightarrow ip_{\textrm{h}}(u-\ft{i}{2})
\, .
\ee
The extra minus sign generated along the way compensates for the change of the circulation of the integral such that
 both contours (\ref{RepReal}) and~(\ref{RepMirr}) are traversed in the anticlockwise direction.

Before computing (\ref{RepMirr}) at strong coupling, it is interesting to note that the representation~(\ref{RepMirr}) gives a more transparent 
understanding of the large spin structure of $E^{\textrm{alt}}$ at weak coupling. Indeed, up to the choice of the integration contour, the integral
(\ref{RepMirr}) is remarkably similar to the L\"uscher integral for scalar excitations. For this reason their structure are the same and $E^{\textrm{alt}}$ 
and $E^{\rm FS}$ admit similar expansion at weak coupling~(\ref{EaltPol}), (\ref{FSEN}), though the expansion coefficients obviously differ. It is in 
particular quite straightforward to derive a close formula for the leading logs in the weak coupling expansion of $E^{\textrm{alt}}$. It follows from the 
same kind of approximation as for the L\"uscher formula, namely, by performing the substituting
\be
Y^{\textrm{mirror}}_{\textrm{h}}(u) 
\rightarrow 
(-1)^{S}{\pi^2 g^4 \over \cosh^2{(\pi u)}}\exp{\bigg[-4g^2(\psi(\ft{1}{2}+iu) + \psi(\ft{1}{2}-iu)-2\psi(1))\log{\bar{\eta}}\bigg]}\, ,
\ee 
into~(\ref{RepMirr}) together with $p_{\textrm{h}}(u) \rightarrow 2u$.

Getting back to the computation of~(\ref{RepMirr}) in the strong coupling regime, we now deform the contour of integration in~(\ref{RepMirr}) in such 
a manner that the integral splits into two,
\be\label{RepMirr2}
E^{\textrm{alt}} = -{1 \over \pi}\int dp_{\textrm{h}}(u) Y^{\textrm{mirror}}_{\textrm{h}}(u) + {1 \over \pi}\int dp_{\textrm{h}}(u-i) Y^{\textrm{mirror}}_{\textrm{h}}(u-i)\, ,
\ee
where in both cases $u$ runs from $-\infty$ to $+\infty$. Since the right-hand side of~(\ref{RepMirr2}) originates from the contour integral in~(\ref{RepMirr}) 
it  should be clear that the shift $u\rightarrow u-i$ appearing in the second term in~(\ref{RepMirr2}) is performed outside of the cut $[-i/2-2g,-i/2+2g]$ where 
both functions $p_{\textrm{h}}(u)$ and $Y^{\textrm{mirror}}_{\textrm{h}}(u)$ possess discontinuities. This shift is therefore different from the one considered 
previously when the transition from the real to mirror kinematics (for scalars) was discussed. In particular, $p_{\textrm{h}}(u-i) \neq -iE_{\textrm{h}}(u)$ and 
$Y^{\textrm{mirror}}_{\textrm{h}}(u-i) \neq Y_{\textrm{h}}(u)$.%
\footnote{This is easily visualized in Fig.~\ref{ScalarDWR} where the cut separating real and mirror kinematics is drawn outward and where the expressions 
involved in the second integrals in~(\ref{RepMirr2}) are obtained on the second sheet of the Riemann surface, which is reached from the upper-half plane by passing below the cut. This obviously does not correspond to getting to the real line that lies on the same sheet as the mirror one on Fig~\ref{ScalarDWR}.} These remarks are 
important because they clarify why the two integrals in~(\ref{RepMirr2}) have a radically different scaling at strong coupling, the second one being much 
more suppressed as a function of the spin.%
\footnote{This contrasts with the weak coupling regime where the two integrals yield comparable contributions.}

To estimate the two integrals in~(\ref{RepMirr2}) we can use the saddle point approximation as before. The first integral is actually identical up to a 
numerical factor to the scalar contribution to the finite-size energy. We already know therefore that it scales as $1/\eta^{2m} \sim 1/S^{2m}$ at large spin 
with $m$ given in~(\ref{mass}). For the second integral we use the fact that the scaling at large spin is controlled by $E_{\textrm{h}}(u-i)$ as $u \to 0$ 
and apply the recurrence relation%
\footnote{With the shift $u\rightarrow u-i$ performed outside of the strip $-2g < \Re \textrm{e}(u) < 2g$.}
\be\label{RecRel}
E_{\textrm{h}}(u-i) = E_{\textrm{gf}}(u) - i(p_{\textrm{sf}}(u+\ft{i}{2})+p_{\textrm{sf}}(u-\ft{i}{2}))\, ,
\ee
which is easily derived from the representation obtained in~\cite{B10}. Note that the second term in~(\ref{RecRel}) is purely imaginary since the 
momentum of a small fermion is odd in $u$, i.e., $p_{\textrm{sf}}(-u) = -p_{\textrm{sf}}(u)$, while the first term is real and even. Now since 
$E_{\textrm{h}}(-i) = E_{\textrm{gf}}(0) = m_{\textrm{gf}}$ where $m_{\textrm{gf}}$ is the mass of the gauge field and because $m_{\textrm{gf}} = \sqrt{2}$ 
to leading order at strong coupling, we deduce that the second integral in~(\ref{RepMirr2}) is suppressed as $1/\eta^{2\sqrt{2}} \sim 1/S^{2\sqrt{2}}$ at 
large spin, which is thus negligible compared to the first contribution.

Putting all pieces together, we conclude that the alternating ABA energy is given at strong coupling by
\be\label{Ealtmcosh}
E^{\textrm{alt}} = -{2(-1)^S \over \pi} K_{1}(mR) + \ldots\, .
\ee
Comparing this result with the finite-size energy~(\ref{FSmcosh}), we observe that the ABA prediction does not properly account for the right number 
of scalar excitations. The ABA energy seems to detect only $2$ out of $6$ scalar excitations of the GKP string. This might not be totally surprising when we remember 
that the ABA equations were designed for the BMN vacuum which singles out $2$ scalars out of the $6$ and breaks the O(6) symmetry down to O(4).%
\footnote{These two scalars are the fields $Z$ and $\bar{Z}$, which are the complex conjugate of one another, with $Z$ being the building block of the 
BMN vacuum $\textrm{Tr} \, Z^L$ and $\bar{Z}$ being a mass $2$ excitations typically unstable in the latter background.} To some extent, the 
result~(\ref{Ealtmcosh}) and its interpretation gives more credit to our substitution rule~(\ref{SubtRule}). Stepping away from the $g \to \infty$ limit, i.e, at 
lower value of the 't Hooft coupling, one has to include fermions and gauge fields into the game, so tha the replacement~(\ref{SubtRule}) can be 
interpreted as restoring the proper account of fluctuations on top of the GKP string.

Finally, let us come back to the difficulties previously encountered when comparing~(\ref{Esc}) with the perturbative string theory computations
\cite{SNZ, BDFPT, GRRT}. It is clear that the formula we derived in~(\ref{Esc}) for the twist-two energy at strong coupling is essentially non-perturbative 
and thus cannot be compared directly against any perturbative string predictions. The point is that our result~(\ref{Esc}) holds at strong coupling iff $mR \gg 1$ 
while the string perturbative analysis always assumes that $mR \ll 1$. The interpolation between these two regimes cannot be achieved by accounting only 
for the first L\"uscher correction alone: we recall indeed that the (complete) finite-size energy should include also higher L\"uscher corrections, which 
incorporate the multi-particle exchanges of virtual scalars, in particular. These should scale as $1/S^{2nm}$ (with $n > 1$) at large spin and they are all 
therefore of the same order of magnitude as the leading L\"uscher correction~(\ref{FSmcosh}) provided $mR \sim 2m\log{S}$ is not big enough. This can 
only happen at strong coupling when $m$ becomes arbitrarily small since otherwise $mR \sim R \sim 2\log{S}$ which is automatically large. As our 
consideration originated at weak coupling where $m\sim 1$, it was quite natural to neglect those higher L\"uscher corrections because they enter at order 
$1/S^4$ and higher, and are hence much more difficult to analyze in our framework. What we observe here is that this hierarchy gets lost when the coupling 
gets strong enough due to the quasi-masslessness of the scalar excitations. On the other hand, on the string theory side, the perturbative analysis always 
proceed by expanding in strong coupling prior to taking the large spin limit. It means that the strong inequality $mR \gg 1$ cannot be met since $m$ is exactly 
zero to any order in the strong coupling expansion.%
\footnote{For the string theory calculations $mR = 0$ which we interpret as saying that the perturbative analysis is valid for $mR\ll 1$.} All of what we have just 
said is in accord with standard observations once we adopt the low-energy perspective and think in terms of the O(6) sigma model. Then, the first L\"uscher 
correction is good approximation in the massive regime corresponding to $mR \gg 1$, while the perturbative analysis applies in the UV regime associated with 
$mR \ll 1$, that is, when the Compton wavelength of the massive excitations can be regarded as infinite compared to the system size. In the latter `phase' the 
correlation length is then infinite and the finite-size corrections exhibits the scaling $\sim 1/R \sim 1/\log{S}$ typical of massless theory, in agreement with the 
stringy perturbative predictions~\cite{SNZ, BDFPT, GRRT, GSSV}. The details of the transition require the full-fledged TBA equations for the O(6) sigma model, 
analyzed in~\cite{BH}, but some of its consequences on the structure of the large-spin expansion on the stringy side can be derived by renormalization group 
arguments only. These were already given at the end of the introduction, and we refer the reader there instead of repeating them here.

These last remarks conclude our analysis of the subleading large-spin corrections to the twist-two anomalous dimension. As we have established,  
the weak-to-strong transition is to a large extent due to the non-trivial dynamics in the scalar sector, which is responsible for the change of scale 
$1/S^2 \rightarrow 1/\log{S}$ as one flows between the two extremes, i.e., the weak- and strong-coupling regimes, characterized by $g^2 \ll 1/\log{S}$ 
and $g \gg \ft{1}{\pi}\log{\log{S}}$, respectively.

\section{Acknowledgements}

We are most grateful to Gregory Korchemsky for interesting discussions and valuable comments. One of us (B.B.) would like to acknowledge  illuminating discussions 
with Patrick Dorey, S\'ebastien Leurent, and Vladimir Kazakov about finite-size corrections, and to Didina Serban and Dmytro Volin, for stimulating discussions. 
This work (A.B.) was supported by the National Science Foundation under Grant No.\ PHY-0757394.

\appendix

\section{Comment on singular part of the function $t(u)$}\label{s-coefficients}

\begin{figure}[t]
\begin{center}
\mbox{
\begin{picture}(0,200)(170,0)
\put(0,0){\insertfig{7}{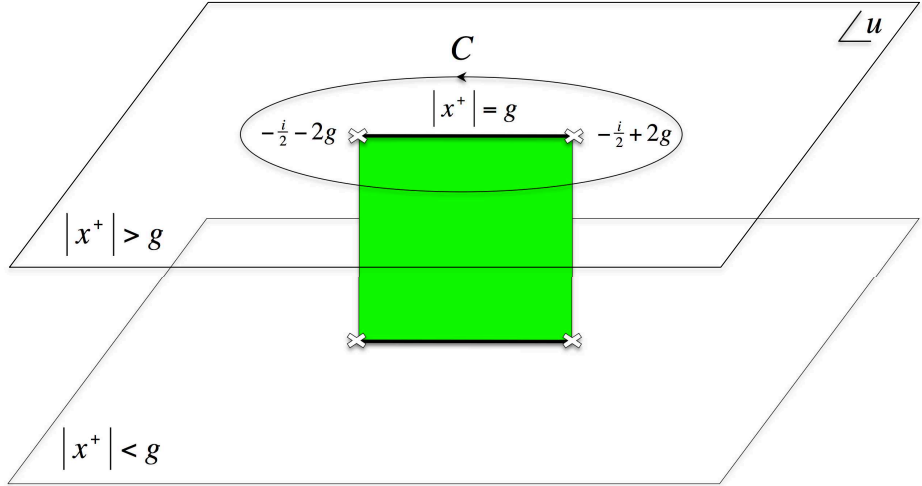}}
\end{picture}
}
\end{center}
\caption{ \label{Riemann} The sketch of the complex $u$-plane for the function $f(u^+)$. The crosses denote the square-root branch points
with a cut running between them.}
\end{figure}

In this appendix, we give more details on the derivation of the expression~(\ref{sn}) for the non-polynomial contribution, $s(u)$, to the function $t(u)$ appearing in the right-hand side of the Baxter equation~(\ref{LRBaxterEq}). We discuss in particular the choice of the integration contour used in the integrals entering Eq.~\re{sn}.

To start with, let us recall that the function $s(u)$ is given in~(\ref{Sintu}) as a series in inverse powers of the shifted Zhukowski rapidities $x^{\pm} \equiv x(u\pm i/2)$, which we reproduce here for convenience,
\be\label{suApp}
s(u) = 2\sum_{n\ge 1}s_{n} \bigg({ig \over x^{+}}\bigg)^{n} + 2\sum_{n\ge 1}\bar{s}_{n} \bigg({g \over ix^{-}}\bigg)^{n}\, .
\ee
The expansion coefficients $s_{n}, \bar{s}_{n},$ are nonzero due to singularities of the dressing factors $\Delta_{\pm}(u)$ in the $u$ plane. Indeed, when regarded as a perturbative expansion at weak coupling, these factors develop poles at $u=0$ of increasing order with each power of the 't Hooft parameter $g^2$. This is the origin for the introduction of the $s(u)$ component of the transfer matrix to compensate 
non-polynomial effects of both sides of the long-range Baxter equation. Therefore, the coefficients $s_{n}, \bar{s}_{n}$ 
are chosen in a manner for the sum \re{suApp} defining $s(u)$ to be a collection of the polar parts in the Laurent expansion 
around $u = \pm \ft{i}{2}$ of the left-hand side of the Baxter equation \re{LRBaxterEq}. Let us see it in more details.

Consider a function $f$ that admits a Laurent expansion around $u^{+} \equiv u+i/2 = 0$
\be
f(u^{+}) = \sum_{n\ge 1}{f_{n} \over ( u^+)^n} + \textrm{regular}\, .
\ee
In the particular case under study, we are interested in resumming the polar parts of the Baxter equation~\re{LRBaxterEq} around 
$u= \pm \ft{i}{2}$. We shall treat the analysis around $u=-i/2$ as a specific example. Then $f$ can be chosen as
\be
f(u^{+}) = \Delta_{+}(u^{+}){Q(u^{+}+\ft{i}{2}) \over Q(u^{+}-\ft{i}{2})}\, ,
\ee
as this is the only source of singularities around $u=- \ft{i}{2}$ on the left-hand side of the Baxter equation.%
\footnote{For convenience we divide both sides of the Baxter equation~(\ref{LRBaxterEq}) by $Q(u)$. We also assume that all the roots of the latter polynomial are real, which is the case in the $\mathfrak{sl}(2)$ sector we are considering here.} For self-consistency of the Baxter equation, the singular part of $f(u^+)$ ought to be exactly counterbalanced by $s(u)$ on its right-hand side, implying that $s(u)$ should contain the sum
\be\label{hatsu}
\hat{s}(u) \equiv f(u^+)\big|_{\textrm{sing}} = \sum_{n\ge 1}{f_{n} \over (u^+)^n}\, .
\ee
Analogous contribution comes from singularities at $u=\ft{i}{2}$, with obvious substitutions in all formulas, and $s(u)$ is the sum of the former and latter. To get to the 
expression~\ref{sn}, we need to re-expand $\hat{s}(u)$ in the basis of inverse powers of $x^+$ 
rather than $u^+$. To perform this step let us first observe that
\be\label{fn}
f_{n} = {1\over 2i\pi}\oint {dv \over v} v^{n} f(v)\, ,
\ee
where the integration contour encloses $v=0$ in the counterclockwise direction. Then, after plugging~(\ref{fn}) into~(\ref{hatsu}) and interchanging summation
and integration, we find
\be
\hat{s}(u) = {1\over 2i\pi}\oint {dv \over u^+ - v} f(v)\, .
\ee
Next, we proceed to the rapidites $x^+$ and $y$ via the Zhukowski maps
\be
x^+ + g^2/x^+ = u^+ \, , \qquad y + g^2/y = v\, ,
\ee
yieding
\be
\hat{s}(u) = {1\over 2i\pi}\oint {dy \over (x^+-y)(1-g^2/x^+y)}(1-g^2/y^2) f(v)\, . 
\ee
Its expansion in inverse powers of $x^{+}$ is straightforward and reads
\be\label{yInt}
\hat{s}(u) 
= 
{1\over 2i\pi}\sum_{n\ge 1}\left({1\over x^+}\right)^{n}\oint_C {dy \over y}\left[y^{n}-\left({g^2 \over y}\right)^n\right] f(v)\, . 
\ee
This is the sought expression, in agreement with the result quoted in~(\ref{sn}) for the coefficient $s_{n}$.%
\footnote{A similar analysis around $u=i/2$ would provide the representation for the coefficient $\bar{s}_{n}$ given in~(\ref{sn}).}

We note that the contour of integration should be defined such that it encloses all possible 
singularities of $f(u^+)$ emanating from $u=-\ft{i}{2}$ as the coupling increases. The poles condensing at this point transform into a 
square-root cut $[-\ft{i}{2} - 2g, -\ft{i}{2} + 2g]$ associated with the Zhukowski map for finite 't Hooft coupling. Thus the
integration path should encircle this cut in the rapidity plane, see Fig.\ \ref{Riemann}. It means that the contour $C$ in the $y$ integral~(\ref{yInt}) should encompass the disk $|y|\leqslant g$. Finally, there are potentially extra singularities stemming from the series $\sim \sum_{n\ge 1}\gamma_{n}(g/x^+)^n$ defining the dressing factors $\Delta_+(u^+)$. In the weak coupling expansion, these singularities are confined to the second sheet $|x^+| < g$. But after resummation, some singularities, like extra poles, can migrate to the first sheet and should be encircled by $C$ in the integral~(\ref{yInt}). This does not happen however in the 
case we are discussing%
\footnote{In our case, indeed, these singularities appear in the form of an infinite sequence of cuts in the domain $|x^+| < g$. They occupy the same position in the associated $u$ plane at any coupling and thus never get to the sheet with $|x^+| > g$. These conclusion can be easily drawn from the analysis of~\cite{KSV08}.} such that the contour of integration $C$ in~(\ref{yInt}) can be chosen to be the circle $|y| = g$, for instance, at all time.

\section{Twist-one excitations in ABA}\label{TO}

\label{ExcitationsABA}

\begin{figure}[t]
\begin{center}
\mbox{
\begin{picture}(0,75)(230,0)
\put(0,0){\insertfig{2.7}{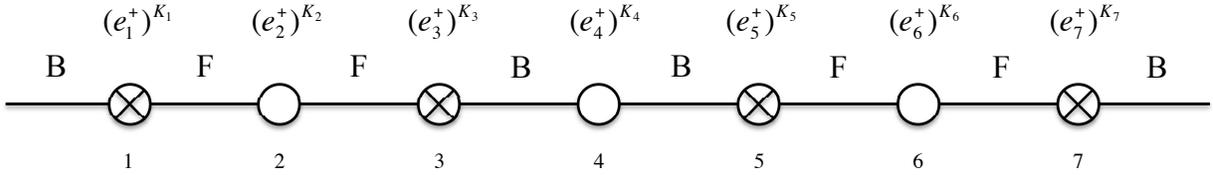}}
\end{picture}
}
\end{center}
\caption{ \label{Dynkin} Dynkin diagram for $\mathfrak{psu} (2,2|4)$ spin chain in the BFFBBFFB grading with excitation numbers $K_j$
generated by the step-up operators $e^+_j$ \re{StepUp} acting on its nodes.}
\end{figure}

Let us briefly recall the identification of elementary twist-one excitations emerging in the consideration of degrees of freedom 
propagating on the long-range spin chain. The latter can be encoded in the Dynkin diagram displayed in Fig.\ \ref{Dynkin}. Each node is 
associated with $K_j$ Bethe roots. These are excited by the acting $K_j$ times, respectively, with step-up operators $e^+_j$ of the 
Serre-Chevalley basis of the superconformal algebra on the vacuum state $\ket{\Omega}$. The ground state  $\ket{\Omega}$ of 
the magnet corresponds to the single trace operator of the $\mathcal{N}=4$ YM theory built from the scalar fields $Z = \bar\phi_{43}$,
\be
\ket{\Omega} = {\tr} \, Z^L
\, .
\ee
So all excitations propagate on this BMN background. The algebra of superconformal generators and their action on the vacuum state
can be very concisely encoded by an oscillator formalism of Ref.\ \cite{GunSac82} (see also Refs.\ \cite{BS05} and, in particular, Ref.\ 
\cite{Bel08} for conventions adopted here\footnote{Compared to Ref.\ \cite{Bel08} we set Bethe-root numbers $n_j$ to $K_j$ as in 
\cite{BS05}.} and more details) such that the step-up operators $e^+$ acting on the nodes of the Dynkin diagram read
\begin{align}
\label{StepUp}
e^+_1 = c_1 a_2^\dagger \, , \qquad  & e^+_2 = c^{1 \dagger}  c_2 \, , \qquad e^+_3 = c^{2 \dagger}  a^1 \, ,
\nonumber
\\
&
e^+_4 = a_1^\dagger b_{\dot 1}^\dagger \, , \qquad 
\\
e^+_5 = - c_3 b^{\dot 1} \, , \qquad & e^+_6 = c^{3 \dagger}  c_4 \, , \qquad  e^+_7 = c^{4 \dagger}  b_{\dot 2}^\dagger
\, ,
\nonumber
\end{align}
in terms of bosonic $(a^\alpha, a^\dagger_\alpha, b^{\dot\beta}, b^\dagger_{\dot\beta})$ and fermionic $(c_A, c^{\dagger A})$ creation 
and annihilation operators corresponding to $\mathfrak{sl} (2)\otimes \mathfrak{sl} (2)$ and $\mathfrak{su} (4)$ subalbegras of 
$\mathfrak{psu} (2,2|4)$. Then a generic state of the spin chain reads
\be
\ket{ \mathcal{O}_{K_1, \dots , K_7} }
=
\left( \prod_{j = 1,2,3}^{\curvearrowleft} (e^+_{4 + j})^{K_{4 + j}} (e^+_{4 - j})^{K_{4 - j}} \right)
(e^+_4)^{K_4} \ket{ \Omega }
\ee
where the vacuum state is 
\be
| \Omega \rangle = (c^{3 \dagger} c^{4 \dagger})^L | 0 \rangle 
\, .
\ee
The quantum numbers of the resulting state can be easily determined by finding the eigenvalues of Cartan generators $h_p$ on them
making use of the conventional commutation relations $[h_p, e^\pm_q] = \pm A_{pq} e^\pm_q$ with the Cartan matrix $A_{pq}$.

The hole excitations were already introduced in the main text. From the point of view of the Dynkin diagram the holes arise as ``duals'' to 
Bethe roots of the central 4-th node with all other excitation numbers set to zero $K_j = 0$, $j \neq 4$. This ABA corresponds to a long-range 
$\mathfrak{sl} (2)$ spin chain with $K_4 = S$ magnons, represented by the light-cone covariant derivatives $D_+ \sim D_{1 \dot{1}}$, 
propagating on it,
\be
(e_4^+ )^{K_4} \ket{\Omega} = \tr D_+^S Z^L
\, .
\ee
The resulting ABA equations were given in the main text and we quote them again below,
\be
- 1 = \frac{\Delta_- (u^-_{4,k})}{\Delta_+ (u^+_{4,k})} \frac{Q_4 (u_{4,k} - i)}{Q_4 (u_{4,k} + i)}
\, .
\ee
Here we used the notation $u_{4,k}$ for the Bethe roots of the central node encoded in the Baxter polynomial
$Q_4 (u) = \prod_{j=1}^{K_4} (u - u_{4,j})$. This polynomial and Bethe roots were stripped from the subscript $4$ in
the body of the paper since they were redundant. 

Below we demonstrate how the remaining twist-one excitations emerge from the all-loop ABA equations. To this end 
we have to know the coupling of nested Bethe roots $u_{k \neq 4}$ with $u_4$. This is easily deduced from the ABA equation for
the central node \cite{BS05}
\be
- 1 = \frac{\Delta_- (u^-_{4,k})}{\Delta_+ (u^+_{4,k})} \frac{Q_4 (u_{4,k} - i)}{Q_4 (u_{4,k} + i)}
\prod_{j=1}^{K_5} S_{54} (u_{5,j}, u_{4,k}) 
\prod_{j=1}^{K_7} S_{74} (u_{7,j}, u_{4,k}) 
\, .
\ee
where we set all excitation numbers for the left wing of the Dynkin diagram to zero, i.e, $K_1 = K_2 = K_3 = 0$. The reason for this is that all 
twist-one excitations can be extracted from either half of Bethe roots (left or right). In the above equation equation, we introduced the scattering
matrices of nested Bethe roots off the main momentum carrying root $u_4$
\be
S_{54} (u_5, u_4) = \frac{x_5 - x^+_4}{ x_5 - x^-_4}
\, , \qquad
S_{74} (u_7, u_4) = \frac{ 1 - g^2/x_7 x_4^+ }{ 1 - g^2/x_7 x_4^-}
\, .
\ee

\subsection{Fermions}

To find the scattering matrix for propagation of twist-one fermionic excitations on the background of covariant derivatives, we have to 
include one of the two adjacent fermionic nodes, either 3 or 5. Let us focus on the latter. An analysis identical to the one that follows
will yield the same results for the mirror-symmetric node 3 giving other twist-one fermions. The above identification can be easily verified by 
realizing the a fermion in the background of $Z$s corresponds to the following state
\be
e_5^+ e_4^+ \ket{\Omega} 
= [b^{\dot 1} b_{\dot 1}^\dagger] (a_1^\dagger c^{4 \dagger}) (c^{3 \dagger} c^{4 \dagger})^{L-1} \ket{0} = \tr \psi_1^4 Z^{L - 1}
\, ,
\ee
where we ignored the overall diagonal Cartan generator $b^{\dot 1} b_{\dot 1}^\dagger$ acting on the state since it does no affect its quantum 
numbers. Thus if one sets all $K_j = 0$ but for $K_{4,5}$ then one attains the autonomous $\mathfrak{sl} (2|1)$ sector where in addition 
to the light-cone derivatives, the operators contain insertions of gauginos. The corresponding equations found from the nested 
$\mathfrak{psu} (2,2|4)$ Bethe Ansatz then read
\ba
- 1 
&=& 
\frac{\Delta_- (u^-_{4,k})}{\Delta_+ (u^+_{4,k})} \frac{Q_4 (u_{4,k} - i)}{Q_4 (u_{4,k} + i)} 
\prod_{j=1}^{K_5} S_{54} (u_{5,j}, u_{4,k})
\, , \\
1 
&=&  
\prod_{j=1}^{K_4} S_{54} (u_{5,k}, u_{4,j})
\, .
\ea
From the above identification of quantum numbers it is clear that the scattering of fermions off magnons is described $S_{54}$,
\be
S_{{\rm f}4} (u_{5,k}, u_{4,j})
=
S_{54} (u_{5,k}, u_{4,j})
\, .
\ee
Thus we see that the second equation of the above system of ABA equation for $\mathfrak{sl} (2|1)$ is the Bethe-Yang equation for the phase factor 
\be
Y_{\rm f} (u) = - \prod_{k=1}^{K_4} S_{{\rm f} 4} (u, u_{4,k})
\ee
accumulated by the fermion with rapidity $u = u_5$ when it scatters off magnons, see Eq.\ \re{Yf}.

\subsection{Gauge fields and bound states}

Let us finally turn to the gauge field. Again we can consider only half of the Dynkin diagram. Namely, the twist-one gauge field 
$F_{+\perp} \sim F_{11}$ corresponds to the following set of oscillator excitation numbers
\be
K_4 = 2 \, , \qquad K_5 = 2 \, , \qquad K_6 = 1
\, ,
\ee
such that
\be
e_6^+ (e_5^+)^2 (e_4^+)^2 \ket{\Omega} = [ (b^{\dot 1} b_{\dot 1}^\dagger)^2 (c_3 c^{3 \dagger}) ] (a_1^\dagger)^2 
(c^{3 \dagger} c^{4 \dagger})^{L-1} \ket{0}
=
\tr F_{11} Z^{L - 1}
\, .
\ee
So the scattering of the gauge field excitation on the background of $K_4$ derivatives is described by the ABA equations
\ba
- 1 
&=& 
\frac{\Delta_- (u^-_{4,k})}{\Delta_+ (u^+_{4,k})} \frac{Q_4 (u_{4,k} - i)}{Q_4 (u_{4,k} + i)} 
\prod_{j=1}^{K_5} \frac{x_{5,j} - x^+_{4,k}}{x_{5,j} - x^-_{4,k}}
\, , \\
1 
&=&  
\prod_{j=1}^{K_6} \frac{u_{6,j}^+ - u_{5,k}}{u_{6,j}^- - u_{5,k}}
\prod_{j=1}^{K_4} \frac{x_{5,k} - x^+_{4,j}}{x_{5,k} - x^-_{4,j}}
\, , \\
- 1 
&=&  
\prod_{j=1}^{K_5} \frac{u_{6,k}^+ - u_{5,j}}{u_{6,k}^- - u_{5,j}}
\, .
\ea
One can immediately see that the trivial solutions to the last nested equation are
\be
u_{5,1} = u_{6,1} + \ft{i}{2}
\, , \qquad
u_{5,2} = u_{6,1} - \ft{i}{2}
\, .
\ee
These are the stacks describing the gluon field found in the analysis of Ref.\ \cite{FRZ09}. The scattering matrix of gauge fields off magnons
can be found by fusion of elementary scattering matrices $S_{54}$ on stacks
\be
\label{GaugeFieldSmatrix}
S_{{\rm gf}, 4} (u_{\rm gf}, u_{4,j})
=
\prod_{k=1}^{K_5} \frac{x_{5,k} - x^+_{4,j}}{x_{5,k} - x^-_{4,j}}
\, ,
\ee
where we introduced the center-of-mass rapidity  $u_{\rm gf} = u_{6,1}$. A short calculation immediately yields
\be
S_{{\rm gf}, 4} (u_{\rm gf}, u)
=
\frac{u^+ - u_{\rm gf}^-}{u^- - u_{\rm gf}^+}
\frac{1 - g^2/x^- x_{\rm gf}^-}{1 - g^2/x^+ x_{\rm gf}^+}
\frac{1 - g^2/x^- x_{\rm gf}^+}{1 - g^2/x^+ x_{\rm gf}^-}
\, .
\ee

Finally, though the follow-up discussion goes beyond the scope of leading twist excitation, let us address for completeness of the picture to the 
bound states of gauge fields that are given by the operators (again for half of the Dynkin diagram only)
\be
D_\perp^{\ell - 1} F_{+ \perp}
\, .
\ee
They are encoded into the following quantum numbers
\be
K_4 = \ell + 1
\, , \qquad
K_5 = \ell + 1
\, , \qquad
K_6 = \ell 
\, , \qquad
K_7 = \ell - 1
\, ,
\ee
reducing to the previously studied single gauge field for $\ell = 1$. Solving the nested ABA equation one finds the stack structure of the 
nested Bethe roots which describe the above gluon bound states \cite{FRZ09}
\be
u_{5,j} = u_\ell + i \left( j - 1 - \ft12 \ell \right)
\, , \qquad
u_{7,j} = u_\ell + i \left( j - \ft12 \ell \right)
\, ,
\ee
where $u_\ell$ stands for the center-of-mass of the bound state excitation (with $u_{\ell = 1} = u_{\rm gf}$) related to $u_6$ via $u_{6, j} = u_\ell
+ i (j - \ft{1}{2} - \ft{1}{2}\ell )$. Fusing the $S$-matrices of elementary Bethe roots, one finds the one for scattering of gluon bound states off magnons
\be
S_{\ell 4} (u_\ell, u_{4,j})
=
\prod_{k=1}^{\ell + 1} \frac{x_{5,k} - x^+_{4,j}}{x_{5,k} - x^-_{4,j}}
\prod_{k=1}^{\ell - 1} \frac{ 1 - g^2/x_{7,k} x_{4,j}^+ }{ 1 - g^2/x_{7,k} x_{4,j}^-}
\, .
\ee
Computing the latter yields the result quoted in Eq.\ \re{YellGen} of the main body of the paper,
\be
S_{\ell 4} (u, u_{4,j})
=
{x^{[- \ell]}-x^{+}_{4, j}\over x^{[+ \ell]}-x^{-}_{4,j}}{1-g^2/x^{[- \ell]}x^{-}_{4, j} \over 1-g^2/x^{[+ \ell]}x^{+}_{4, j}}
\, ,
\ee
where $x^{[\pm \ell]} \equiv x (u \pm \ft{i}{2} \ell)$.

\section{Mirror $Y$-functions at weak coupling}\label{MYWK}

In this appendix, we compile various results concerning the amplitude of propagation of excitations in the vacuum 
in the mirror kinematics. As we advocated in the main text, the latter admits the following representation
\be\label{YMApp}
Y^{\textrm{mirror}}_{\star}(u) = (-1)^{S}\exp{\bigg[-2E_{\star}(u)\log{\bar{\eta}}-2\delta E_{\star}(u) + O(1/\eta^2)\bigg]}
\, ,
\ee
where $S$ stands for the spin, $E_{\star}(u)$ is the energy of the excitation with the $\star$-flavor, and $\bar{\eta} = 
S  \e^{\gamma_{\rm E}}$ at leading order in the large spin. Both $E_{\star}(u)$ and $\delta E_{\star}(u)$ are solely 
functions of the coupling $g$ and rapidity $u$, and, in particular, they are spin-independent. Our goal is to derive explicit
weak coupling expressions for~(\ref{YMApp}) for all twist-one excitations, with, in particular, the fermion in the large 
rapidity domain ($\star=$ lf). The energies $E_{\star}(u)$, with $\star$ = h, lf, gf, are known and their weak coupling 
expansion was computed in Refs.\ \cite{B10,AGMSV10}. Thus, we have to complement this data with power-law series
expansion in 't Hooft coupling $g^2$ for the corresponding $\delta E_{\star}(u)$ making use the representations
determined in Eqs.\ (\ref{Mh}, \ref{Mgf}, \ref{Mlf}). We shall present our results, however, in a form that is different from~(\ref{YMApp}) but more convenient for the evaluation of the L\"uscher formula at weak coupling. It would not be difficult to extract from them the two components $E_{\star}(u)$ and $\delta E_{\star}(u)$ if needed. Note also that to the accuracy in the large-spin expansion, the $O(1/\eta^2)$ 
corrections in~(\ref{YMApp}) are irrelevant for our current analysis.

In order to avoid redundancy in the representation of the $Y$-functions for the various twist-two excitations, it is
instructive to cast them in the following factorized form
\be\label{Yfy}
Y^{\textrm{mirror}}_{\star}(u) = (-1)^{S}f_{\star}(u)y^{\textrm{mirror}}_{\star}(u)\, ,
\ee
where $f_{\star}(u)$ is a $i$-periodic function of $u$, while $y^{\textrm{mirror}}_{\star}(u)$ is a reduced $Y$-functions, 
henceforth called the essential factor for the reason that once it is set to one for all excitations, the L\"uscher formula 
\re{Luscher} vanishes identically. By definition, the periodic factor satisfies the functional relation
\be\label{fPer}
f_{\star}(u+i) = f_{\star}(u)\, ,
\ee
for any of the twist-one excitations. Here the shift $u\rightarrow u+i$ is performed outside of the strip $-2g < \Re{\rm e} 
(u) < 2g$, where the function $f_{\star}(u)$ happens to have branch-point singularities. One of the reasons to factor out 
the latter quantity in $Y$-functions (\ref{Yfy}) is that it possesses the following interesting property
\be\label{RecPer}
f_{\textrm{h}}(u) = f_{\textrm{gf}}(u) = -f_{\textrm{lf}}(u\pm \ft{i}{2})\, .
\ee
Here, we emphasize again, the fermionic expression is obtained for the large fermion with $|x| > g$. The identity
(\ref{RecPer}), which can be immediately verified to hold true for leading contribution at weak coupling making
use of the explicit results
\be
f_{\textrm{h}}(u) 
= 
f_{\textrm{gf}}(u) = {\pi^2 g^4 \over \eta^2\cosh^2{(\pi u)}}(1+O(g^2))
\, , \qquad 
f_{\textrm{lf}}(u) = {\pi^2 g^4 \over \eta^2\sinh^2{(\pi u)}}(1+O(g^2))\, ,
\ee
is responsible for numerous cancellations between bosonic and fermionic contributions in the L\"uscher formula.
When evaluating L\"uscher corrections up to five loops ($=O(g^{10})$) the established conspiracy between different 
flavors makes it sufficient to know $f_{\star}(u)$ one order lower, i.e., up to $O(g^8)$. This property also relies on the 
fact that $y^{\textrm{mirror}}_{\star}(u) = 1+O(g^2)$ at weak coupling, as we shall see momentarily. Moreover, thanks 
to the relations~(\ref{RecPer}), it is enough to compute the case of the scalar excitation only. On the other hand, the 
factor $y^{\textrm{mirror}}_{\star}(u)$ has to be determined independently for all twist-one excitations anew up to 
$O(g^{6})$. Explicit derivations and final expressions both for $y^{\textrm{mirror}}_{\star}(u)$ and $f_{\textrm{h}}(u)$ 
can be found below. Finally, we shall also include, at the end of this appendix, the weak coupling expansion for the 
momenta of various excitations, since they enter the L\"uscher formula on equal footings with the $Y$-factors.

\subsection{Periodic factor}

Let us start with the calculation of the periodic factor shared by all $Y$-functions. As we stressed above, we can restrict 
our consideration to the case of the hole, since the gauge field and (large) fermion excitations follow from Eq.\ 
(\ref{RecPer}). We split the complete expression~(\ref{Mh}) as $\mathcal{E}_{\rm h} = -\log f_{\rm h} - \log 
y^{\rm mirror}_{\rm h}$ and introduce the first term in the sum as
\be
\begin{aligned}\label{fhfirst}
\log{f_{\textrm{h}}(u)} &= 2\log{\left(g^2/ \eta\right)} - 2\psi(1) + 4\int_{0}^{\infty}{dt \over t}{\left(\cos{(ut)}\e^{t/2}-J_{0}(2gt)\right)J_{0}(2gt)-t/2 \over \e^{t}-1}\\
&  - 2\int_{0}^{\infty}{dt \over t}{\cos{(ut)}\e^{t/2}-J_{0}(2gt) \over \e^{t}-1}\gamma_{+}(2gt)- 2\int_{0}^{\infty}{dt \over t}{\gamma_{-}(2gt)J_{0}(2gt)-2g\gamma_{1}t \over \e^{t}-1} \, .
\end{aligned}
\ee
It can be equivalently rewritten as
\be\label{fh}
\log{f_{\textrm{h}}(u)} = 2\log{\left({\pi g^2 \over \eta \cosh{(\pi u)}}\right)} + 4\int_{0}^{\infty}{dt \over t}{\cos{(ut)}\e^{t/2} \over \e^{t}-1}\bigg[J_{0}(2gt)-1-\ft{1}{2}\gamma_{+}(2gt) \bigg]  + \alpha\, ,
\ee
where $\alpha$ is constant w.r.t.\ $u$,
\be\label{alpha}
\alpha = -4\int_{0}^{\infty}{dt \over t}{J_{0}^2(2gt)-1 \over \e^{t}-1} + 2\int_{0}^{\infty}{dt \over t}{J_{0}(2gt)\gamma(-2gt)+2g\gamma_{1}t \over \e^{t}-1}\, ,
\ee
and we used the following explicit solution for the integral
\be
2\int_{0}^{\infty}{dt \over t}{\cos{(ut)}\e^{t/2}-1-t/2 \over \e^{t}-1} = \log{\left({\pi \over \cosh{(\pi u)}}\right)} + \psi(1)\, .
\ee
To get the representation \re{alpha} for $\alpha$ we used the relation $\gamma(-2gt) = \gamma_{+}(-2gt)+\gamma_{-}
(-2gt) = \gamma_{+}(2gt) - \gamma_{-}(2gt)$. 

Before we move to the analysis of the weak coupling expansion, let us verify first that it has the periodicity~(\ref{fPer}). 
To demonstrate this, we notice that making use of the decomposition \re{BessDec}, the expression in the square 
brackets in Eq.\ \re{fh}, can be rewritten as a series over the even Bessel functions,
\be\label{Integrand}
J_{0}(2gt)-1-\ft{1}{2}\gamma_{+}(2gt) = -\sum_{n\ge 1}(2+2n\gamma_{2n})J_{2n}(2gt)\, .
\ee
Then we observe that each integral contributing to the series
\be\label{IntPer}
\int_{0}^{\infty}{dt \over t}{\cos{(ut)}\e^{t/2} \over \e^{t}-1}J_{2n}(2gt)\, , \qquad n=1, 2, \ldots\, , 
\ee
is $i$-periodic. This is evident by shifting the argument by half the period $\pm \ft{i}{2}$
\be\label{shifti}
\int_{0}^{\infty}{dt \over t}{\cos{((u\pm\ft{i}{2})t)}\e^{t/2} \over \e^{t}-1}J_{2n}(2gt) =  \int_{0}^{\infty}{dt \over t}{\cos{(ut)} \over \e^{t}-1}J_{2n}(2gt) + {1 \over 2}\int_{0}^{\infty}{dt \over t}\e^{\mp iut}J_{2n}(2gt) \, ,
\ee
and recognizing that the last term
\be
\int_{0}^{\infty}{dt \over t}\e^{\mp iut}J_{2n}(2gt)
\ee
is merely two representations of the same function, namely $(ig/x)^{2n}/(2n)$, for $u$ taken in the lower- and upper-half 
plane, respectively, as was exhibited by Eq.\ \re{BesselRep1overX}. Thus the right-hand sides of the above equations 
coincide independently of whether the shift is positive or negative%
\footnote{Note that there is also an alternative representation for the integral \re{IntPer}
\be
\int_{0}^{\infty}{dt \over t}{\cos{(ut)}\e^{t/2} \over \e^{t}-1}J_{2n}(2gt) 
= (-1)^n\sum_{k\geq 1}{(-1)^k\over 2k}I_{2n}(4\pi kg)\e^{-2\pi k u}\, ,
\ee
with $I_{2n}(z) = (-1)^{n}J_{2n}(iz)$ being the modified Bessel's function. The representation is valid for $\Re{\rm e}(u) > 2g$ 
and obtained by closing the integration contour at $\infty$ and performing integration by the method of residues. The 
expression in the symmetric domain $\Re{\rm e}(u) < -2g$ can be easily obtained by the substitution $u\rightarrow -u$. 
The periodicity then becomes manifest.}. More precisely, the two above integrals are related to 
one another through analytic continuation for $u$ away from the cut $u^2 <(2g)^2$. So the periodicity of~(\ref{IntPer}) is 
guaranteed if the shift $u\rightarrow u+i$ is performed outside of the strip $-2g < \Re{\rm e}(u) < 2g$, where the integrals
(\ref{IntPer}) have an infinite number of cuts running between $u=\pm 2g +in/2$ with $n =\pm 1, \pm 2, \ldots\, $. One may 
worry about the commutativity of operations of summation and integration. However, the summation does not move away 
or change the nature of existing singularities, but it induces extra ones though they are confined to the cuts in the strip.

Let us come back now to the perturbative expansion of~(\ref{fh}). Note first that the second and last terms in the right-hand 
side of~(\ref{fh}) appear explicitly suppressed at weak coupling compared to the leading term. Thus, dominant contribution
for small 't Hooft coupling reads
\be
\log{f_{\textrm{h}}(u)} = 2\log{\left({\pi g^2 \over \eta \cosh{(\pi u)}}\right)} + O(g^2)\, .
\ee
As we previously explained, we only need to know $\log{f_{\textrm{h}}(u)}$ up to order $O(g^4)$, i.e., $f_{\textrm{h}}(u)$ 
up to $O(g^8)$. This result can be easily obtained by Taylor expanding the integrand in~(\ref{Integrand}) and then 
plugging it back into Eq.\ (\ref{fh}). We then find that
\be
f_{\textrm{h}}(u) 
= 
{\pi^2 g^4 \over \eta^2\cosh^2{(\pi u)}}\exp{\left[\alpha + \beta{\pi^2 \over \cosh^2{(\pi u)}} 
+ 
\gamma{\pi^4(\cosh{(2\pi u)} -2) \over \cosh^4{(\pi u)}}  + O(g^6)\right]}\, ,
\ee
with the coefficients being
\be
\beta = -2g^2(1+\gamma_{2})\, , \qquad \gamma = -{g^{4} \over 3}(3+4\gamma_{2}-2\gamma_{4})\, ,
\ee
and where we used the value for the integral
\be
2\int_{0}^{\infty}dt {t\e^{t/2}\cos{(ut)} \over \e^{t}-1} = {\pi^2 \over \cosh^2{(\pi u)}}\, , 
\ee
and its derivatives. Finally, to determine the constant $\alpha$, one can Taylor expand the numerators of each integrand 
in~(\ref{alpha}), and get
\be
\alpha =  
{2\pi^2 \over 3}g^2(2+\gamma_{2}) +4\zeta_{3} g^3(3\gamma_{1}-\gamma_{3}) 
- 
{2\pi^4 \over 45}g^4(9+8\gamma_{2}-\gamma_{4}) + \ldots\, .
\ee
To find the numerical values for coefficients accompanying powers of the 't Hooft coupling for these $u$-independent
constant, we use the known decomposition $\gamma_{n} = \gamma_{n}^{\o} \log{\bar{\eta}}+\delta \gamma_{n}^{\o}$
valid up to $1/\eta^2$-suppressed corrections and substitute the available expressions~(\ref{gammn}, \ref{dgammn}) 
for the coefficients $\gamma_{n}^{\o}$, $\delta \gamma_{n}^{\o}$ to obtain
\be
\alpha = 24\zeta_{3}g^4\log{\bar{\eta}} +{4\pi^2\over 3}g^2-{13\pi^4\over 45}g^4 + O(g^6) \, ,
\ee
and
\be
\beta = -2g^2-{\pi^2 \over 3}g^4 + O(g^6) \, , \qquad \gamma = -g^4  + O(g^6)\, .
\ee

\subsection{Essential factor}

Now we move on to discuss the second ingredient of the $Y$-functions, the essential factor $y^{\textrm{mirror}}$ in
the decomposition \re{Yfy}. The latter reads
\be\label{EssP}
\log{y^{\textrm{mirror}}_{\star}(u)} 
= 
2\int_{0}^{\infty}{dt \over t}{\e^{-(s_{\star}-1)t}\gamma_{-}(2gt)\cos{(ut)}-2g\gamma_{1} t \over \e^{t}-1}\, ,
\ee
for the excitations of $\star$-flavor encoded in the conformal spin $s_{\star}$, i.e., $s_{\star} = 1/2, 1, 3/2, $ for the scalar, 
fermion and gauge field, respectively.

As a consistency check, let us verify that the essential factor~(\ref{EssP}) put together with the periodic one~(\ref{fhfirst}) 
reproduce the complete expression for $Y$-function in the mirror kinematics according to (\ref{Yfy}) with (\ref{RecPer}). 
This is quite straightforward for the hole where the sum of~(\ref{EssP}) and~(\ref{fhfirst}) immediately reproduces the 
total expression~(\ref{Mh}). The same holds for the (large) fermion, where the shift involved in~(\ref{RecPer}) can be 
performed making use of the representation (\ref{shifti}). Finally, for the gauge field,  the matching is recovered with the 
help of the identity
\be
\int_{0}^{\infty}{dt \over t}J_{0}(2gt)\gamma_{+}(2gt) = 0\, ,
\ee
which can be understood as emerging from the orthogonality condition for the Bessel functions
\be
\int_{0}^{\infty}{dt \over t}J_{0}(2gt)J_{2n}(2gt) = 0\, , \qquad  n = 1, 2, \ldots\, ,
\ee
in the Neumann series representation for $\gamma_+ (t)$, i.e., $\gamma_{+}(2gt) = \sum_{n\ge 1}2(2n)\gamma_{2n}
J_{2n}(2gt)$. It is assumed that the commutation of the summation and integration is legitimate due to tamed large-$n$ 
asymptotic behavior of the expansion coefficients $\gamma_{2n}$.

At leading order in weak coupling, the essential part~(\ref{EssP}) is controlled by the one-loop energy of excitations. 
Indeed, this follows from $\gamma_{-}(2gt) \sim 4g^2t\log{\bar{\eta}}$, together with $2g\gamma_{1} \sim 4g^2\log{\bar{\eta}}$, 
which leads to
\be
y^{\textrm{mirror}}_{\star}(u) = \exp{\bigg[-4g^2(\psi(s_{\star} + iu) + \psi(s_{\star} - iu)-2\psi(1))\log{\bar{\eta}} + O(g^4)\bigg]} \, ,
\ee
with $\psi(z)$ the digamma function. It is straightforward to determine further terms in the perturbative expansion. Namely,
with the help of the series representation for $\gamma_{-}(2gt)$ in odd Bessel functions,
\be\label{GmApp}
\gamma_{-}(2gt) = 2\sum_{n\ge 1}(2n-1)\gamma_{2n-1} J_{2n-1}(2gt)\, ,
\ee
and the scaling behavior $J_{2n-1}(2gt) \sim (gt)^{2n-1}$ at weak coupling (with $t$ fixed), one can truncate the Neumann 
series~(\ref{GmApp}) and find the representation
\be\label{GmAppbis}
y^{\textrm{mirror}}_{\star}(u) = \exp{\bigg[a \psi_{0}(s_{\star}, u) + b \psi_{2}(s_{\star}, u) + c \psi_{4}(s_{\star}, u)+ O(g^8)\bigg]}
\, .
\ee
Here we introduced the notation
\be\label{PsiNot}
\psi_{n}(s_{\star}, u) = \psi_{n}(s_{\star}+iu) + \psi_{n}(s_{\star}-iu) -2\psi(1)\delta_{n, 0}\, ,
\ee
with polygamma functions $\psi_{n}(z) = \partial_{z}^{n+1}\log{\Gamma(z)}$. The coefficients $a, b, c, $ in~(\ref{GmAppbis}) 
are given by
\be\label{abc}
a = -2g\gamma_{1}\, , \qquad b = g^3(\gamma_{1}-\gamma_{3})\, , \qquad c = -{g^5 \over 12}(2\gamma_{1}-3\gamma_{3}+\gamma_{5})\, .
\ee
This result is obtained by Taylor expanding~(\ref{GmApp}) up to $O(t^5)$, plugging it back into~(\ref{EssP}), and evaluating 
the resulting integrals with the help of the relation
\be
\label{PsiFourier}
2\int_{0}^{\infty}dt \, {t^{2n}\e^{-(s-1)t}\cos{(ut)}-\delta_{n, 0}t \over \e^{t}-1} = -\psi_{2n}(s, u)\, ,
\ee
with the right-hand side being (\ref{PsiNot}).

Finally, the weak coupling expansion of the coefficient (\ref{abc}) is found by substituting  the expressions obtained before 
for $\gamma_{n} = \gamma_{n}^{\o} \log{\bar{\eta}}+\delta \gamma_{n}^{\o}$, neglecting $O(1/\eta^2)$ corrections. 
Using~(\ref{gammn}, \ref{dgammn}), we get
\be
a = -\left(4g^2-{4\pi^2 \over 3}g^4+{44\pi^4 \over 45}g^6\right)\log{\bar{\eta}} 
+ 
12\zeta_{3}g^4 -\left({8\pi^2 \zeta_{3}\over 3} +80\zeta_{5}\right)g^6 + O(g^8) \, ,
\ee
and
\be
b = \left(2g^4-{2\pi^2 \over 3}g^6\right)\log{\bar{\eta}} -{20\zeta_{3} \over 3}g^6 + O(g^8) 
\, , \qquad 
c = -{g^6 \over 3}\log{\bar{\eta}} + O(g^8)\, ,
\ee
where we also used the fact that $\gamma_{5} \sim g^5$ when evaluating $c$.

\subsection{Momenta}

The final ingredient in the L\"uscher formula \re{Luscher} is the momentum $p_\star$ of the $\star$-excitation.
Analogously to the previous consideration, we can decompose these into periodic and essential components as well, 
\be
p_{\star}(u) =  p_{\star}^{\textrm{per}}(u) + p_{\star}^{\textrm{ess}}(u)  \, ,
\ee
with the periodicity condition $p_{\star}^{\textrm{per}}(u+ i) = p_{\star}^{\textrm{per}}(u)$ for $u$ outside the strip 
$-2g < \Re{\rm e}{(u)} < 2g$. Then, similarly to~(\ref{RecPer}), we have the following relations between the periodic 
components of the twist-one excitations,
\be
p_{\textrm{h}}^{\textrm{per}}(u) = p_{\textrm{gf}}^{\textrm{per}}(u)
\, , \qquad 
p_{\textrm{lf}}^{\textrm{per}}(u) = p_{\textrm{h}}^{\textrm{per}}(u\pm \ft{i}{2})\, .
\ee
As alluded to before, a purely periodic integrand does not contribute to the L\"uscher formula, thus we merely need to 
know $p_{\textrm{h}}^{\textrm{per}}(u)$ up to $O(g^4)$ only. On the other hand, the essential part $p_{\star}^{\textrm{ess}}(u)$ 
is needed up to $O(g^6)$.

The expression for the periodic component of the scalar momentum is given by
\be\label{MomPer}
p_{\textrm{h}}^{\textrm{per}}(u) = -\int_{0}^{\infty}{dt \over t}{\sin{(ut)}\e^{t/2}\gamma^{\o}_{-}(2gt) \over \e^{t}-1}\, .
\ee
Note that it involves the solution to the BES equation only, i.e. $\gamma^{\o}$, as opposed to full function 
$\gamma = \gamma^{\o} \log{\bar{\eta}} + \delta\gamma^{\o}+O(1/\eta^2)$ entering the $Y$-functions, see Eq.~(\ref{EssP}) 
for instance. So this portion of the momentum depends  on the rapidity $u$ and 't Hooft coupling $g$ but not
on the spin of the excitation. The weak coupling expansion of~(\ref{MomPer}) is constructed along the same 
lines as before. Below, we present the result for the derivative of the momentum instead, since this is the quantity 
that enters the L\"uscher formula. The former reads
\be
{dp^{\textrm{per}}_{\textrm{h}}(u) \over du} 
= 
-2\pi^2g^2{1 \over \cosh^2{(\pi u)}}\left(1-{\pi^2g^2 \over 3}\right) -2\pi^4 g^4{\cosh{(2\pi u)}
-
2 \over \cosh^4{(\pi u)}}\left(1-{\pi^2g^2 \over 3}\right) + O(g^6)\, .
\ee

The essential part of the momentum admits the following general expression
\be
p_{\star}^{\textrm{ess}}(u) = 2u - \int_{0}^{\infty}{dt \over t}{\sin{(ut)}\gamma^{\o}_{+}(2gt) \over \e^{t}-1}\e^{-(s_{\star}-1)t}
\, ,
\ee
and explicitly depends on the conformal spin $s_{\star}$ of the  $\star$-excitation. At weak coupling it is simply given by
\be
{dp^{\textrm{ess}}_{\star}(u) \over du} = 2+4\zeta_{3}g^6 \psi_{2}(s_{\star}, u) + O(g^8)\, ,
\ee
with the notation for the combination of the polygamma functions introduced in Eq.\ (\ref{PsiNot}).

\section{L\"uscher formula at weak coupling}
\label{LFWC}

In this appendix, we present details on the computation of the perturbative expansion of the L\"uscher formula by
assembling all results derived in the previous appendices.  Starting from the expression for the $Y$-functions in the 
mirror kinematics given in Appendix~\ref{MYWK}, it is quite straightforward to perform the weak coupling expansion 
of Eq.\ (\ref{Luscher}). In Sect.~\ref{LF}, we found that it admits the following general representation
\be
E^{\rm FS} = {(-1)^{S} \over \eta^2}\sum_{n \ge 2}\mathcal{P}_{n}\, g^{2n} + o(1/\eta^2)\, ,
\ee
where $\mathcal{P}_{n}$ are polynomials of degree $n-2$ in the variable $\log{\bar{\eta}} \equiv \log{\eta} + 
\gamma_{\rm E}$. In the main text we provided explicit expressions for each of the polynomials $\mathcal{P}_{n}$ 
up to five loops, i.e., for $n=2, \ldots, 5$, see Eq.~(\ref{FinalRes}). Below we provide intermediate expression obtained 
with the help of the mirror $Y$-functions listed in Appendix~\ref{MYWK} as well as evaluation of all relevant integrals.

Making use of the results in Appendix~\ref{MYWK}, we find the following expressions for the polynomials $\mathcal{P}_{n}$, 
\be
\begin{aligned}\label{PolIS}
\mathcal{P}_{2} &= 0\, , \\
\mathcal{P}_{3} &=  4 \mathcal{I}^{(1)}\log{\bar{\eta}}\, , \\
\mathcal{P}_{4} &= -12\zeta_{3} \mathcal{I}^{(1)} + \left(-2\mathcal{K}^{(0)}-12\mathcal{J}^{(1)}
+
4\pi^2 \mathcal{I}^{(1)}\right)\log{{\bar{\eta}}}-8 \mathcal{I}^{(2)}\log^2{\bar{\eta}}\, , \\
\mathcal{P}_{5} &= \left({14\zeta_{3}\over 3}\mathcal{K}^{(0)}+36\zeta_{3}\mathcal{J}^{(1)}
+
\left(80\zeta_{5}-{40\pi^2\zeta_{3} \over 3}\right)\mathcal{I}^{(1)}\right) \\
&\, \, +\left({1\over 3}\mathcal{L}+6\mathcal{M}+32\mathcal{N}-8\mathcal{O}
-
2\pi^2\mathcal{K}^{(0)}-12\pi^2\mathcal{J}^{(1)}+{8\pi^4 \over 5}\mathcal{I}^{(1)}
+
48\zeta_{3}\mathcal{I}^{(2)}\right)\log{\bar{\eta}} \\
&\, \, +\left(8\mathcal{K}^{(1)}+24\mathcal{J}^{(2)}+96\zeta_{3}\mathcal{I}^{(1)}
-
{16\pi^2\over 3}\mathcal{I}^{(2)}\right)\log^2{\bar{\eta}} +{32\over 3}\mathcal{I}^{(3)}\log^3{\bar{\eta}}\, ,
\end{aligned}
\ee
in terms of the integrals $\mathcal{I}^{(n)}, \mathcal{J}^{(n)}, \ldots$ presented below. The latters encode contributions both from bosonic and 
fermionic excitations. The representation~(\ref{PolIS}) already accommodates simplifications obtained after discarding all 
terms involving purely periodic integrands, since they cancel out when adding bosons and fermions together. The result 
(\ref{FinalRes}) quoted in Sec.~\ref{LF} follows immediately from Eq.\ (\ref{PolIS}) upon substitutions of explicit 
expressions (\ref{I}, \ref{J}, \ref{K}, \ref{L}, \ref{M}, \ref{N}, \ref{O}) for the integrals $\mathcal{I}^{(n)}, \mathcal{J}^{(n)}, \ldots\, $. 
The general structure of the latters reads
\be\label{GenInt}
\sum_{\star}\int {du} \, M_{\star}(u) \, \psi^{k}_{n}(s_{\star}, u)\, ,
\ee
where summation runs over all twist-one excitations  $\star =$ h, gf, f.%
\footnote{The fermionic integrands arising here are always implied to take the form of the ones in the large rapidity domain. 
Nevertheless, in order to simplify the notation, we label all the former with `f' only,  instead of `lf'.}  Here the measure factor 
$M_{\star}(u)$ is expressed in terms of the hyperbolic functions, e.g.,
\be
M_{\textrm{h}, \textrm{gf}}(u) \sim {1\over \cosh^2{(\pi u)}}\, , \qquad M_{\textrm{f}}(u) \sim {1\over \sinh^2{(\pi u)}}\, ,
\ee
which are suppressed at least as $\exp{(-2\pi |u|)}$ at large rapidity. The combination of polygamma functions
$\psi^{k}_{n}(s_{\star}, u)$ was introduced in Eq.\ (\ref{PsiNot}) where $s_{\star}$ stands for the conformal spin of the 
$\star$-excitation, $s = 1/2, 1, 3/2,$ for the hole, fermion and gauge field, respectively. Finally, all integrals in~(\ref{GenInt}) 
are taken over the entire real axis, with the $\pm i0$ prescription for the integration around $u=0$ in the fermionic integrand.%
\footnote{We recall that the integration contour for the fermion passes above, or equivalently below, the cut 
$u^2 <(2g)^2$. At weak coupling, it implies that one should avoid all the singularities of the type $\sim g^{2n}/u^{2n}$ which 
accumulate at $u\sim 0$.} With the latter specification, all integrals of the type~(\ref{GenInt}) that we are going to consider, 
are (separately) well defined.

Before listing all relevant integrals and their explicit values, let us briefly comment on the procedure of their computation. 
Most of the integrals can be evaluated using the Fourier representation for $\psi_{n}(s_{\star}, u)$ given in~(\ref{PsiFourier}). 
This immediately applies for $k=1$ in~(\ref{GenInt}). However when the integrand displays higher powers of $k$, we did not 
find a better way of computing the integrals but by closing the integration contour at infinity and then performing the resulting
integrals by the method of residues. This procedure typically yields an infinite series representation for them that can be
calculated for low values of $k$. Notice however, that for the type-$\mathcal{I}$ integrals, it is necessary to first use the 
integration by parts in order to make the sum over the residues convergent. This reflects the fact that these integrals involve 
the digamma function $\psi(s+iu)$ which possesses growing asymptotics at large rapidity, $\psi(s+iu) \sim \log{u}$. Finally,
it is interesting to notice the similarity of these integrals to the ones appearing in the analysis of the OPE for light-like Wilson 
loops~\cite{AGMSV10, GMSV, GMSVbis} and thus one could adopt the same technique exploring their motivic structure and finding corresponding
symbols as a systematic framework for their calculation (see~\cite{BLP} for explicit results).

\subsection{Integrals for leading logs}

The only integrals that we need for evaluating the leading log behavior are of the type
\be\label{LLInt}
\begin{aligned}
\mathcal{I}^{(k)}_{\textrm{h}} &= \pi\int du\, {\left(\psi(\ft{1}{2}+iu) + \psi(\ft{1}{2}-iu)-2\psi(1)\right)^k \over \cosh^2{(\pi u)}}
\, , \\
\mathcal{I}^{(k)}_{\textrm{gf}} &= \pi\int du\, {\left(\psi(\ft{3}{2}+iu) + \psi(\ft{3}{2}-iu)-2\psi(1)\right)^k \over \cosh^2{(\pi u)}}
\, , \\
\mathcal{I}^{(k)}_{\textrm{f}} &= \pi\int du\, {\left(\psi(1+iu) + \psi(1-iu)-2\psi(1)\right)^k \over \sinh^2{(\pi u)}}
\, , \\
\end{aligned}
\ee
where $k$ is an positive integer, which runs over $k=0, 1, 2, 3,$ if one aims at computing all the way up to five loops. They 
are obviously related to the integrals~(\ref{Ihgff}), introduced in Sect.~\ref{LF}, by
\be
I^{(k)}_{\star} = {(-4)^{k} \over k!}\mathcal{I}_{\star}^{(k)}\, .
\ee
For the lowest three $k$s, they read
\be\label{FirstI}
\begin{aligned}
&\mathcal{I}^{(0)}_{\textrm{h}} = 2
\, , \, \, \qquad \qquad \qquad  \, \, \, \mathcal{I}^{(0)}_{\textrm{gf}} = 2
\, , \qquad \qquad \qquad \qquad \, \, \, \, \, \, \, \mathcal{I}^{(0)}_{\textrm{f}} = -2 \, ,\\
&\mathcal{I}^{(1)}_{\textrm{h}} = -4
\, , \, \, \qquad \qquad \, \, \, \, \, \, \, \, \, \, \mathcal{I}^{(1)}_{\textrm{gf}} =  -4 + {2\pi^2 \over 3} 
\, , \qquad \qquad \, \, \, \, \, \, \, \, \,  \mathcal{I}^{(1)}_{\textrm{f}} = 4- {\pi^2 \over 3}\, , \\
&\mathcal{I}^{(2)}_{\textrm{h}} =  16-{2\pi^2 \over 3}
\, , \, \, \qquad \, \, \, \, \, \, \mathcal{I}^{(2)}_{\textrm{gf}} = 16 + {2\pi^2 \over 3}-16\zeta_{3}
\, , \qquad \, \, \, \, \mathcal{I}^{(2)}_{\textrm{f}} = -16+{2\pi^2\over 3} + 8\zeta_{3}\, . \\
\end{aligned}
\ee
The $k=3$ case, which is required for the computation of the five-loop expression, is much more involved. As we are only 
interested in a particular weighted sum of different flavors, we circumvented the complication by computing suitable linear 
combinations of the integrals (\ref{LLInt}). For instance, by applying the identity
\be
\psi(\ft{3}{2}+iu) + \psi(\ft{3}{2}-iu) = \psi(\ft{1}{2}+iu) + \psi(\ft{1}{2}-iu) + {1\over u^2+\ft{1}{4}}\, ,
\ee
one can relate gauge-field integrals to scalar ones up to integrals involving lower powers of digamma functions. The latter 
integral can then be taken by the method of residues and yields
\be\label{GFPS}
\mathcal{I}^{(3)}_{\textrm{gf}}-\mathcal{I}^{(3)}_{\textrm{h}} = 4\pi^2 + {7\pi^4 \over 15}-48\zeta_{3} \, .
\ee
Similar manipulations can be done for the fermionic integral. Namely, by deforming the contour of integration, 
$u\rightarrow u-i/2$, such that
\be
{1\over \sinh^2{(\pi (u-\ft{i}{2}))}}  = -{1\over \cosh^2{(\pi u)}}\, , 
\ee
and using furthermore the identity
\be
\psi(1+i(u-\ft{i}{2})) + \psi(1-i(u-\ft{i}{2})) = \psi(\ft{1}{2}+iu) + \psi(\ft{1}{2}-iu) -{i \over u-\ft{i}{2}}\, ,
\ee
one finds along these lines
\be\label{FPS}
\mathcal{I}^{(3)}_{\textrm{f}}+\mathcal{I}^{(3)}_{\textrm{h}} = -{7\pi^4 \over 30}\, .
\ee
It turns out that the two expressions~(\ref{GFPS}) and (\ref{FPS}) are all we need for our purposes. Summarizing
our findings, we can compute the weighted sums immediately from~(\ref{FirstI}), (\ref{GFPS}) and (\ref{FPS}). They 
read
\be\label{I}
\begin{aligned}
&\mathcal{I}^{(0)} \equiv 6\,\mathcal{I}^{(0)}_{\textrm{h}} + 2\,\mathcal{I}^{(0)}_{\textrm{gf}}+8\,\mathcal{I}^{(0)}_{\textrm{f}} 
= 0\, , \\
&\mathcal{I}^{(1)} \equiv 6\,\mathcal{I}^{(1)}_{\textrm{h}} + 2\,\mathcal{I}^{(1)}_{\textrm{gf}}+8\,\mathcal{I}^{(1)}_{\textrm{f}} 
= -{4\pi^2 \over 3}\, , \\
&\mathcal{I}^{(2)} \equiv 6\,\mathcal{I}^{(2)}_{\textrm{h}} + 2\,\mathcal{I}^{(2)}_{\textrm{gf}}+8\,\mathcal{I}^{(2)}_{\textrm{f}} 
= {8\pi^2 \over 3}+32\zeta_{3}\, , \\
&\mathcal{I}^{(3)} \equiv 6\,\mathcal{I}^{(3)}_{\textrm{h}} + 2\,\mathcal{I}^{(3)}_{\textrm{gf}}+8\,\mathcal{I}^{(3)}_{\textrm{f}} 
= 8\pi^2-96\zeta_{3}-{14\pi^4 \over 15} \, .
\end{aligned}
\ee
Plugging these values into~(\ref{PolIS}), one immediately reproduces the result~(\ref{FinalRes}) obtained in Sect.~\ref{LF} 
for the leading logs.

\subsection{Integrals for subleading logs}

Turning to the subleading logs, we find two kinds of integrals. The first class has the following representatives
\be\label{NLLInt1}
\begin{aligned}
\mathcal{J}^{(k)}_{\textrm{h}} &= \pi^3\int du\, {\left(\psi(\ft{1}{2}+iu) + \psi(\ft{1}{2}-iu)-2\psi(1)\right)^k \over \cosh^4{(\pi u)}}\, , \\
\mathcal{J}^{(k)}_{\textrm{gf}} &= \pi^3\int du\, {\left(\psi(\ft{3}{2}+iu) + \psi(\ft{3}{2}-iu)-2\psi(1)\right)^k \over \cosh^4{(\pi u)}}\, , \\
\mathcal{J}^{(k)}_{\textrm{f}} &= \pi^3\int du\, {\left(\psi(1+iu) + \psi(1-iu)-2\psi(1)\right)^k \over \sinh^4{(\pi u)}}\, , \\
\end{aligned}
\ee
where $k$ is a positive integer, running over $k=0, 1, 2$ for three, four and five loops, respectively. Their values are
\be
\begin{aligned}
&\mathcal{J}^{(0)}_{\textrm{h}} = {4\pi^2 \over 3}
\, , \qquad \qquad \qquad \qquad \, \, \, \, \, \, \, \mathcal{J}^{(0)}_{\textrm{gf}} = {4\pi^2 \over 3}
\, , \qquad \qquad \qquad \qquad \qquad \mathcal{J}^{(0)}_{\textrm{f}} = {4\pi^2 \over 3} \, ,\\
&\mathcal{J}^{(1)}_{\textrm{h}} = -{8\pi^2 \over 3}-4\zeta_{3}
\, , \qquad \qquad \, \, \, \, \, \, \, \, \, \, \mathcal{J}^{(1)}_{\textrm{gf}} = -{8\pi^2 \over 3}-4\zeta_{3}+{22\pi^4 \over 45}
\, , \qquad \, \, \, \, \, \, \, \, \mathcal{J}^{(1)}_{\textrm{f}} =-{8\pi^2 \over 3}-4\zeta_{3}+{11\pi^4 \over 45}\, , \\
&\mathcal{J}^{(2)}_{\textrm{h}} =  {32\pi^2 \over 3}+16\zeta_{3}-{22\pi^4 \over 45}\,, \qquad \mathcal{J}^{(2)}_{\textrm{gf}} 
= {32\pi^2 \over 3}+16\zeta_{3}+{22\pi^4 \over 45}-{32\pi^2\zeta_{3}\over 3}-24\zeta_{5}\, , \\
&\mathcal{J}^{(2)}_{\textrm{f}} = {32\pi^2 \over 3}+16\zeta_{3}-{22\pi^4 \over 45}-{16\pi^2\zeta_{3} \over 3}-12\zeta_{5}\, . \\
\end{aligned}
\ee
In the present case the weighted sums have an explicit minus sign for the fermion contribution, yielding
\be\label{J}
\begin{aligned}
&\mathcal{J}^{(0)} \equiv 6\, \mathcal{J}^{(0)}_{\textrm{h}} +2\, \mathcal{J}^{(0)}_{\textrm{gf}} - 8\, \mathcal{J}^{(0)}_{\textrm{f}} 
= 0 \, ,\\
&\mathcal{J}^{(1)} \equiv 6\, \mathcal{J}^{(1)}_{\textrm{h}} +2\, \mathcal{J}^{(1)}_{\textrm{gf}} - 8\, \mathcal{J}^{(1)}_{\textrm{f}} 
= - {44\pi^4 \over 45}\, , \\
&\mathcal{J}^{(2)} \equiv 6\, \mathcal{J}^{(2)}_{\textrm{h}} +2\, \mathcal{J}^{(2)}_{\textrm{gf}} - 8\, \mathcal{J}^{(2)}_{\textrm{f}} 
= {88\pi^4 \over 45}+{64\pi^2\zeta_{3} \over 3}+48\zeta_{5}\, . \\
\end{aligned}
\ee

The second class of integrals is of the form
\be\label{NLLInt2}
\begin{aligned}
\mathcal{K}^{(k)}_{\textrm{h}} 
&= \pi\int du\, {\left(\psi_{2}(\ft{1}{2}+iu) + \psi_{2}(\ft{1}{2}-iu)\right)\left(\psi(\ft{1}{2}+iu) 
+ \psi(\ft{1}{2}-iu)-2\psi(1)\right)^k\over \cosh^2{(\pi u)}}\, , \\
\mathcal{K}^{(k)}_{\textrm{gf}} &= \pi\int du\, {\left(\psi_{2}(\ft{3}{2}+iu) 
+ \psi_{2}(\ft{3}{2}-iu)\right)\left(\psi(\ft{3}{2}+iu) + \psi(\ft{3}{2}-iu)-2\psi(1)\right)^k \over \cosh^2{(\pi u)}}\, , \\
\mathcal{K}^{(k)}_{\textrm{f}} &= \pi\int du\, {\left(\psi_{2}(1+iu) + \psi_{2}(1-iu)\right)\left(\psi(1+iu) 
+ \psi(1-iu)-2\psi(1)\right)^k \over \sinh^2{(\pi u)}}\, . \\
\end{aligned}
\ee
Here we need only $k=0, 1$ at four and five loop order, respectively. We found by an explicit computation 
\be
\begin{aligned}
&\mathcal{K}^{(0)}_{\textrm{h}} = -24\zeta_{3}
\, , \qquad \qquad \, \, \, \, \mathcal{K}^{(0)}_{\textrm{gf}} = -24\zeta_{3}+{4\pi^4 \over 15}
\, , \qquad \qquad  \, \, \, \mathcal{K}^{(0)}_{\textrm{f}} = 24\zeta_{3} -{2\pi^4 \over 15} \, ,\\
&\mathcal{K}^{(1)}_{\textrm{h}} =  48\zeta_{3} + {2\pi^4 \over 9}
\, , \qquad \, \, \, \mathcal{K}^{(1)}_{\textrm{gf}} = {8\pi^2 \over 3}+64\zeta_{3}+{22\pi^4\over 45}
-{16\pi^2\zeta_{3}\over 3}-88\zeta_{5} \, , \\
&\mathcal{K}^{(1)}_{\textrm{f}} = -48\zeta_{3} - {2\pi^4 \over 9}+{8\pi^2\zeta_{3} \over 3}+44\zeta_{5}\, ,
\end{aligned}
\ee
such that we weighted sums read
\be\label{K}
\begin{aligned}
&\mathcal{K}^{(0)} \equiv 6\, \mathcal{K}^{(0)}_{\textrm{h}} + 2\, \mathcal{K}^{(0)}_{\textrm{gf}} 
+ 8\, \mathcal{K}^{(0)}_{\textrm{f}} =  -{8\pi^4 \over 15}\, , \\
&\mathcal{K}^{(1)} \equiv 6\, \mathcal{K}^{(1)}_{\textrm{h}} +2\, \mathcal{K}^{(1)}_{\textrm{gf}}
+8\, \mathcal{K}^{(1)}_{\textrm{f}}  =  {16\pi^2 \over 3}+32\zeta_{3} + {8\pi^4 \over 15}+{32\pi^2\zeta_{3} \over 3}+176\zeta_{5}\, .
\end{aligned}
\ee

\subsection{Other types of integrals}

Last but not least, for the five-loop computation of the next-to-next-to-leading logs one requires computing of various 
miscellaneous integrals. We list them below with their final values. First,
\be\label{NNLLInt1}
\begin{aligned}
&\mathcal{L}_{\textrm{h}} = \pi\int du\, {\psi_{4}(\ft{1}{2}+iu) + \psi_{4}(\ft{1}{2}-iu) \over \cosh^2{(\pi u)}} 
= -480\zeta_{5}\, , \\
&\mathcal{L}_{\textrm{gf}} = \pi\int du\, {\psi_{4}(\ft{3}{2}+iu) + \psi_{4}(\ft{3}{2}-iu) \over \cosh^2{(\pi u)}} 
= -480\zeta_{5}+{32\pi^6 \over 63}\, , \\
&\mathcal{L}_{\textrm{f}} = \, \, \pi\int du\, {\psi_{4}(1+iu) + \psi_{4}(1-iu) \over \sinh^2{(\pi u)}} 
\, = \, \, \, \, \, 480\zeta_{5} - {16\pi^6 \over 63} \, , \\
\end{aligned}
\ee
with the sum equal to
\be\label{L}
\mathcal{L} \equiv 6\, \mathcal{L}_{\textrm{h}} + 2\, \mathcal{L}_{\textrm{gf}} +8\, \mathcal{L}_{\textrm{f}} = -{64\pi^6 \over 63}\, .
\ee
Next,
\be\label{NNLLInt2}
\begin{aligned}
&\mathcal{M}_{\textrm{h}} = \pi^3\int du\,  {\psi_{2}(\ft{1}{2}+iu) + \psi_{2}(\ft{1}{2}-iu) \over \cosh^4{(\pi u)}} 
= -16\pi^2\zeta_{3}-80\zeta_{5}\, , \\
&\mathcal{M}_{\textrm{gf}} = \pi^3\int du\,  {\psi_{2}(\ft{3}{2}+iu) + \psi_{2}(\ft{3}{2}-iu) \over \cosh^4{(\pi u)}} 
= -16\pi^2\zeta_{3}-80\zeta_{5}+{248\pi^6 \over 945}\, , \\
&\mathcal{M}_{\textrm{f}} = \, \, \pi^3\int du\,  {\psi_{2}(1+iu) + \psi_{2}(1-iu) \over \sinh^4{(\pi u)}}  
\, =\,  -16\pi^2\zeta_{3}-80\zeta_{5}+{124\pi^6 \over 945}\, , \\
\end{aligned}
\ee
with the weighted superposition where fermion are accompanied by an explicit minus sign given by
\be\label{M}
\mathcal{M} \equiv 6\, \mathcal{M}_{\textrm{h}} + 2\, \mathcal{M}_{\textrm{gf}} -8\, \mathcal{M}_{\textrm{f}} 
= -{496\pi^6 \over 945}\, .
\ee
Third,
\be\label{NNLLInt3}
\begin{aligned}
&\mathcal{N}_{\textrm{h}}  = \pi^5\int du\,  {\left(\psi(\ft{1}{2}+iu) + \psi(\ft{1}{2}-iu)-2\psi(1)\right) \over \cosh^6{(\pi u)}} 
= -4\pi^2\zeta_{3}-{32\pi^4 \over 15}-4\zeta_{5}\, , \\
&\mathcal{N}_{\textrm{gf}}  = \pi^5\int du\,  {\left(\psi(\ft{3}{2}+iu) + \psi(\ft{3}{2}-iu)-2\psi(1)\right) \over \cosh^6{(\pi u)}} 
= -4\pi^2\zeta_{3}-{32\pi^4 \over 15}-4\zeta_{5}+{382\pi^6 \over 945}\, , \\
&\mathcal{N}_{\textrm{f}}  =\, \,  \pi^5\int du\,  {\left(\psi(1+iu) + \psi(1-iu)-2\psi(1)\right) \over \sinh^6{(\pi u)}}  
\, \, = \, \, \, \, \, 4\pi^2\zeta_{3}+ {32\pi^4 \over 15}+4\zeta_{5}-{191\pi^6 \over 945}\, ,
\end{aligned}
\ee
resulting in
\be\label{N}
\mathcal{N} \equiv 6\, \mathcal{N}_{\textrm{h}} + 2\, \mathcal{N}_{\textrm{gf}} +8\, \mathcal{N}_{\textrm{f}} 
= -{764\pi^6 \over 945}\, .
\ee
Finally, we have the set of integrals
\be\label{NNLLInt4}
\begin{aligned}
&\mathcal{O}_{\textrm{h}} = \pi^5\int du\,  {\cosh{(2\pi u)}\left(\psi(\ft{1}{2}+iu) 
+ \psi(\ft{1}{2}-iu)-2\psi(1)\right) \over \cosh^6{(\pi u)}} = -{16\pi^4 \over 5}-4\pi^2\zeta_{3}+4\zeta_{5}\, , \\
&\mathcal{O}_{\textrm{gf}} = \pi^5\int du\,  {\cosh{(2\pi u)}\left(\psi(\ft{3}{2}+iu) 
+ \psi(\ft{3}{2}-iu)-2\psi(1)\right) \over \cosh^6{(\pi u)}} = -{16\pi^4 \over 5}-4\pi^2\zeta_{3}+4\zeta_{5}+{542\pi^6\over 945}\, , \\
&\mathcal{O}_{\textrm{f}} = \, \, \pi^5\int du\,  {\cosh{(2\pi u)}\left(\psi(1+iu) 
+ \psi(1-iu)-2\psi(1)\right) \over \sinh^6{(\pi u)}}  \, \, = -{16\pi^4 \over 5}-4\pi^2\zeta_{3}+4\zeta_{5}+{271\pi^6\over 945}\, ,
\end{aligned}
\ee
yielding (again with a minus sign for the fermion in this composition)
\be\label{O}
\mathcal{O} \equiv 6\, \mathcal{O}_{\textrm{h}} + 2\, \mathcal{O}_{\textrm{gf}} -8\, \mathcal{O}_{\textrm{f}} 
= -{1084\pi^6 \over 945}\, .
\ee

\end{document}